\def\be{\begin{equation}}
\def\ee{\end{equation}}
\def\bea{\begin{eqnarray*}}
\def\eea{\end{eqnarray*}}

\def\Spitzer{\textit{Spitzer}}
\def\Chandra{\textit{Chandra}}
\def\Herschel{\textit{Herschel}}

\documentclass[useAMS,usenatbib]{mn2e}
\usepackage{subfigure}
\usepackage{graphicx}
\usepackage{flafter}
\usepackage{lscape}
\usepackage{amssymb}
\begin{document}

\title[Redshift and Nature of AzTEC GOODS-South sources]{Deep 1.1\,mm-wavelength imaging of the GOODS-South field by AzTEC/ASTE -- II. Redshift distribution and nature of the submillimetre galaxy population}

\author[M.~S.~Yun et al.]{
Min S.~Yun,$^1$
K.~S.~Scott,$^2$
Yicheng~Guo,$^1$
I.~Aretxaga,$^3$ 
M.~Giavalisco,$^1$
\newauthor
J. E.~Austermann,$^4$ 
P.~Capak,$^5$
Yuxi~Chen,$^1$
H.~Ezawa,$^6$ 
B.~Hatsukade,$^7$ 
\newauthor
D. H.~Hughes,$^3$ 
D.~Iono,$^8$ 
S.~Johnson,$^1$
R.~Kawabe,$^8$ 
K.~Kohno,$^9$ 
\newauthor
J.~Lowenthal,$^{10}$
N.~Miller,$^{11}$
G.~Morrison,$^{12,13}$
T.~Oshima,$^8$ 
T. A.~Perera,$^{14}$ 
\newauthor
M.~Salvato,$^{15}$
J.~Silverman,$^{16}$
Y.~Tamura,$^9$ 
C. C.~Williams,$^1$ 
and
G. W.~Wilson,$^1$ 
\\
$^1$Department of Astronomy, University of Massachusetts, Amherst, MA 01003, USA\\
$^2$Department of Physics \& Astronomy, University of Pennsylvania, 209 South 33rd Street, Philadelphia, PA 19104, USA\\
$^3$Instituto Nacional de Astrof\'{i}sica, \'{O}ptica y Electr\'{o}nica, Tonantzintla, Puebla, M\'{e}xico\\
$^4$Center for Astrophysics and Space Astronomy, University of Colorado, Boulder, CO 80309, USA\\
$^5$Spitzer Science Center, 314-6 California Institute of Technology, 1200 East California Boulevard, Pasadena, CA 91125, USA\\
$^6$ALMA Project Office, National Astronomical Observatory of Japan, 2-21-1 Osawa, Mitaka, Tokyo 181-8588, Japan\\
$^7$Department of Astronomy, Kyoto University, Kyoto 606-8502, Japan\\
$^8$Nobeyama Radio Observatory, National Astronomical Observatory of Japan, Minamimaki, Minamisaku, Nagano 384-1305, Japan\\
$^9$Institute of Astronomy, University of Tokyo, 2-21-1 Osawa, Mitaka, Tokyo 181-0015, Japan\\
$^{10}$Smith College, Northampton, MA 01063, USA\\
$^{11}$Department of Astronomy, University of Maryland, College Park, MD 20742, USA\\
$^{12}$Institute for Astronomy, University of Hawaii, Manoa, HI 96822, USA\\
$^{13}$Canada-France-Hawaii Telescope Corp., Kamuela, HI 96743, USA\\
$^{14}$Department of Physics, Illinois Wesleyan University, Bloomington, IL 61701, USA\\
$^{15}$Max-Planck Institute for Plasma Physics \& Cluster of Excellence, Boltzmann Strasse 2, 85748 Garching Germany\\
$^{16}$Institute for the Physics and Mathematics of the Universe (IPMU), University of Tokyo, Kashiwa 277-8582, Japan\\
}

\date{\today}

\pagerange{\pageref{firstpage}--\pageref{lastpage}} \pubyear{2011}

\maketitle

\label{firstpage}

\begin{abstract}
We report the results of the counterpart identification and a detailed analysis of the physical properties of the 48 sources discovered in our deep 1.1mm wavelength imaging survey of the GOODS-South field using the AzTEC instrument on the Atacama Submillimeter Telescope Experiment (ASTE).  One or more robust or tentative counterpart candidate is found for 27 and 14 AzTEC sources, respectively, by employing deep radio continuum, \Spitzer\ MIPS \& IRAC, and LABOCA 870 \micron\ data.  Five of the sources (10\%) have two robust counterparts each, supporting the idea that these galaxies are strongly clustered and/or heavily confused. 
Photometric redshifts and star formation rates ($SFRs$) are derived by analyzing UV-to-optical and IR-to-radio SEDs. The median redshift of $z_{med}\sim 2.6$ is similar to other earlier estimates, but we show that 80\% of the AzTEC-GOODS sources are at $z\ge2$, with a significant high redshift tail (20\% at $z\ge 3.3$).  
Rest-frame UV and optical properties of AzTEC sources are extremely diverse, spanning 10 magnitude in the $i-$ and $K-$band photometry (a factor of $10^4$ in flux density) with median values of $i=25.3$ and $K=22.6$ and a broad range of red colour ($i-K=$0-6) with an average value of $i-K\approx3$.
These AzTEC sources are some of the most luminous galaxies in the rest-frame optical bands at $z\ge2$, with inferred stellar masses $M_*=$ (1-30) $\times 10^{10} M_\odot$ and UV-derived star formation rates of $SFR_{UV} \gtrsim 10^{1-3} M_\odot$ yr$^{-1}$.  The IR-derived $SFR$, 200-2000 $M_\odot$ yr$^{-1}$, is independent of $z$ or $M_*$.
The resulting specific star formation rates, $SSFR \approx$ 1-100 Gyr$^{-1}$, are 10-100 times higher than similar mass galaxies at $z=0$, and they extend the previously observed rapid rise in the $SSFR$ with redshift to $z=2-5$.  
These galaxies have a $SFR$ high enough to have built up their {\em entire} stellar mass within their Hubble time.  
We find only marginal evidence for an AGN contribution to the near-IR and mid-IR SEDs, even among the X-ray detected sources, and the
derived $M_*$ and $SFR$ show little dependence on the presence of an X-ray bright AGN.

\end{abstract}


\section{INTRODUCTION}
\label{sec:intro} 

Early studies of the far--infrared (FIR) cosmic background indicated that up to half of the cosmic energy density is generated by dusty starbursts and active galactic nuclei \citep{fixsen98,pei99}.  Deep, wide field surveys at 850\micron\ \citep{smail97,barger98,hughes98,eales99,eales00,cowie02,scott02,webb03a,serjeant03,wang04,coppin06} with the Submillimeter Common--User Bolometric Array \citep[SCUBA;][]{holland99} on the James Clerk Maxwell Telescope (JCMT), and later surveys at millimetre wavelengths  \citep{borys03,greve04,laurent05,bertoldi07,greve08,perera08,scott08,austermann10,scott10}, revealed that this IR background is resolved into a large population of discrete individual sources.

Identifying and understanding the nature of these discrete FIR sources (``submillimetre galaxies'' or SMGs) has proven to be challenging because of the low angular resolution of these instruments and the faintness of counterparts in the rest-frame optical and UV-bands \citep[see review by][]{blain02}.  Utilizing sub-arcsec astrometry of interferometric radio continuum data and sensitive spectroscopy using the Keck telescopes, \citet{chapman05} reported spectroscopic redshifts of 73 SMGs culled from earlier SCUBA surveys and concluded that they are massive, young objects seen during their formation epoch, with very high star formation rates at $z>1$. Deep 24 \micron\ band imaging using the Multiband Imaging Photometer for \textit{Spitzer} \citep[MIPS;][]{rieke04} on the \textit{Spitzer} Space Telescope and follow-up spectroscopy using the Infrared Spectrograph \citep[IRS;][]{houck04} have also provided useful insights on the nature and redshifts of additional SMGs \citep{lutz05,menendez07,valiante07,pope08,menendez09,huang09}. However, the use of high resolution radio continuum and MIPS 24 \micron\ images for the counterpart identification suffers from a well-known systematic bias against high redshift ($z\gtrsim3$) sources.  Indeed, a large fraction of the counterpart sources identified using direct interferometric imaging in the mm/submm wavelengths are shown to be extremely faint in nearly all other wavelength bands ($r>26, K>24$) {\it with little or no radio or MIPS 24 \micron\ emission} \citep{iono06,wang07,younger07,younger09}, and high redshift SMGs may have been missed or mis-identified with a foreground source in the earlier studies.  

Obtaining a more complete understanding of the SMG population requires a study of a larger, more uniform sample identified utilizing the deepest available multiwavelength complementary data and a robust counterpart identification method that is less prone to a redshift bias.  In this paper we present the identification of the 48 AzTEC $1.1\,\rm{mm}$ sources found in the deepest survey at mm wavelengths ever carried out in the Great Observatories Origins Deep Survey-South (GOODS-S) field by \citet[][``Paper I'' hereafter]{scott10}.  Several different identification methods are employed simultaneously to complement and to calibrate each other.  A thorough analysis of the counterpart properties and redshift distribution is also carried out as the GOODS-S field represents one of the most widely studied regions of sky with some of the deepest multi-wavelength data: X-ray data from \textit{Chandra} \citep{luo08,xue11,johnson11}, optical to near-IR photometry from the \textit{Hubble Space Telescope} \citep[\textit{HST};][]{giavalisco04a}, \textit{\textit{Spitzer}} IRAC (Dickinson et al. in prep.) and MIPS (Chary et al. in prep.) imaging in the mid-IR, submm imaging at $250-500$~\micron~with the Balloon-borne Large Aperture Submillimeter Telescope \citep[BLAST;][]{devlin09}, and $1.4~\rm{GHz}$ interferometric imaging with the Very Large Array \citep[VLA;][]{kellermann08,miller08}. Extensive spectroscopy of optical sources in this field is also available \citep{lefevre04,szokoly04,mignoli05,vanzella05,vanzella06,norris06,kriek08,vanzella08,popesso09,treister09,wuyts09,balestra10,silverman10,casey11}.  
Including the AzTEC GOODS-North field sources \citep{perera08,chapin09}, our combined AzTEC-GOODS sample includes $\sim80$ SMGs identified using a uniform set of criteria and the deepest multiwavelength data available and offers the best opportunity yet to examine the nature of the SMGs as a population and to verify the conclusions of earlier studies of mostly smaller and often radio-selected samples \citep{lilly99,fox02,ivison02,webb03b,borys04,chapman05,ivison07,clements08}.

\section{COUNTERPART IDENTIFICATION}
\label{sec:ID}

Here we describe the methods of identifying multiwavelength counterparts to the 48 AzTEC GOODS-S (AzTEC/GS hereafter) sources reported by \citet{scott10}.  We adopt the updated AzTEC source positions and photometry derived using the improved point source kernel by \citet{downes11}.   Our counterpart identification relies primarily on three observed multi-wavelength properties: (1) high resolution radio continuum; (2) \textit{Spitzer} MIPS 24 \micron\ photometry; and (3) red colors in the \textit{Spitzer} IRAC bands.  A robust counterpart is identified using a combination of these criteria for most AzTEC sources, and proposed identifications and multi-wavelength photometry for each of the AzTEC 1.1\,mm sources are summarized in Table~\ref{tab:ID} and Table~\ref{tab:fluxes}.  A more detailed discussion of the individual identification and the nature of the individual counterpart candidates are discussed in Appendix~\ref{sec:notes}.

\subsection{Methods} \label{sec:methods}

Since the origin of the millimetre continuum emission detected by the AzTEC instrument is likely reprocessed radiation from dust-obscured starburst or AGN activity, the main data sets we examine for the multi-wavelength counterpart identification are mid-IR data from the \textit{Spitzer} MIPS 24 \micron\ (full width at half maximum angular resolution of $\theta_{\mathrm{FWHM}}\sim 6\arcsec$) and IRAC 3.6 to 8.0 \micron\ band ($\theta_{\mathrm{FWHM}}\sim 2\arcsec$) and deep radio continuum data obtained using the VLA ($\theta_{\mathrm{FWHM}}\sim 2\arcsec$), exploiting the well-known radio-IR correlation for star-forming galaxies \citep[see review by][]{condon92}.  The \textit{Spitzer} IRAC and MIPS images and catalogues used come from the \textit{Spitzer} GOODS\footnote{http://www.stsci.edu/science/goods/}, the FIDEL\footnote{http://ssc.spitzer.caltech.edu/legacy/abs/dickinson2.html}, and the SIMPLE\footnote{http://www.astro.yale.edu/dokkum/simple/} Legacy Surveys.  The radio continuum data used come from the VLA 1.4 GHz deep imaging survey \citep[$\sigma \sim8\mu$Jy;][]{miller08,kellermann08}.  Given their high resolution, the astrometric accuracy of these catalogues are sufficient to identify unique optical and near-IR counterparts in the deep ground-based telescope or HST images when such a counterpart is present.  The $i-$band and $K-$band photometry of the counterpart candidates are also reported from the band-merged GOODS team photometry catalogue (Grogin et al., in prep.) constructed using a template fitting software package TFIT \citep{laidler07} and the MUSYC survey \citep{gawiser06}.

Unlike most previous works, we employed a variable search radius based on the beam size ($\theta_{\mathrm{FWHM}}\sim 30\arcsec$) and the $S/N$ of the AzTEC detection.  Given the modest $S/N$ (typically $\le10$), the positional offset between an AzTEC source and its counterpart is expected to be dominated by the map noise.  This means we can exploit the measured $S/N$ of each detection to constrain the counterpart identification.  
We derive the search radius, $R_S$, listed in Table~\ref{tab:ID} empirically through simulations by injecting artificial sources into the signal map one at a time and measuring the distribution in the input to output source positions as described in Paper~I. For each AzTEC source we select $R_S$ such that there is a 95\% probability that the true position of the source (assumed to be the position of the radio and/or Spitzer\ counterpart) will be within $R_s$ of the AzTEC centroid.

The primary method of identifying AzTEC counterparts is the ``$P-$statistic'' described by \citet{downes86}. This method computes the likelihood of a chance coincidence for each source in the comparison catalogue from the measured catalogue source density and the distance to a given AzTEC source position.  Following previous work, a counterpart with a $P-$statistic less than 0.05 is deemed a ``robust'' identification, while a counterpart with 0.05 $< P <$ 0.20 is considered a ``tentative'' identification. Unlike most works, however, we compute all $P$-statistics based on the number density of {\it all} sources in the comparison catalog, rather than the number density of sources {\it brighter} than the candidate counterpart in question. This means that all candidate counterparts equidistant from the AzTEC centroid will have the same $P$-statistic. This avoids biasing the identifications to the brightest radio and mid-IR sources, which could result in more misidentifications with low-redshift galaxies.

For the radio data, we created a $\ge4\sigma$ catalog using the SAD program in the Astronomical Image Processing System (AIPS)\footnote{http://www.aips.nrao.edu/}. This program builds a catalog iteratively by first identifying bright pixels and then quantitatively testing their significance by fitting the PSF to the surrounding pixel brightness distribution.  By allowing for a collection of connected sources as an acceptable model, this algorithm also provides a good estimate of the radio flux for extended objects as well.  
Submm/mm galaxies are almost always associated with IRAC galaxies with faint but visible radio emission.  The extremely deep Spitzer\ data in these GOODS fields ensures that radio sources without an IRAC counterpart are rare as reported by \citet{kellermann08}, who find that only three out of 266 cataloged radio sources have no apparent counterpart at any other wavelengths.
Taking advantage of this fact, we probe deeper into the radio data by creating a combined IRAC+VLA radio catalog by using the positions of IRAC sources detected with $(S/N)_{3.6}\ge4$ as prior positions. For each IRAC source, we fit a 2-D Gaussian to the radio map at the IRAC position, fixing the FWHM to 1.6\arcsec and 2.8\arcsec in RA and Dec, respectively, based on the best-fit Gaussian to the synthesized beam \citep{miller08}.  We limit the location of the peak to within 2\arcsec of the initial IRAC position. If the best-fit 1.4\,GHz peak emission is $>3\sigma$ of the rms noise in the surrounding region, we include this in our combined IRAC+VLA catalog. This list is cross-checked with the $>4\sigma$ radio catalog created by the SAD program, and we use the SAD catalogue flux estimates where available. The number density of IRAC+VLA sources in this catalog is 8330\,deg$^{-2}$ for the GOODS-S+VLA catalog, and 7860\,deg$^{-2}$ for the shallower SIMPLE+VLA catalog.

 For the MIPS 24 \micron\ catalogues, we use the number density of $(S/N)_{24}\ge4$ sources to compute the P-statistics, which are 45700\,deg$^{-2}$ and 25600\,deg$^{-2}$ for the GOODS-S and FIDEL 24\,\micron\ catalogs, respectively.

The third and an entirely new method we use for identifying SMG counterparts employs their characteristic red IRAC color.   Interferometric imaging studies of SMGs in submillimetre continuum \citep{iono06,wang07,wang11,younger07,younger08a,younger09,hatsukade10,tamura10,ikarashi11} have shown that every source is detected in the IRAC 3.6 \micron\ and 4.5 \micron\ bands at the $\ge1\mu$Jy level, while their radio and MIPS 24 \micron\ counterparts are not always detected in the best available data.  By examining the spectral energy distribution (SED) of these IRAC counterparts, \citet{yun08} showed that SMGs as a population have characteristic red IRAC colors, similar to dust obscured AGN as proposed by \citet{lacy04} and \citet{stern05}.  These SMGs are systematically offset from the color region associated with power-law AGN, however.  Citing theoretical color tracks of dust obscured starbursts, Yun et al. advocated a dust-obscured young stellar population as the origin of the red IRAC color (see their Figs.~1 \& 2).  Objects with red IRAC color are rare ($\sim1$ arcmin$^{-2}$ for $[3.6]-[4.5]\ge0.0$) and distinct from the large number of foreground galaxies with characteristically {\it blue} IRAC colors.  Both of these qualities can be successfully exploited for distinguishing the SMG counterpart candidates.  Several color combinations are proposed by Yun et al., and we adopt here the simplest form, $[3.6]-[4.5]\ge0.0$, since these two bands are the most sensitive and the most robust among the four IRAC bands. In computing P-statistics for the IRAC counterparts, we thus use the number density of IRAC sources with $[3.6]-[4.5]\ge0.0$, $(S/N)_{3.6}>4$, and $S_{3.6}>1\mu$Jy. These are 36900\,deg$^{-2}$ and 31400\,deg$^{-2}$ for the GOODS-S and SIMPLE IRAC catalogs, respectively. 

Examining the \Spitzer\ IRAC and MIPS properties of 73 radio-selected SMGs, \citet{hainline09} reported that an IRAC color selection method similar to what we adopted is more successful in identifying correct counterparts than the IRAC color-magnitude selection method described by \citet{pope06}, but they caution that the density of sources meeting the \citet{yun08} color selection criteria is high enough to diminish the utility of this method.  We adopt a more selective limit of $[3.6]-[4.5]\ge0.0$, which is more effective in reducing the foreground confusion.  In addition, we also employ a $P-$statistic analysis to give our method a more discriminative power.  

\begin{figure*}
\includegraphics[width=16cm]{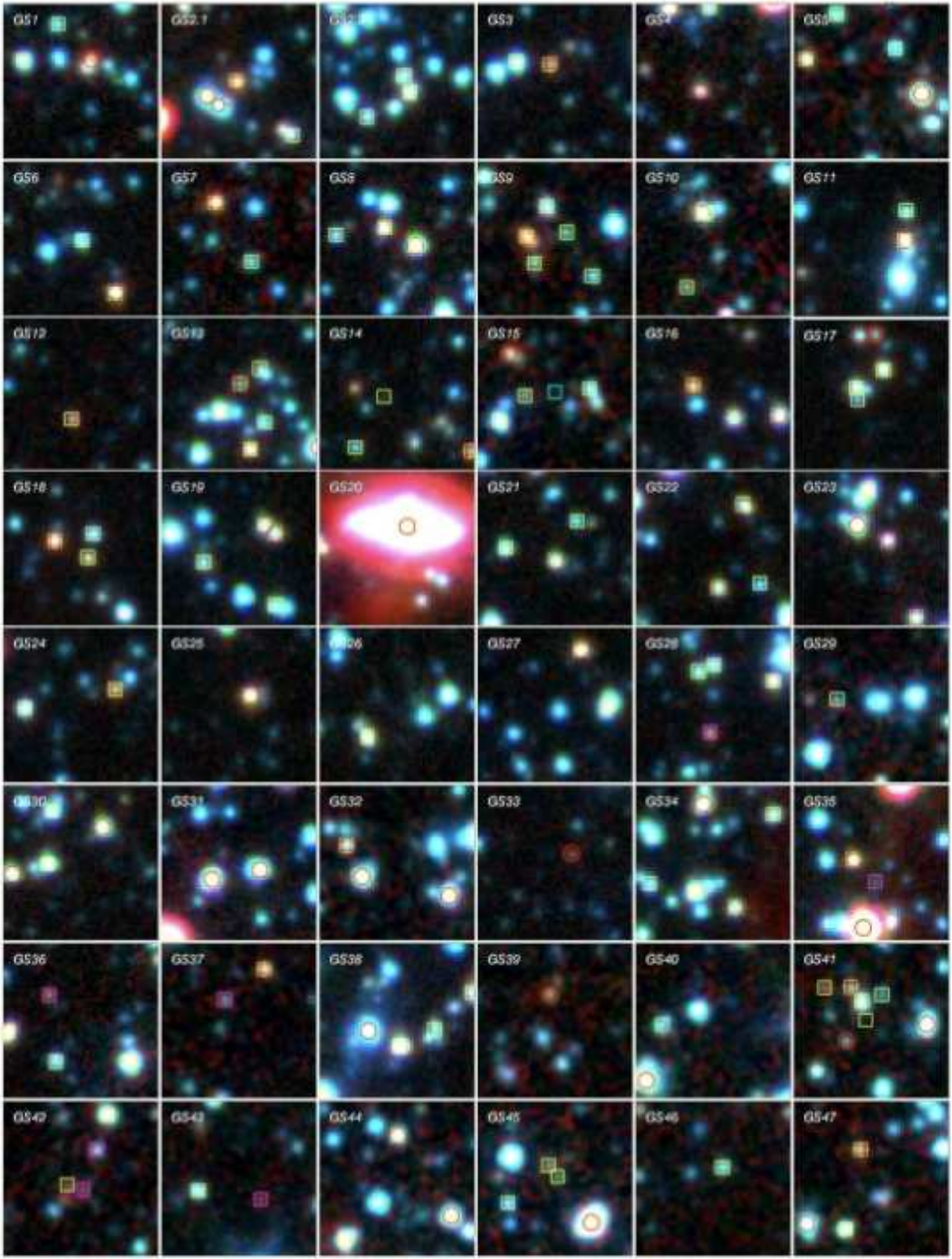}
\caption{Finding charts for the 48 AzTEC/GS sources. These false color images are $30\arcsec \times 30\arcsec$ in size and produced using the IRAC 3.6 \micron\ (blue), 4.5 \micron\ (green), and 8.0 \micron\ (red) band images.  Red circles mark the 1.4 GHz radio sources while yellow squares are MIPS 24 \micron\ sources.  For sources without a plausible radio or MIPS 24 \micron\ counterpart, IRAC sources with red IRAC color ($[3.6]-[4.5]>0$) are identified using magenta squares.}
\label{fig:fchart}
\end{figure*}

\subsection{Counterpart Identification Results} \label{sec:results}

Finding charts for the 48 AzTEC/GS sources (GS2 is split into two) are shown in Fig.~\ref{fig:fchart} in the order they appeared in \citet{scott10}.  Sources meeting the radio, MIPS 24 \micron, and red IRAC color selection criteria are identified in each 30\arcsec\ $\times$ 30\arcsec\ image centered on the AzTEC source position.  A unique counterpart is easily identifiable in about 50\% of the cases while two or more candidates are present in others, requiring a more systematic and quantitative analysis. 

Candidate radio and \textit{Spitzer} counterparts and their computed $P-$statistics are given in Table~\ref{tab:ID}.  All ``robust'' counterparts satisfying $P \le 0.05$ in any of the 3 bands are highlighted in bold-faced letters based on the analysis of the radio ($P_{1.4}$), MIPS 24 \micron\ ($P_{24\mu}$), or the IRAC color ($P_{color}$) properties.  For 21 out of 48 cases, an 870 \micron\ LABOCA source \citep{weiss09} is found within a 10\arcsec\ radius.  Given the extremely low source density at 870 \micron, the likelihood of a chance coincidence is essentially zero.  Therefore, we elevate the status of the 13 AzTEC/GS sources initially classified as only as a ``tentative'' identification based on the 3 bands analysis to ``robust'' by folding in the astrometry of the 19\arcsec\ resolution 870 \micron\ LABOCA Survey data (see Appendix A and Table~\ref{tab:ID}) -- the remaining AzTEC-LABOCA sources are already classified as ``robust''.  We note that the low rate of coincidence between the LABOCA and AzTEC surveys (21/48 = 44\%) can be largely accounted by the low S/N detections of sources in both surveys, although the presence of high redshift ($z>3$) sources detected by AzTEC at 1.1mm \citep[e.g.,][]{eales03} may play a role.
Taking advantage of the available rich multiwavelength database, we provide complete photometry for each source in Table~\ref{tab:fluxes}.

\begin{table*}
\caption{Radio and {\it Spitzer\/} identifications of AzTEC sources (procedure described in Section~\ref{sec:ID}). The counterpart search radius $R_S$ and the likelihood $P$ values are described in detail in the text (\S~\ref{sec:ID}), and  robust counterparts are emphasized in boldface.  Spectroscopic redshifts are given in the column labeled $z_{\mathrm{spec}}$ (references for these measurements are given in Appendix~\ref{sec:notes}).  
}
  \begin{tabular}{lcccccccccc}
   \hline
 AzTEC & $R_S$ & radio coordinate & Dist. & \Spitzer\ coordinate  & Dist. & $P_{1.4}$ & $P_{24\mu}$ & $P_{color}$ & [3.6]$-$[4.5] & $z_{\mathrm{spec}}$ \\
 ID   & ($^{\prime\prime}$) & (J2000) & ($^{\prime\prime}$) & (J2000) & ($^{\prime\prime}$) & & & &(mag)& \\
 \hline
{\bf GS1a}   &  4.7 & {\bf J033211.37-275212.1} &  4.8 &      J033211.36-275213.0  &  4.0 & {\bf 0.045}     & 0.161    & 0.133    & +0.37 & ... \\
{\bf GS2.1a} &  4.5 & {\bf J033219.06-275214.6} &  0.8 & {\bf J033219.05-275214.3} &  0.7 & {\bf 0.001}     & {\bf 0.006 }    & {\bf 0.005}     & +0.38 & ... \\
{\bf GS2.1b} &      & {\bf J033219.14-275218.1} &  3.9 &            ...            &  ... & {\bf 0.030}     &  ...      &  ...      &  ...  & ... \\
     GS2.1c  &      &            ...            &  ... &      J033218.75-275212.7  &  3.9 &  ...      &  0.154    &  ...      & -0.32 & 0.644 \\
     GS2.2a  &  8.7 &            ...            &  ... &      J033216.62-275243.3  &  4.6 &  ...      & 0.212     &  ...      & -0.23 & 1.046 \\
     GS2.2b  &      &            ...            &  ... &      J033216.52-275246.5  &  7.4 &  ...      & 0.457     & 0.390     & +0.26 & ... \\
     GS2.2c  &      &            ...            &  ... &      J033216.75-275249.5  &  8.0 &  ...      &  ...      & 0.439     & +0.10 & ... \\
{\bf GS3a}   &  5.9 & {\bf J033247.99-275416.4} &  4.8 &      J033247.96-275416.3  &  4.6 & {\bf 0.045}$^a$ & 0.211     & 0.174     & +0.37 & ... \\
     GS3b    &      &            ...            &  ... &      J033247.70-275423.5  &  3.8 &  ...      &  ...      & 0.123     & +0.14 & ... \\
{\bf GS4a}   &  6.5 & {\bf J033248.97-274252.0} &  3.2 &      J033248.96-274251.6  &  2.8 & {\bf 0.021}$^a$ &  ...      & 0.070     & +0.28 & ... \\
{\bf GS5a}$^d$    &  7.1 &      J033151.11-274437.5  &  6.4 &      J033151.08-274437.0  &  6.5 & 0.075    & 0.233$^b$ & 0.274$^c$ & +0.23 & {\bf 1.599} \\
     GS5b    &      &      J033152.81-274430.3  & 17.4 &      J033152.80-274429.6  & 17.5 & 0.438     & 0.850$^b$ & 0.903$^c$ & +0.43 & ... \\
{\bf GS6a}$^d$    &  7.5 &      J033225.27-275230.6  & 12.4 &      J033225.25-275230.2  & 12.2 & 0.268$^a$ & 0.809     & 0.737     & +0.45 & ... \\
{\bf GS6b}   &      &            ...            &  ... & {\bf J033225.76-275220.0} &  0.4 &  ...      & {\bf 0.002}     &  ...      & -0.23 & {\bf 1.102} \\
{\bf GS7a}$^d$    &  8.7 &      J033213.84-275600.2  &  8.4 &      J033213.85-275559.9  &  8.7 & 0.126    & 0.366$^b$ & 0.439$^c$ & +0.46 & {\bf 2.676} \\
     GS7b     &      &            ...            &  ... &      J033213.31-275611.5  &  4.9 &  ...      & 0.151$^b$ & 0.168$^c$ & +0.04 & ... \\
{\bf GS8a}   &  8.7 & {\bf J033204.90-274647.4} &  4.4 &      J033204.87-274647.3  &  4.5 & {\bf 0.038}  & 0.203     & 0.168     & +0.33 & {\bf 2.252} \\
     GS8b    &      &            ...            &  ... &      J033205.35-274644.0  &  2.9 &  ...      & 0.089     & 0.072     & +0.17 & ... \\
     GS9a     &  8.7 &      J033303.02-275146.5  &  6.2 &      J033302.99-275146.2  &  5.9 & 0.070     & 0.140$^b$ & 0.232$^c$ & +0.52 & ... \\
     GS9b     &      &            ...            &  ... &      J033302.44-275145.3  &  3.5 &  ...      & 0.090$^b$ & 0.089$^c$ & +0.29 & ... \\
     GS9c    &      &            ...            &  ... &      J033302.90-275151.0  &  5.1 &  ...      &  ...      & 0.179$^c$ & +0.32 & ... \\
{\bf GS10a}$^d$   &  9.0 &      J033207.30-275120.8  &  5.3 &      J033207.27-275120.1  &  5.9 & 0.053     & 0.181$^b$ & 0.233$^c$ & +0.14 & {\bf 2.035} \\
     GS10b   &      &            ...            &  ... &      J033207.09-275128.9  &  3.2 &  ...      &  ...      & 0.077$^c$ & +0.04 & ... \\
     GS11a   &  9.0 &      J033215.33-275037.6  &  6.5 &      J033215.29-275038.3  &  6.8 & 0.081    & 0.404     &  ...      & -0.02 & ... \\
{\bf GS12a}  &  9.0 & {\bf J033229.30-275619.9} &  4.0 &      J033229.29-275619.2  &  3.3 & {\bf 0.032}$^a$ & 0.113    & 0.092     & +0.10 & {\bf 4.762} \\
{\bf GS13a}  &  9.0 & {\bf J033211.94-274615.3} &  2.1 & {\bf J033211.92-274615.2} &  2.2 & {\bf 0.009}$^a$ & {\bf 0.050}     & {\bf 0.041}     & +0.24 & ... \\
     GS13b   &      &      J033211.60-274613.0  &  5.7 &      J033211.56-274613.0  &  6.1 & 0.065$^a$ & 0.338     & 0.283     & +0.02 & 1.039 \\
     GS13c   &      &      J033212.23-274621.6  &  6.3 &      J033212.22-274620.6  &  5.5 & 0.076$^a$ & 0.285     &  ...      & -0.25 & 1.033 \\
{\bf GS14a}$^d$   &  9.0 &            ...            &  ... &      J033234.73-275217.3  &  3.1 &  ...      &  ...      & 0.083     & +0.04 & {\bf 3.640} \\
{\bf GS15a}$^d$   &  9.0 &      J033151.61-274552.1  & 12.7 &      J033151.54-274553.1  & 11.3 & 0.264$^a$ &  ...      & 0.619$^c$ & +0.44 & ... \\
     GS15b   &      &            ...            &  ... &      J033151.36-274601.0  &  5.6 &  ...      &  ...      & 0.215$^c$ & +0.05 & ... \\
     GS15c   &      &            ...            &  ... &      J033150.97-274554.7  &  6.4 &  ...      &  ...      & 0.264$^c$ & +0.00 & ... \\
     GS16a   & 10.5 &      J033238.00-274400.8  &  6.1 &      J033238.00-274400.6  &  6.2 & 0.072$^a$ & 0.345     & 0.290     & +0.53 & 1.719 \\
     GS16b   &      &      J033237.35-274407.8  &  7.9 &      J033237.40-274407.0  &  7.0 & 0.119$^a$ & 0.419     &  ...      & -0.29 & 1.017 \\
{\bf GS17a}  & 10.5 & {\bf J033222.54-274818.2} &  1.8 & {\bf J033222.54-274817.6} &  1.2 & {\bf 0.007}$^a$ & {\bf 0.017}     &  ...      & -0.27 & ... \\
{\bf GS17b}  &      &            ...            &  ... & {\bf J033222.54-274814.9} &  1.5 &  ...      & {\bf 0.026}     & {\bf 0.021}     & +0.19 & ... \\
     GS17c   &      &            ...            &  ... &      J033222.15-274811.3  &  7.0 &  ...      & 0.415     & 0.351     & +0.36 & ... \\
     GS17d   &      &      J033222.53-274804.6  & 11.8 &      J033222.51-274804.6  & 11.8 & 0.245     &  ...      & 0.710     & +0.23 & ... \\
     GS17e   &      &      J033222.26-274804.8  & 12.1 &      J033222.26-274804.3  & 12.5 & 0.254$^a$ & 0.824     & 0.755     & +0.32 & ... \\
{\bf GS18a}  &  9.3 & {\bf J033243.48-274639.5} &  4.2 &      J033243.52-274639.1  &  3.7 & {\bf 0.035}$^a$ & 0.138     & 0.113     & +0.47 & ... \\
     GS18b   &      &      J033243.98-274635.9  &  5.2 &      J033244.01-274635.2  &  5.5 & 0.053$^a$ & 0.288     & 0.240     & +0.31 & 2.688 \\
     GS18c   &      &            ...            &  ... &      J033243.45-274634.3  &  2.3 &  ...      & 0.058     &  ...      & -0.39 & ... \\
     GS19a   & 10.5 &      J033222.92-274125.4  &  7.3 &      J033222.87-274124.9  &  8.0 & 0.102     & 0.509     & 0.438     & +0.22 & ... \\
     GS19b   &      &      J033222.70-274126.7  &  8.7 &      J033222.70-274126.4  &  8.8 & 0.143     &  ...      & 0.501     & +0.36 & ... \\
     GS19c   &      &            ...            &  ... &      J033223.76-274131.5  &  6.6 &  ...      & 0.181$^b$ & 0.280$^c$ & +0.17 & ... \\
{\bf GS20a}  & 10.5 & {\bf J033235.09-275532.6} &  4.6 &      J033235.06-275532.7  &  4.5 & {\bf 0.042}     & 0.200 &  ... & -0.43 & {\bf 0.0369} \\
     GS21a   & 10.4 &      J033247.58-274452.4  &  7.5 &      J033247.59-274452.2  &  7.4 & 0.108     & 0.452     & 0.385     & +0.24 & 1.910 \\
     GS21b   &      &            ...            &  ... &      J033247.29-274444.3  &  2.5 &  ...      & 0.065     & 0.053     & +0.13 & ... \\
     GS22a   & 13.0 &      J033212.56-274305.9  &  7.8 &      J033212.54-274306.1  &  7.9 & 0.116$^a$ & 0.502     & 0.431     & +0.30 & 1.794 \\
     GS22b   &      &            ...            &  ... &      J033212.56-274252.9  &  5.4 &  ...      &  ...      & 0.231     & +0.13 & ... \\
{\bf GS23a}  & 12.2 & {\bf J033221.14-275626.6} &  3.9 &      J033221.12-275626.5  &  4.2 & {\bf 0.030}$^a$ & 0.176     & 0.145     & +0.45 & ... \\
{\bf GS23b}$^d$   &      &      J033221.61-275623.7  &  5.5 &      J033221.58-275623.5  &  5.4 & 0.058     & 0.274     & 0.228     & +0.19 & {\bf 2.277} \\
     GS24a   & 12.2 &      J033234.29-274941.1  &  8.7 &      J033234.26-274940.1  &  9.7 & 0.141$^a$ & 0.649     & 0.571     & +0.35 & ... \\
{\bf GS25a}$^d$   & 13.6 &      J033246.84-275121.0  &  6.8 &      J033246.82-275120.8  &  7.0 & 0.089     & 0.416     & 0.352     & +0.33 & {\bf 2.292} \\
 \hline
\end{tabular}
\label{tab:ID}
(a) Radio sources identified with IRAC positions priors; 
(b) MIPS 24\,$\mu$m flux and P-statistic determined from the FIDEL catalog;
(c) IRAC fluxes and P-statistic determined from SIMPLE catalog;
(d) robust identification based on AzTEC+LABOCA analysis.
\end{table*}

\begin{table*}
  \contcaption{}
  \begin{tabular}{lcccccccccc}
   \hline
 AzTEC & $R_S$ & radio coordinate & Dist. & \Spitzer\ coordinate  & Dist. & $P_{1.4}$ & $P_{24\mu}$ & $P_{color}$ & [3.6]$-$[4.5] & $z_{\mathrm{spec}}$ \\
 ID   & ($^{\prime\prime}$) & (J2000) & ($^{\prime\prime}$) & (J2000) & ($^{\prime\prime}$) & & & &(mag)& \\
 \hline
     GS26a   & 12.2 &            ...            &  ... &      J033215.56-274335.5  &  5.5 &  ...      &  ...      & 0.237     & +0.22 & ... \\
     GS26b   &      &            ...            &  ... &      J033216.41-274341.1  &  7.1 &  ...      &  ...      & 0.366     & +0.25 & 2.331 \\
     GS26c   &      &            ...            &  ... &      J033215.42-274339.7  &  7.2 &  ...      &  ...      & 0.372     & +0.10 & ... \\
     GS27a   & 13.0 &      J033242.09-274141.7  & 13.0 &      J033242.06-274141.3  & 13.6 & 0.291$^a$ & 0.870     & 0.807     & +0.38 & 2.577 \\
     GS28a   & 13.0 &            ...            &  ... &      J033242.78-275212.6  &  2.9 &  ...      & 0.089     & 0.073     & +0.55 & ... \\
     GS28b   &      &            ...            &  ... &      J033242.53-275216.9  &  4.4 &  ...      &  ...      & 0.136$^c$ & +0.07 & ... \\
{\bf GS29a}  & 13.0 &            ...            &  ... & {\bf J033158.67-274500.2} &  3.8 &  ...      & {\bf 0.050}$^b$ &  ...      & -0.43 & {\bf 0.577} \\
     GS29b   &      &            ...            &  ... &      J033159.30-274500.4  &  4.6 &  ...      & 0.122$^b$ &  ...      & -0.10 & 2.340 \\
     GS30a   & 13.5 &      J033220.65-274235.3  &  6.5 &      J033220.66-274234.5  &  7.2 & 0.082$^a$ & 0.439     & 0.373     & +0.19 & ... \\
     GS30b   &      &      J033221.52-274242.5  &  9.0 &      J033221.48-274241.7  &  8.4 & 0.152$^a$ & 0.540     & 0.466     & +0.26 & ... \\
     GS30c   &      &            ...            &  ... &      J033220.90-274236.9  &  4.4 &  ...      &  ...      & 0.160     & +0.29 & ... \\
{\bf GS31a}  & 13.6 & {\bf J033242.76-273927.4} &  2.7 & {\bf J033242.81-273927.1} &  2.1 & {\bf 0.015}$^a$ & {\bf 0.046}     &  ...      & -0.25 & {\bf 1.843} \\
     GS31b   &      &      J033243.47-273929.3  &  7.9 &      J033243.49-273929.1  &  7.9 & 0.118$^a$ & 0.502     &  ...      & -0.36 & 0.733 \\
     GS32a   & 13.5 &      J033308.60-275134.8  &  9.6 &      J033308.61-275134.4  &  9.2 & 0.162     & 0.421$^b$ &  ...      & -0.42 & 0.734 \\
     GS32b   &      &      J033309.93-275131.4  & 10.5 &      J033309.88-275131.0  &  9.8 & 0.191     & 0.456$^b$ &  ...      & -0.12 & ... \\
     GS32c   &      &      J033310.13-275125.1  & 13.4 &      J033310.12-275124.7  & 13.3 & 0.291     & 0.683$^b$ & 0.741$^c$ & +0.08 & ... \\
     GS33a   & 13.0 &      J033248.78-275314.8  &  7.4 &      J033248.78-275314.4  &  7.4 & 0.104$^a$ & 0.457     & 0.390     & +0.34 & ... \\
     GS34a   & 13.5 &      J033229.94-274301.6  & 11.5 &      J033229.95-274301.7  & 11.5 & 0.235$^a$ & 0.768     & 0.693     & +0.09 & 1.356 \\
     GS34b   &      &            ...            &  ... &      J033229.85-274317.7  &  5.8 &  ...      & 0.311     &  ...      & -0.15 & 1.097 \\
     GS34c   &      &            ...            &  ... &      J033229.74-274306.7  &  6.0 &  ...      & 0.326     & 0.273     & +0.14 & ... \\
     GS34d   &      &            ...            &  ... &      J033230.07-274306.8  &  7.9 &  ...      &  ...      & 0.430     & +0.12 & ... \\
     GS34e   &      &            ...            &  ... &      J033229.47-274322.2  &  9.9 &  ...      & 0.664     & 0.586     & +0.15 & 1.609 \\
{\bf GS35a}  & 13.0 & {\bf J033227.21-274052.1} &  2.0 & {\bf J033227.17-274051.7} &  1.6 & {\bf 0.008}     & {\bf 0.027}     & {\bf 0.022}     & +0.37 & ... \\
     GS35b   &      &            ...            &  ... &      J033226.84-274056.1  &  4.9 &  ...      &  ...      & 0.191     & +0.37 & ... \\
     GS36a   & 13.5 &            ...            &  ... &      J033214.42-275515.1  &  6.4 &  ...      &  ...      & 0.304     & +0.68 & ... \\
     GS37a   & 15.0 &      J033256.83-274627.8  & 13.3 &      J033256.79-274626.8  & 12.2 & 0.285$^a$ & 0.631$^b$ & 0.675$^c$ & +0.08 & ... \\
     GS37b   &      &            ...            &  ... &      J033256.79-274612.1  &  4.3 &  ...      &  ...      & 0.132$^c$ & +0.15 & ... \\
     GS38a   & 13.5 &      J033209.71-274248.6  &  7.8 &      J033209.70-274248.2  &  7.5 & 0.116     & 0.463     &  ...      & -0.55 & 0.735 \\
     GS38b   &      &            ...            &  ... &      J033208.74-274248.6  &  7.4 &  ...      & 0.453     & 0.386     & +0.15 & ... \\
{\bf GS39a}$^d$   & 15.0 &      J033154.44-274531.6  &  6.7 &      J033154.39-274530.3  &  7.9 & 0.083     &  ...      & 0.379$^c$ & +0.55 & ... \\
     GS39b   &      &            ...            &  ... &      J033154.54-274539.5  &  2.8 &  ...      & 0.053$^b$ &  ...      & -0.10 & ... \\
     GS40a   & 15.0 &            ...            &  ... &      J033201.15-274635.9  & 10.2 &  ...      & 0.479$^b$ &  ...      & -0.01 & ... \\
     GS41a   &  6.7 &      J033302.78-275653.1  &  8.2 &      J033302.78-275652.8  &  8.0 & 0.120$^a$ &  ...      & 0.386$^c$ & +0.23 & ... \\
{\bf GS41b}$^d$   &      &      J033302.71-275642.5  &  8.6 &      J033302.68-275642.6  &  8.3 & 0.132     & 0.348$^b$ & 0.408$^c$ & +0.30 & ... \\
     GS41c   &      &            ...            &  ... &      J033302.23-275651.4  &  2.7 &  ...      &  ...      & 0.053$^c$ & +0.23 & ... \\
     GS41d   &      &            ...            &  ... &      J033302.55-275644.8  &  5.5 &  ...      & 0.173$^b$ & 0.208$^c$ & +0.15 & ... \\
{\bf GS42a}$^d$   &  6.9 &            ...            &  ... &      J033314.16-275612.0  &  4.6 &  ...      &  ...      & 0.146$^c$ & +0.04 & ... \\
{\bf GS43a}$^d$   &  8.6 &            ...            &  ... &      J033302.90-274432.9  &  4.7 &  ...      &  ...      & 0.157$^c$ & +0.23 & ... \\
     GS44a   & 10.4 &      J033240.84-273752.3  &  9.3 &      J033240.84-273752.6  &  8.9 & 0.151     & 0.389$^b$ &  ...      & -0.07 & ... \\
     GS45a   & 12.2 &      J033218.65-273743.3  & 12.1 &      J033218.58-273742.3  & 12.0 & 0.244$^a$ & 0.620$^b$ &  ...      & -0.32 & ... \\
{\bf GS45b}  &      &            ...            &  ... & {\bf J033219.09-273733.5} &  0.9 &  ...      & {\bf 0.004}$^b$ & {\bf 0.006}$^c$ & +0.18 & ... \\
{\bf GS45c}  &      &            ...            &  ... & {\bf J033219.21-273731.5} &  1.9 &  ...      & {\bf 0.016}$^b$ & {\bf 0.026}$^c$ & +0.21 & ... \\
     GS45d   &      &            ...            &  ... &      J033218.94-273730.0  &  4.3 &  ...      &  ...      & 0.129$^c$ & +0.20 & ... \\
     GS46a   & 13.0 &            ...            &  ... &      J033157.27-275656.2  &  6.2 &  ...      & 0.226$^b$ & 0.255$^c$ & +0.17 & ... \\
{\bf GS47a}$^d$  & 12.2 &      J033208.27-275814.0  &  7.6 &      J033208.23-275813.9  &  7.8 & 0.105$^a$ & 0.280$^b$ & 0.371$^c$ & +0.46 & ... \\
\hline
\end{tabular}
\end{table*}

\begin{table*}
  \caption{Photometry data listed in the same order as the
    identifications in Table~\ref{tab:ID}. All upper-limits are given at a 
    significance of 3-$\sigma$. De-boosted AzTEC 1.1\,mm flux densities are 
    taken from \citet{downes11}. The LABOCA 870 \micron\ photometry comes from \citet{weiss09}.}
  \begin{tabular}{lrrrrrrrrcc}
    \hline
    AzTEC & 1.4\,GHz & 1.1\,mm & 870\,\micron\ & 24\,\micron\ & 
    8\,\micron\ & 5.8\,\micron\ & 4.5\,\micron\ & 3.6\,\micron\ & {\it i}& {\it K}\\ 
    ID & ($\mu$Jy) & (mJy) & (mJy) & ($\mu$Jy) & ($\mu$Jy) & ($\mu$Jy) & 
    ($\mu$Jy) & ($\mu$Jy) & (mag) & (mag) \\  
    \hline
{\bf GS1a}   & $ 32.0 \pm   6.3$ & $6.7^{+0.6}_{-0.7}$ & $ 9.2 \pm 1.2$ & $ 122.0 \pm   5.2$ & $  28.2 \pm   0.7$ & $  20.0 \pm   0.6$ & $  14.60 \pm   0.09$ & $  10.39 \pm   0.06$ & $>25.3$ & $>22.9$ \\ 
{\bf GS2.1a} & $ 50.7 \pm   6.2$ & $6.4^{+0.7}_{-0.6}$ & $ 9.1 \pm 1.2$ & $ 148.0 \pm   4.1$ & $  21.4 \pm   0.5$ & $  14.0 \pm   0.4$ & $  10.62 \pm   0.06$ & $   7.50 \pm   0.04$ & $ 26.1$ & $ 24.2$ \\ 
{\bf GS2.1b} & $ 44.1 \pm   6.2$ &        ...          &       ...      & $          < 13.9$ & $           < 1.3$ & $           < 1.2$ & $            < 0.19$ & $            < 0.11$ & $>27.5$ & $>24.5$ \\ 
     GS2.1c  & $           < 18$ &        ...          &       ...      & $  16.0 \pm   3.2$ & $   6.9 \pm   0.5$ & $   7.8 \pm   0.4$ & $   7.73 \pm   0.06$ & $  10.36 \pm   0.04$ & $ 22.1$ & $ 21.8$ \\ 
     GS2.2a  & $           < 18$ & $4.0^{+0.6}_{-0.7}$ &       ...      & $  62.9 \pm   3.9$ & $  16.2 \pm   0.5$ & $  15.6 \pm   0.4$ & $  17.98 \pm   0.07$ & $  22.28 \pm   0.04$ & $ 22.9$ & $ 21.1$ \\ 
     GS2.2b  & $           < 18$ &       ...           &       ...      & $  83.2 \pm   4.2$ & $  23.8 \pm   0.5$ & $  24.8 \pm   0.4$ & $  20.46 \pm   0.07$ & $  16.10 \pm   0.04$ & $ 26.4$ & $ 23.3$ \\ 
     GS2.2c  & $           < 18$ &       ...           &       ...      & $          < 12.8$ & $  14.3 \pm   0.5$ & $  19.9 \pm   0.4$ & $  16.75 \pm   0.07$ & $  15.30 \pm   0.04$ & $ 26.6$ & $ 21.3$ \\ 
{\bf GS3a}   & $ 40.7 \pm   6.5$ & $4.8^{+0.6}_{-0.5}$ & $ 8.8 \pm 1.2$ & $  49.2 \pm   2.8$ & $  13.0 \pm   0.5$ & $   8.7 \pm   0.4$ & $   6.17 \pm   0.06$ & $   4.38 \pm   0.04$ & $>25.3$ & $>22.9$ \\ 
     GS3b    & $           < 19$ &       ...           &       ...      & $          < 12.6$ & $   3.1 \pm   0.5$ & $   4.2 \pm   0.4$ & $   3.13 \pm   0.07$ & $   2.76 \pm   0.04$ & $>25.3$ & $>22.9$ \\ 
{\bf GS4a}   & $ 25.4 \pm   6.5$ & $5.1^{+0.6}_{-0.6}$ & $ 8.8 \pm 1.2$ & $          < 16.7$ & $  18.6 \pm   0.9$ & $  13.6 \pm   0.7$ & $   9.32 \pm   0.13$ & $   7.21 \pm   0.06$ & $ 26.8$ & $>24.5$ \\ 
{\bf GS5a}   & $ 96.4 \pm   6.7$ & $4.8^{+0.6}_{-0.7}$ & $ 3.9 \pm 1.4$ & $ 521.7 \pm  10.9$ & $  58.2 \pm   0.9$ & $  62.6 \pm   0.9$ & $  73.70 \pm   0.18$ & $  59.67 \pm   0.11$ & $ 23.1$ & $ 20.7$ \\ 
     GS5b    & $111.7 \pm   6.7$ &       ...           &        ...     & $ 282.3 \pm   7.7$ & $  16.1 \pm   0.9$ & $  17.9 \pm   0.9$ & $  14.12 \pm   0.18$ & $   9.50 \pm   0.11$ & $ 25.7$ & $ 22.6$ \\
{\bf GS6a}   & $ 31.0 \pm   6.3$ & $3.6^{+0.5}_{-0.6}$ & $ 5.8 \pm 1.4$ & $ 141.0 \pm   4.0$ & $  27.4 \pm   0.5$ & $  25.6 \pm   0.4$ & $  17.70 \pm   0.06$ & $  11.72 \pm   0.04$ & $ 28.3$ & $ 22.8$ \\ 
{\bf GS6b}   & $           < 18$ &       ...           &       ...      & $  94.2 \pm   3.7$ & $  11.4 \pm   0.5$ & $  12.1 \pm   0.4$ & $  17.00 \pm   0.06$ & $  21.09 \pm   0.04$ & $ 23.5$ & $ 21.1$ \\ 
{\bf GS7a}   & $ 51.2 \pm   6.4$ & $3.8^{+0.6}_{-0.7}$ & $ 9.1 \pm 1.2$ & $ 103.0 \pm   9.3$ & $  22.4 \pm   0.6$ & $  17.7 \pm   0.6$ & $  12.03 \pm   0.12$ & $   7.89 \pm   0.08$ & $>25.3$ & $>22.9$ \\ 
GS7b    & $           < 19$ &       ...           &       ...      & $  60.0 \pm   9.5$ & $   5.0 \pm   0.9$ & $   7.6 \pm   0.7$ & $   9.33 \pm   0.23$ & $   9.03 \pm   0.09$ & $ 24.2$ & $>22.9$ \\ 
{\bf GS8a}   & $ 71.4 \pm   6.6$ & $3.4^{+0.6}_{-0.6}$ & $ 7.5 \pm 1.2$ & $ 620.0 \pm   6.5$ & $  42.9 \pm   0.7$ & $  62.6 \pm   0.6$ & $  50.07 \pm   0.10$ & $  36.85 \pm   0.06$ & $ 24.7$ & $ 21.6$ \\ 
GS8b    & $           < 19$ &       ...           &       ...      & $ 164.0 \pm   4.8$ & $  27.5 \pm   0.6$ & $  20.6 \pm   0.5$ & $  15.73 \pm   0.09$ & $  13.44 \pm   0.05$ & $ 26.6$ & $>22.9$ \\ 
GS9a    & $ 86.8 \pm   6.6$ & $3.6^{+0.6}_{-0.6}$ &       ...      & $ 228.9 \pm  10.3$ & $  27.3 \pm   0.9$ & $  14.9 \pm   0.9$ & $  12.37 \pm   0.17$ & $   7.69 \pm   0.13$ & $ 25.3$ & $>22.9$ \\ 
GS9b    & $           < 19$ &       ...           &       ...      & $ 117.4 \pm   9.7$ & $   3.8 \pm   0.9$ & $   6.3 \pm   0.9$ & $   7.27 \pm   0.17$ & $   5.58 \pm   0.12$ & $>27.5$ & $ 22.6$ \\ 
     GS9c    & $           < 19$ &       ...           &       ...      & $          < 23.1$ & $   8.7 \pm   0.9$ & $   8.0 \pm   0.9$ & $   8.06 \pm   0.17$ & $   5.99 \pm   0.13$ & $>25.3$ & $>22.9$ \\ 
{\bf GS10a}  & $ 89.3 \pm   6.4$ & $3.8^{+0.7}_{-0.7}$ & $ 7.6 \pm 1.3$ & $ 214.0 \pm   8.4$ & $  32.0 \pm   0.7$ & $  40.5 \pm   0.9$ & $  47.03 \pm   0.22$ & $  41.21 \pm   0.15$ & $ 23.6$ & $ 21.3$ \\ 
     GS10b   & $           < 19$ &       ...           &       ...      & $          < 22.7$ & $   4.8 \pm   1.0$ & $   6.9 \pm   1.2$ & $   5.73 \pm   0.22$ & $   5.52 \pm   0.15$ & $ 26.8$ & $>22.9$ \\ 
     GS11a   & $ 46.0 \pm   6.4$ & $3.3^{+0.6}_{-0.6}$ &       ...      & $ 117.0 \pm   4.5$ & $  32.5 \pm   0.4$ & $  23.8 \pm   0.3$ & $  22.45 \pm   0.05$ & $  22.89 \pm   0.04$ & $>25.3$ & $>22.9$ \\ 
{\bf GS12a}  & $ 21.0 \pm   6.5$ & $3.1^{+0.6}_{-0.6}$ & $ 5.1 \pm 1.4$ & $  31.6 \pm   5.1$ & $  12.2 \pm   0.5$ & $   7.2 \pm   0.5$ & $   3.89 \pm   0.07$ & $   3.54 \pm   0.04$ & $ 25.2$ & $>22.9$ \\ 
{\bf GS13a}  & $ 22.8 \pm   6.3$ & $3.1^{+0.6}_{-0.6}$ &       ...      & $  34.7 \pm   3.6$ & $   6.2 \pm   0.5$ & $   6.0 \pm   0.4$ & $   5.32 \pm   0.07$ & $   4.24 \pm   0.04$ & $ 26.8$ & $ 22.5$ \\ 
     GS13b   & $ 24.0 \pm   6.3$ &         ...         &       ...      & $ 112.0 \pm   3.8$ & $  13.8 \pm   0.5$ & $  14.5 \pm   0.4$ & $  14.03 \pm   0.07$ & $  13.72 \pm   0.04$ & $ 23.2$ & $ 22.0$ \\ 
     GS13c   & $ 23.7 \pm   6.3$ &         ...         &       ...      & $ 224.0 \pm   3.7$ & $  31.9 \pm   0.5$ & $  33.1 \pm   0.4$ & $  42.70 \pm   0.07$ & $  53.65 \pm   0.05$ & $ 22.5$ & $ 20.2$ \\ 
{\bf GS14a}   & $           < 18$ & $2.9^{+0.6}_{-0.5}$ &  $3.3\pm 1.2$      & $          < 12.8$ & $           < 1.3$ & $           < 1.2$ & $   1.24 \pm   0.06$ & $   1.19 \pm   0.04$ & $ 25.2$ & $ 23.9$ \\ 
{\bf GS15a}  & $ 27.6 \pm   6.5$ & $3.9^{+0.7}_{-0.8}$ & $ 4.2 \pm 1.4$ & $          < 24.2$ & $  16.0 \pm   1.0$ & $  13.2 \pm   0.9$ & $   9.29 \pm   0.18$ & $   6.17 \pm   0.12$ & $>25.3$ & $>22.9$ \\ 
     GS15b   & $           < 19$ &         ...         &       ...      & $          < 24.6$ & $   5.8 \pm   1.0$ & $   5.7 \pm   0.9$ & $   5.71 \pm   0.18$ & $   5.47 \pm   0.12$ & $>25.3$ & $>22.9$ \\ 
     GS15c   & $           < 19$ &         ...         &       ...      & $          < 25.0$ & $           < 2.9$ & $           < 2.6$ & $   1.12 \pm   0.18$ & $   1.12 \pm   0.12$ & $>25.3$ & $>22.9$ \\ 
     GS16a   & $ 30.7 \pm   6.4$ & $2.7^{+0.5}_{-0.6}$ &       ...      & $  46.4 \pm   3.3$ & $  16.4 \pm   0.5$ & $  10.9 \pm   0.4$ & $   8.12 \pm   0.07$ & $   4.98 \pm   0.04$ & $ 26.5$ & $ 25.8$ \\ 
     GS16b   & $ 22.1 \pm   6.4$ &         ...         &       ...      & $ 144.0 \pm   3.5$ & $  15.7 \pm   0.5$ & $  15.9 \pm   0.4$ & $  21.74 \pm   0.07$ & $  28.44 \pm   0.04$ & $ 23.0$ & $ 20.6$ \\ 
{\bf GS17a}  & $ 26.1 \pm   6.3$ & $2.9^{+0.6}_{-0.6}$ &       ...      & $  71.3 \pm   9.6$ & $   5.1 \pm   0.4$ & $   4.6 \pm   0.3$ & $   4.21 \pm   0.05$ & $   5.42 \pm   0.03$ & $ 23.6$ & $ 22.6$ \\ 
{\bf GS17b}  & $           < 18$ &         ...         &       ...      & $  61.9 \pm   7.2$ & $  21.2 \pm   0.4$ & $  26.3 \pm   0.3$ & $  20.19 \pm   0.05$ & $  16.94 \pm   0.03$ & $ 26.0$ & $ 22.8$ \\ 
     GS17c   & $           < 18$ &         ...         &       ...      & $ 200.0 \pm   5.3$ & $  20.9 \pm   0.4$ & $  23.9 \pm   0.3$ & $  16.50 \pm   0.05$ & $  11.79 \pm   0.03$ & $ 26.7$ & $ 22.7$ \\ 
     GS17d   & $ 42.1 \pm   6.2$ &         ...         &       ...      & $          < 12.3$ & $  11.3 \pm   0.4$ & $  12.0 \pm   0.3$ & $   9.07 \pm   0.05$ & $   7.37 \pm   0.03$ & $ 25.2$ & $ 23.6$ \\ 
     GS17e   & $ 37.9 \pm   6.2$ &         ...         &       ...      & $  23.5 \pm   4.1$ & $   9.7 \pm   0.4$ & $   7.1 \pm   0.3$ & $   5.23 \pm   0.05$ & $   3.88 \pm   0.03$ & $ 27.6$ & $ 23.5$ \\ 
{\bf GS18a}  & $ 25.1 \pm   6.3$ & $3.1^{+0.6}_{-0.6}$ & $ 6.4 \pm 1.3$ & $  84.6 \pm   2.9$ & $  13.1 \pm   0.4$ & $  11.7 \pm   0.3$ & $   7.56 \pm   0.05$ & $   4.91 \pm   0.03$ & $ 28.1$ & $>24.5$ \\ 
     GS18b   & $ 20.2 \pm   6.3$ &         ...         &       ...      & $ 126.0 \pm   4.0$ & $  22.2 \pm   0.4$ & $  16.0 \pm   0.3$ & $  10.85 \pm   0.05$ & $   8.19 \pm   0.03$ & $ 24.5$ & $ 23.2$ \\ 
     GS18c   & $           < 18$ &         ...         &       ...      & $  91.4 \pm   2.8$ & $  11.2 \pm   0.4$ & $  12.6 \pm   0.3$ & $  11.84 \pm   0.05$ & $  16.97 \pm   0.03$ & $ 22.8$ & $ 20.8$ \\ 
     GS19a   & $ 34.0 \pm   6.5$ & $2.6^{+0.5}_{-0.6}$ &       ...      & $ 317   \pm  17  $ & $  18.9 \pm   0.5$ & $  25.8 \pm   0.4$ & $  22.97 \pm   0.07$ & $  18.74 \pm   0.04$ & $ 24.2$ & $ 21.7$ \\ 
     GS19b   & $ 40.0 \pm   6.5$ &         ...         &       ...      & $ 149   \pm  17  $ & $  15.9 \pm   0.5$ & $  22.7 \pm   0.4$ & $  17.67 \pm   0.07$ & $  12.72 \pm   0.04$ & $ 26.6$ & $ 21.7$ \\ 
     GS19c   & $           < 19$ &         ...         &       ...      & $ 105.1 \pm  10.7$ & $   9.1 \pm   0.3$ & $  12.3 \pm   0.3$ & $  11.61 \pm   0.06$ & $   9.97 \pm   0.04$ & $26.9$ & $ 21.6$ \\ 
{\bf GS20a}  & $793   \pm  99  $ & $2.7^{+0.6}_{-0.6}$ &       ...      & $4030.0 \pm  44.1$ & $3499.5 \pm   0.6$ & $1113.8 \pm   0.6$ & $ 759.28 \pm   0.11$ & $1131.36 \pm   0.16$ & $ 16.1$ & $ 15.0$ \\ 
     GS21a   & $ 43.6 \pm   6.3$ & $2.7^{+0.6}_{-0.7}$ &       ...      & $ 299.0 \pm   4.0$ & $  16.6 \pm   0.5$ & $  24.0 \pm   0.4$ & $  19.97 \pm   0.08$ & $  15.98 \pm   0.04$ & $ 25.3$ & $ 22.5$ \\ 
     GS21b   & $           < 18$ &          ...        &       ...      & $  27.8 \pm   2.6$ & $   4.8 \pm   0.5$ & $   6.9 \pm   0.4$ & $   8.44 \pm   0.08$ & $   7.51 \pm   0.04$ & $25.9$ & $ >24.5$ \\ 
     GS22a   & $ 34.6 \pm   6.5$ & $2.1^{+0.6}_{-0.6}$ &       ...      & $ 290.0 \pm   4.4$ & $  13.7 \pm   0.5$ & $  15.3 \pm   0.4$ & $  13.23 \pm   0.07$ & $  10.02 \pm   0.04$ & $ 26.4$ & $ 21.5$ \\ 
     GS22b   & $           < 19$ &          ...        &       ...      & $          < 11.4$ & $           < 1.4$ & $   2.7 \pm   0.4$ & $   2.55 \pm   0.07$ & $   2.26 \pm   0.04$ & $ 25.9$ & $ 22.8$ \\ 
{\bf GS23a}  & $ 23.4 \pm   6.5$ & $2.3^{+0.6}_{-0.6}$ &       ...      & $  42.4 \pm   5.9$ & $  19.7 \pm   0.5$ & $  18.9 \pm   0.5$ & $  12.19 \pm   0.07$ & $   8.06 \pm   0.05$ & $ 24.5$ & $>22.9$ \\ 
{\bf GS23b}  & $ 35.2 \pm   6.5$ &          ...        & $ 4.7 \pm 1.4$ & $ 586.0 \pm   6.2$ & $  36.4 \pm   0.5$ & $  47.1 \pm   0.5$ & $  46.86 \pm   0.07$ & $  39.37 \pm   0.05$ & $ 24.0$ & $ 21.2$ \\ 
     GS24a   & $ 18.5 \pm   6.1$ & $2.3^{+0.6}_{-0.6}$ &       ...      & $  90.8 \pm   3.8$ & $  16.4 \pm   0.4$ & $   9.8 \pm   0.3$ & $   6.61 \pm   0.05$ & $   4.79 \pm   0.03$ & $ 25.5$ & $ 24.1$ \\ 
{\bf GS25a}  & $ 89.5 \pm   6.2$ & $1.9^{+0.5}_{-0.6}$ & $ 5.9 \pm 1.3$ & $ 140.0 \pm   3.6$ & $  32.2 \pm   0.5$ & $  24.5 \pm   0.4$ & $  18.81 \pm   0.06$ & $  13.88 \pm   0.04$ & $ 23.9$ & $ 22.6$ \\ 
    \hline
\end{tabular}
\label{tab:fluxes}
\end{table*}

\begin{table*}
  \contcaption{}
  \begin{tabular}{lrrrrrrrrcc}
    \hline
    AzTEC & 1.4\,GHz & 1.1\,mm & 870\,\micron\ & 24\,\micron\ & 
    8\,\micron\ & 5.8\,\micron\ & 4.5\,\micron\ & 3.6\,\micron\ & {\it i}& {\it K}\\ 
    ID & ($\mu$Jy) & (mJy) & (mJy) & ($\mu$Jy) & ($\mu$Jy) & ($\mu$Jy) & 
    ($\mu$Jy) & ($\mu$Jy) & (mag) & (mag) \\  
    \hline
     GS26a   & $           < 18$ & $2.2^{+0.5}_{-0.6}$ &       ...      & $          < 11.3$ & $   2.4 \pm   0.5$ & $   2.7 \pm   0.4$ & $   3.64 \pm   0.07$ & $   2.98 \pm   0.04$ & $25.2$ & $24.6$ \\ 
     GS26b   & $           < 18$ &          ...        &       ...      & $          < 12.1$ & $   9.0 \pm   0.5$ & $   7.1 \pm   0.4$ & $   6.71 \pm   0.07$ & $   5.35 \pm   0.04$ & $ 24.9$ & $ 23.3$ \\ 
     GS26c   & $           < 18$ &          ...        &       ...      & $          < 11.0$ & $  11.5 \pm   0.5$ & $  18.5 \pm   0.4$ & $  23.55 \pm   0.07$ & $  21.52 \pm   0.04$ & $ 25.9$ & $ 21.3$ \\ 
     GS27a   & $ 23.6 \pm   6.5$ & $2.2^{+0.6}_{-0.6}$ &       ...      & $ 171.0 \pm   5.6$ & $  25.3 \pm   0.5$ & $  23.7 \pm   0.5$ & $  16.57 \pm   0.07$ & $  11.67 \pm   0.04$ & $ 24.8$ & $ 24.2$ \\ 
     GS28a   & $           < 18$ & $2.1^{+0.6}_{-0.5}$ &       ...      & $  17.3 \pm   2.6$ & $  10.8 \pm   0.5$ & $   6.9 \pm   0.4$ & $   4.33 \pm   0.06$ & $   2.62 \pm   0.04$ & $ 26.5$ & $ 23.6$ \\ 
     GS28b   & $           < 18$ &          ...        &       ...      & $          < 26.4$ & $           < 1.0$ & $           < 1.0$ & $   1.12 \pm   0.06$ & $   1.05 \pm   0.04$ & $ 26.0$ & $ 24.6$ \\ 
{\bf GS29a}  & $           < 19$ & $2.3^{+0.6}_{-0.6}$ &       ...      & $  54.4 \pm  10.7$ & $  15.3 \pm   1.0$ & $  27.4 \pm   0.9$ & $  32.93 \pm   0.20$ & $  49.07 \pm   0.13$ & $ 20.4$ & $ 19.8$ \\ 
     GS29b   & $           < 19$ &          ...        &       ...      & $  47.4 \pm  10.1$ & $   5.2 \pm   1.0$ & $   6.2 \pm   1.0$ & $   5.51 \pm   0.19$ & $   6.04 \pm   0.13$ & $ 23.8$ & $ 22.5$ \\ 
     GS30a   & $ 37.2 \pm   6.2$ & $1.8^{+0.5}_{-0.6}$ &       ...      & $ 459.0 \pm   6.2$ & $  25.9 \pm   0.5$ & $  35.4 \pm   0.4$ & $  32.99 \pm   0.07$ & $  27.57 \pm   0.04$ & $ 24.6$ & $ 21.0$ \\ 
     GS30b   & $ 24.2 \pm   6.2$ &          ...        &       ...      & $ 316.0 \pm   4.2$ & $  27.6 \pm   0.5$ & $  33.7 \pm   0.4$ & $  34.14 \pm   0.07$ & $  26.96 \pm   0.04$ & $ 26.3$ & $ 21.2$ \\ 
     GS30c   & $           < 18$ &          ...        &       ...      & $          < 10.5$ & $   2.4 \pm   0.5$ & $   2.7 \pm   0.4$ & $   2.20 \pm   0.07$ & $   1.69 \pm   0.04$ & $ 25.7$ & $ 24.2$ \\ 
{\bf GS31a}  & $ 25.1 \pm   6.9$ & $2.2^{+0.7}_{-0.7}$ &       ...      & $ 427.0 \pm   5.6$ & $  31.5 \pm   0.7$ & $  49.3 \pm   0.7$ & $  51.71 \pm   0.11$ & $  64.98 \pm   0.07$ & $ 21.6$ & $ 19.8$ \\ 
     GS31b   & $ 37.5 \pm   6.9$ &          ...        &       ...      & $ 521.0 \pm   6.7$ & $  60.0 \pm   0.8$ & $  75.4 \pm   0.7$ & $  70.27 \pm   0.11$ & $  97.45 \pm   0.07$ & $ 22.5$ & $ 19.6$ \\ 
     GS32a   & $ 30.3 \pm   6.8$ & $2.3^{+0.8}_{-0.8}$ &       ...      & $ 371.1 \pm  11.7$ & $  32.4 \pm   0.9$ & $  47.0 \pm   0.9$ & $  45.47 \pm   0.17$ & $  67.02 \pm   0.12$ & $ 21.6$ & $ 19.8$ \\ 
     GS32b   & $ 34.8 \pm   6.9$ &          ...        &       ...      & $ 265.6 \pm  11.9$ & $  32.8 \pm   0.9$ & $  35.8 \pm   0.9$ & $  57.75 \pm   0.17$ & $  64.39 \pm   0.13$ & $ 22.4$ & $ 20.2$ \\ 
     GS32c   & $ 46.3 \pm   6.9$ &          ...        &       ...      & $ 270.4 \pm  12.2$ & $  12.4 \pm   0.9$ & $  12.2 \pm   0.9$ & $  11.06 \pm   0.17$ & $  10.27 \pm   0.13$ & $ 25.4$ & $>22.9$ \\ 
     GS33a   & $ 28.6 \pm   6.2$ & $2.0^{+0.5}_{-0.6}$ &       ...      & $  16.1 \pm   3.7$ & $   5.1 \pm   0.5$ & $   2.4 \pm   0.4$ & $   1.70 \pm   0.07$ & $   1.24 \pm   0.04$ & $ 26.9$ & $ 24.1$ \\ 
     GS34a   & $ 33.0 \pm   6.3$ & $1.7^{+0.5}_{-0.6}$ &       ...      & $ 173.0 \pm   3.3$ & $  32.9 \pm   0.5$ & $  28.8 \pm   0.4$ & $  35.68 \pm   0.07$ & $  32.72 \pm   0.05$ & $ 23.4$ & $ 20.9$ \\ 
     GS34b   & $           < 19$ &          ...        &       ...      & $  41.4 \pm   3.7$ & $  14.3 \pm   0.5$ & $  16.7 \pm   0.4$ & $  20.34 \pm   0.07$ & $  23.42 \pm   0.04$ & $ 23.5$ & $ 21.2$ \\ 
     GS34c   & $           < 18$ &          ...        &       ...      & $  56.8 \pm   3.3$ & $   5.8 \pm   0.5$ & $   8.1 \pm   0.4$ & $  10.08 \pm   0.07$ & $   8.83 \pm   0.04$ & $ 25.6$ & $ 22.3$ \\ 
     GS34d   & $           < 19$ &          ...        &       ...      & $          < 10.5$ & $   3.5 \pm   0.5$ & $   4.7 \pm   0.4$ & $   4.54 \pm   0.07$ & $   4.05 \pm   0.04$ & $ 24.3$ & $ 23.4$ \\ 
     GS34e   & $           < 19$ &          ...        &       ...      & $  70.1 \pm   3.2$ & $  14.9 \pm   0.5$ & $  17.2 \pm   0.4$ & $  19.86 \pm   0.07$ & $  17.31 \pm   0.04$ & $ 24.4$ & $ 22.1$ \\ 
{\bf GS35a}  & $ 41.3 \pm   6.7$ & $2.1^{+0.6}_{-0.6}$ &       ...      & $ 153.0 \pm   3.8$ & $  22.2 \pm   0.5$ & $  19.8 \pm   0.4$ & $  14.72 \pm   0.07$ & $  10.51 \pm   0.04$ & $ 24.7$ & $ 23.1$ \\ 
     GS35b   & $           < 19$ &          ...        &       ...      & $          < 12.9$ & $   2.4 \pm   0.5$ & $   3.1 \pm   0.4$ & $   2.76 \pm   0.07$ & $   1.96 \pm   0.04$ & $ 25.3$ & $ 23.5$ \\ 
     GS36a   & $           < 19$ & $2.0^{+0.6}_{-0.6}$ &       ...      & $          < 17.4$ & $   7.6 \pm   0.7$ & $   5.2 \pm   0.6$ & $   2.84 \pm   0.10$ & $   1.52 \pm   0.06$ & $>25.3$ & $>22.9$ \\ 
     GS37a   & $ 20.5 \pm   6.4$ & $2.1^{+0.8}_{-0.8}$ &       ...      & $ 112.9 \pm   7.0$ & $  17.1 \pm   1.0$ & $  16.1 \pm   1.0$ & $  15.08 \pm   0.20$ & $  13.95 \pm   0.13$ & $ 25.4$ & $>22.9$ \\ 
     GS37b   & $           < 19$ &          ...        &       ...      & $          < 21.5$ & $           < 3.0$ & $   3.6 \pm   1.0$ & $   2.89 \pm   0.19$ & $   2.52 \pm   0.13$ & $ 25.9$ & $>22.9$ \\ 
     GS38a   & $220.0 \pm   6.5$ & $1.7^{+0.6}_{-0.6}$ &       ...      & $  39.2 \pm   2.6$ & $  34.2 \pm   0.5$ & $  58.1 \pm   0.4$ & $  67.61 \pm   0.08$ & $ 112.41 \pm   0.06$ & $ 21.1$ & $ 18.8$ \\ 
     GS38b   & $           < 19$ &          ...        &       ...      & $ 184.0 \pm   4.9$ & $  15.8 \pm   0.5$ & $  21.2 \pm   0.4$ & $  16.86 \pm   0.07$ & $  14.66 \pm   0.04$ & $>25.3$ & $>22.9$ \\ 
{\bf GS39a}  & $ 37.2 \pm   6.6$ & $1.5^{+0.7}_{-0.7}$ & $ 3.8 \pm 1.4$ & $          < 26.9$ & $   8.4 \pm   1.0$ & $   6.7 \pm   0.9$ & $   4.91 \pm   0.19$ & $   2.96 \pm   0.12$ & $>25.3$ & $>22.9$ \\ 
     GS39b   & $           < 19$ &          ...        &       ...      & $  48.4 \pm   7.4$ & $   5.9 \pm   1.0$ & $   8.5 \pm   0.9$ & $   9.17 \pm   0.18$ & $  10.04 \pm   0.12$ & $ 23.1$ & $ 22.3$ \\ 
     GS40a   & $           < 19$ & $1.8^{+0.6}_{-0.7}$ &       ...      & $ 169.6 \pm   8.3$ & $   7.9 \pm   1.0$ & $  13.7 \pm   1.1$ & $  13.45 \pm   0.20$ & $  13.64 \pm   0.13$ & $ 25.1$ & $ 22.0$ \\ 
     GS41a   & $ 28.2 \pm   7.0$ & $7.2^{+0.9}_{-1.0}$ &      ...       & $          < 22.4$ & $   4.9 \pm   0.9$ & $   6.5 \pm   0.9$ & $   4.35 \pm   0.17$ & $   3.52 \pm   0.12$ & $>25.3$ & $ 23.3$ \\ 
{\bf GS41b}  & $223.0 \pm   7.0$ &       ...           & $12.0 \pm 1.2$ & $  62.0 \pm   6.0$ & $   4.7 \pm   2.4$ & $   6.6 \pm   2.6$ & $   5.35 \pm   0.46$ & $   4.06 \pm   0.34$ & $>25.3$ & $>22.9$ \\ 
     GS41c   & $           < 21$ &       ...           &       ...      & $          < 23.9$ & $           < 2.7$ & $           < 2.8$ & $   1.71 \pm   0.17$ & $   1.39 \pm   0.12$ & $>25.3$ & $>22.9$ \\ 
     GS41d   & $           < 20$ &       ...           &       ...      & $ 212.2 \pm   6.8$ & $  17.1 \pm   0.9$ & $  19.9 \pm   0.9$ & $  18.94 \pm   0.17$ & $  16.55 \pm   0.12$ & $ 24.3$ & $ 22.0$ \\ 
{\bf GS42a}  & $           < 21$ & $9.2^{+1.2}_{-1.4}$ & $14.5 \pm 1.2$ & $          < 26.6$ & $   5.7 \pm   0.9$ & $           < 2.7$ & $   2.50 \pm   0.18$ & $   2.40 \pm   0.13$ & $ 25.4$ & $>22.9$ \\ 
{\bf GS43a}  & $           < 20$ & $6.1^{+1.0}_{-1.1}$ & $ 6.7 \pm 1.3$ & $          < 25.6$ & $   4.0 \pm   0.9$ & $           < 2.7$ & $   1.26 \pm   0.16$ & $   1.02 \pm   0.12$ & $>25.3$ & $>22.9$ \\ 
     GS44a   & $ 43.5 \pm   7.3$ & $3.2^{+0.8}_{-0.8}$ & $^\dagger5.0 \pm 1.4$ & $ 143.1 \pm   9.3$ & $  19.4 \pm   1.0$ & $  17.9 \pm   1.0$ & $  27.54 \pm   0.19$ & $  29.39 \pm   0.14$ & $ 23.7$ & $ 21.5$ \\ 
     GS45a   & $ 33.7 \pm   6.9$ & $4.0^{+1.2}_{-1.1}$ & $ 8.1 \pm 1.2$ & $ 446.3 \pm   9.4$ & $  91.9 \pm   1.1$ & $ 106.6 \pm   1.1$ & $ 140.28 \pm   0.21$ & $ 188.72 \pm   0.15$ & $ 18.6$ & $ 17.8$ \\ 
{\bf GS45b}  & $           < 20$ &       ...           &       ...      & $  64.4 \pm   9.3$ & $   4.3 \pm   1.1$ & $   5.9 \pm   1.1$ & $   5.02 \pm   0.21$ & $   4.24 \pm   0.15$ & $>25.3$ & $>22.9$ \\ 
{\bf GS45c}  & $           < 20$ &       ...           &       ...      & $  48.9 \pm   9.2$ & $   7.3 \pm   1.1$ & $   5.8 \pm   1.1$ & $   5.33 \pm   0.22$ & $   4.41 \pm   0.15$ & $ 26.4$ & $>22.9$ \\ 
     GS45d   & $           < 20$ &       ...           &       ...      & $          < 37.7$ & $   3.8 \pm   1.1$ & $           < 3.3$ & $   2.11 \pm   0.22$ & $   1.75 \pm   0.14$ & $>25.3$ & $ >22.9$ \\ 
     GS46a   & $           < 21$ & $4.8^{+1.4}_{-1.7}$ & $ ^\dagger4.8 \pm 1.4$ & $  74.6 \pm   9.6$ & $   6.7 \pm   0.8$ & $   9.0 \pm   0.9$ & $  10.54 \pm   0.16$ & $   9.01 \pm   0.11$ & $ 25.1$ & $>22.9$ \\ 
{\bf GS47a}  & $ 43.2 \pm   7.0$ & $3.5^{+1.0}_{-1.0}$ & $ 7.3 \pm 1.2$ & $  72.3 \pm  10.6$ & $   9.7 \pm   1.0$ & $   9.6 \pm   0.9$ & $   6.88 \pm   0.18$ & $   4.51 \pm   0.12$ & $>25.3$ & $>22.9$ \\ 
\hline
\end{tabular}
$^\dagger$ The AzTEC and LABOCA centroid positions are offset by a significant amount ($\gtrsim10\arcsec$).
\end{table*}

A robust counterpart is identified for 27 (56\%) out of 48 AzTEC/GS sources using the $P-$statistic analysis combined with the LABOCA comparison. A total of 13, 8, and 5 AzTEC sources have a robust counterpart based solely on the radio, MIPS 24 \micron, or IRAC color analysis, respectively.  An additional 19, 14, and 18 have tentative identifications with $0.05<P<0.20$, respectively.  The robust radio and MIPS 24 \micron\ identification rates of 13/48 (27\%) and 8/48 (17\%) are consistent with other similar studies. 
For example, using similar depth \Spitzer\ data and slightly deeper radio data in the GOODS-North field, \citet{pope06} reported robust identification rates of 21/35 (60\%) and 6/35 (17\%) for the 1.4 GHz radio and MIPS 24 \micron\ data and additional 10 and 6 tentative identifications, respectively.  For the SCUBA Half Degree Extragalactic Survey (SHADES), \citet{ivison07} reported 56\% and 54\% robust identification using much shallower MIPS 24 \micron\ and comparable depth radio data.\footnote{The MIPS 24 \micron\ robust detection rate by \citet{ivison07} is significantly higher than ours or by \citet{pope06}, despite their much shallower data, and this analysis may be in error.  Although the majority of radio-identified sources in Table~\ref{tab:ID} as well as by Pope et al. have a MIPS 24 \micron\ counterpart, the MIPS 24 \micron\ source density is also much higher than the radio, leading to a greater chance-coincidence and thus a higher $P-$value in general.}  The frequency of robust counterpart identification rate using red IRAC color is similar to the MIPS 24 \micron\ identification rate, indicating that their respective candidate source density is comparable.

Five AzTEC/GS sources (10\%) have two robust counterparts each.  This multiple robust candidate identification rate is similar to those found in the GOODS-North field \citep{pope06,chapin09} and the SHADES fields \citep{ivison07,clements08}.  This multiple identification frequency is about 40 times higher than one would expect at random.  A distinct possibility is that AzTEC counterpart sources are intrinsically strongly clustered \citep[see][and references therein]{williams11}, and the $P-$statistic computes implicitly the likelihood that a particular candidate is {\it either the AzTEC counterpart or a close companion}. A strong clustering of SMGs is also expected if they represent a rapid build-up of stellar mass for $\ge M_{*}$ galaxies associated with a $\gtrsim10^{12}M_\odot$ dark matter halo -- see discussions by \citet{blain04}.  Similar SEDs and redshifts of the multiple candidate counterparts for AzTEC/GS19, AzTEC/GS31, and GN19 \citep{pope06} offer further anecdotal evidence for the clustering explanation.  \citet{wang11} has reported two examples where a single SMG is broken up into multiple discrete components when observed at high angular resolution with an interferometer, further supporting the clustering scenario.  Based on the simulations of two large gas-rich galaxies, \citet{hayward11} have suggested that some fraction of SMGs may be such closely interacting pairs just prior to a merger, and such a scenario would certainly boost the pair fraction.  Uncertainties in the parameters chosen for the simulations, such as the details of the progenitors and the microphysics of star formation and gas consumption, make the comparison with the observed statistics difficult.  These new observations should serve as important observational constraints for future modeling studies.

As shown in Table~\ref{tab:fluxes}, only 22 out of 47 AzTEC/GS sources have an 870 \micron\ LABOCA counterpart in the published catalog by \citet{weiss09}.  Utilizing the radio and MIPS 24 micron $P$-statistic and the IRAC 3.6 and 5.8 \micron\ color-magnitude selection by \citet{pope06}, \citet{biggs11} identified 16 secure and 3 tentative counterparts among these 22 sources in common.  In comparison, we identify 16 robust counterparts based on the $P$-statistics alone, and all 19 individual candidates identified by Biggs et al. is either a robust (8) or tentative (11) counterpart in Table~\ref{tab:ID}.  The agreement between our results and theirs is very good mostly because both groups rely heavily on the radio continuum data for the counterpart identification.  

\subsection{Counterpart Identification for SMGs in GOODS-North}
\label{ssec:cpgn}

To improve the statistics of the subsequent analysis, we also apply the same counterpart identification methods to the AzTEC 1.1mm sources identified in the GOODS-North field \citep{perera08}, using the updated positions and photometry for these sources presented in \citet{downes11}. \citet{chapin09} reported one or more robust counterpart to 21 out of 29 AzTEC sources and at least one tentative counterpart for the remainder.  Our analysis, employing slightly different criteria, identifies 16 robust counterpart sources and one or more tentative counterpart to all but two of the remaining sources.  The agreement between Chapin et al. and our work is quite good, as 13 out of 16 robust counterparts we identified were also identified as robust counterparts by Chapin et al.

\section{Redshifts and Spectral Energy Distributions}
\label{sec:zdet}
 
Spectroscopic redshifts, $z_{spec}$, are available for only a small subset ($\sim30\%$) of candidate counterparts despite the extensive redshift surveys that have been conducted in the GOODS-South field over the years \citep{lefevre04,szokoly04,mignoli05,vanzella05,vanzella06,norris06,kriek08,vanzella08,popesso09,treister09,wuyts09,balestra10,silverman10,casey11}.  The primary reason for this is that many of the candidate counterparts are extremely faint in the optical, with a median brightness of $i\sim 25.4$ among those listed in Table~\ref{tab:fluxes} (also see Fig.~\ref{fig:magmag}).  Robustly identified AzTEC counterparts are even fainter as discussed below.  

To learn more about the redshift distribution of these AzTEC sources and their nature, we rely on the extensive database of extremely deep, multi-wavelength broad-band photometry to analyze their SEDs using empirical and theoretical models.  We first examine the optical/near-IR photometry data using standard methods for estimating photometric redshift (``photoz'' hereafter), stellar mass ($M_*$), and star formation rate ($SFR_{UV}$).  We also employ an independent analysis of the IR-to-radio SEDs to derive photometric redshift, IR luminosity, and dust-obscured star formation rate ($SFR_{IR}$).

\begin{figure}
\includegraphics[width=7.4cm]{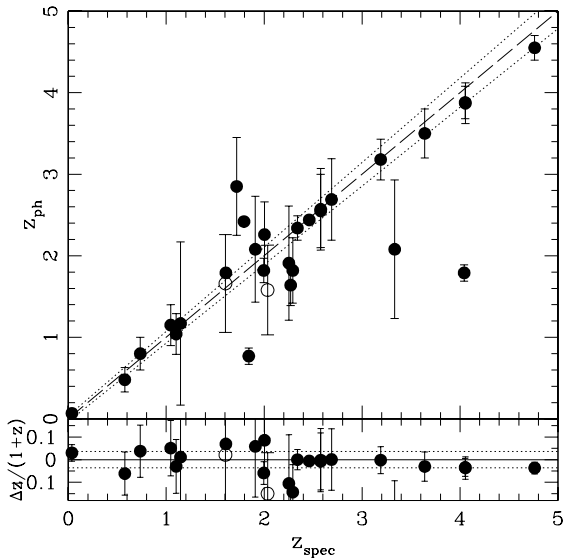}
\caption{Comparison of spectroscopic and photometric redshifts for the candidate AzTEC counterpart sources in the GOODS-South (Table~\ref{tab:ID}) field.  Photometric redshifts based on the shallower MUSYC photometry \citep{gawiser06} are shown in empty symbols.  The two dotted curves represent redshift uncertainties of $\Delta z/(1+z) = 0.036$ (see the text).  The distribution of $\Delta z/(1+z)$ is shown as a function of $z$ in the bottom panel. }
\label{fig:testzopt}
\end{figure}

\subsection{Optical/Near-IR SED Analysis 
\label{sec:OIRSED}}

The photometric redshift of each galaxy is computed by fitting the observed optical and near-IR spectral energy distribution of the galaxies to stellar population synthesis models drawn from the PEGASE 2.0\footnote{http://www2.iap.fr/pegase/} library \citep{fioc97}. The models are shifted in the redshift range of $0<z<7$ with a step size of $\Delta z=0.01$. For each galaxy, the weighted average photoz is derived as: 
\begin{equation}
  z_{photo} = \frac{\int zP(z)dz}{\int P(z)dz}, 
\label{eq:photo-z}
\end{equation}
where $P(z)$ is the probability distribution function of redshift $P(z)\propto exp(-\chi^2(z))$. To evaluate the reliability of our photoz measurements, we compare the photozs with spectroscopic redshifts (specz) of GOODS galaxies with reliable emission line redshifts in Figure~\ref{fig:testzopt}.  We find that the relative error (defined as ${\rm (z_{phot}-z_{spec})/(1+z_{spec})}$) has a zero mean (0.0005) and a very small deviation of 0.036 after $3\sigma$ clipping of the outliers.  The fraction of outliers beyond 3$\sigma$ is 9.9\%.  The means of the relative errors have no significant offset from zero at all redshift bins, especially for our interested range of $1<z<5$. The demonstrated accuracy of our photoz estimation is sufficient to justify a statistical study of the physical properties of our selected galaxies. Derived photometric redshifts of the AzTEC counterpart candidates are listed as $z_{opt}$ in Table~\ref{tab:photoz}.  A blank entry notes that the optical counterpart is undetected or too faint.

Physical properties (stellar mass, SFR, dust reddening) of the galaxies are measured by fitting the observed SEDs with the CB09 (Charlot \& Bruzual, in prep.) theoretical stellar population synthesis models.  The Salpeter IMF with a lower and an upper mass cutoff of ${\rm 0.1 M_{\odot}}$ and ${\rm 100 M_{\odot}}$ are adopted.\footnote{Adopting the Chabrier or Krupa IMF will lower the derived $M_*$ and $SFR$ by a factor of 1.7 -- see \citet{salimbeni09}.}  The galaxy redshift is fixed to its $z_{spec}$ or $z_{opt}$ in this step.  A regular grid of models spanning a wide range of $E(B-V)$, star formation history (characterized by duration $\tau$ and age), and metallicity are examined.  We apply the Calzetti law \citep{calzetti00} for the internal dust extinction correction and follow the method described by \citet{madau95} to account for the IGM opacity. The $\chi^2$ value for each SED model fitting is computed as
\begin{equation}
  \chi^2 = \Sigma_i \frac{(F_{obs,i}-\alpha F_{model,i})^2}{\sigma_{i}^2},
\label{eq:chi2}
\end{equation}
where $F_{obs,i}$, $F_{model,i}$, and $\sigma_{i}$ are the observed flux, model flux, and observational uncertainty in the $i$th band. The normalization factor $\alpha$ is equal to stellar mass if $F_{model,i}$ is normalized to ${\rm 1 M_{\odot}}$ in our pre-computed database. The model with the smallest $\chi^2$ is considered the best-fit model, and its parameters are used to compute the stellar mass and star formation rate $SFR_{UV}$ -- see Table~\ref{tab:photoz}. 

We estimate the systematic uncertainties in the derived quantities using simulations. We generate theoretical SED templates with different redshift, stellar mass, SFR, age, and dust extinction.  In each band, we randomly draw a photometric uncertainty from the error distribution of all sources with the same magnitude from the GOODS parent photometric catalog and perturb the photometry of the template using a Gaussian random deviation with variance equal to the drawn photometric uncertainty.  These steps are repeated 100 times for each template in each band to generate mock SEDs.  Our SED-fitting code is applied to these mock SEDs to derive the systematic uncertainties in the derived quantities.
If redshift and IMF are known, a typical uncertainty in stellar mass from the SED-fitting is about 0.1-0.2 dex for all stellar masses.  The uncertainty in $SFR$ is about 0.1 dex for $SFR>100 M_\odot$ yr$^{-1}$.  If a photoz is used, a typical redshift error of $\delta z/(1+z)=0.05$ translates to a 0.2 dex error in $M_*$ and $SFR$.  Excluding the uncertainty in the IMF, the overall typical uncertainties in $M_*$ and $SFR$ are 0.3 dex and 0.5 dex, respectively.

\begin{table*}
  \caption{
Redshifts and derived properties of the AzTEC GOODS-South sources.}
  \begin{tabular}{lcccccc}
    \hline
    AzTEC ID & $z_{\mathrm{spec}}^a$ & $z_{\mathrm{opt}}^b$ & log $M_*$ & $SFR_{UV}$ & $z_{MR}^c$ & $SFR_{IR}$ \\
    & & & ($M_\odot$) & ($M_\odot$/yr) & & ($M_\odot$/yr) \\
  \hline
{\bf GS1a} & ... & $2.96\pm0.45$ & 11.24 & 94 & $3.56^{+0.66}_{-1.20}$ & 439 \\
{\bf GS2.1a} & ... & $2.13\pm0.60$ & 9.81 & 234 & $3.20^{+0.60}_{-1.10}$ & 500 \\
{\bf GS3a} & ... & ... & ... & ... & $3.09^{+0.55}_{-1.11}$ & \\
{\bf GS4a} & ... & $3.37\pm0.25$ & 10.87 & 75 & $3.53^{+0.57}_{-1.27}$ & 416 \\
{\bf GS5a} & 1.599 & $1.66\pm0.60$ & 11.15 & 1632 & $2.03^{+0.37}_{-0.73}$ & 646 \\
{\bf GS6a} & ... & $2.47\pm0.65$ & 11.37 & 37 & $2.78^{+0.60}_{-0.98}$ & 220 \\
{\bf GS7a} & 2.676 & ... & ... & ... & $2.56^{+0.52}_{-0.92}$ & 638 \\
{\bf GS8a} & 2.252 & $1.91\pm0.70$ & 11.04 & 1553 & $2.11^{+0.41}_{-0.73}$ & 466 \\
GS9a & ... & $3.49\pm0.35$ & 11.01 & 285 & $1.98^{+0.38}_{-0.74}$ & \\
{\bf GS10a} & 2.035 & $1.58\pm0.55$ & 11.27 & 1555 & $2.03^{+0.41}_{-0.75}$ & 350 \\
GS11a & ... & ... & ... & ... & $2.50^{+0.52}_{-0.88}$ & \\
{\bf GS12a} & 4.762 & $4.55\pm0.15$ & 11.28 & 87 & $3.28^{+0.70}_{-1.26}$ & 803 \\
{\bf GS13a} & ... & $2.28\pm0.90$ & 10.20 & 3172 & $2.92^{+0.58}_{-1.10}$ & 250 \\
{\bf GS14a} & 3.640 & $3.50\pm0.30$ & 9.36 & 459 & $>3.0$ & 600 \\
{\bf GS15a} & ... & $3.01\pm0.45$ & 10.00 & 1024 & $3.23^{+0.67}_{-1.13}$ &  416 \\
GS16a & 1.719 & $2.85\pm0.60$ & 9.33 & 114 & $2.67^{+0.55}_{-0.95}$ & 200 \\
{\bf GS17a} & ... & $1.01\pm0.10$ & 9.67 & 3.4 & $2.94^{+0.44}_{-1.08}$ &  \\
{\bf GS17b} & ... & $3.11\pm0.20$ & 11.41 & 14 & $>3.1$ & 330 \\
{\bf GS18a} & ... & ... & ... & ... & $3.00^{+0.56}_{-1.14}$ & \\
GS19a & ... & $1.83\pm0.35$ & 10.93 & 99 & $2.74^{+0.52}_{-1.04}$ & 200 \\
{\bf GS20a} & 0.037 & $0.069\pm0.038$ & 9.52 & 339 & $0.57^{+0.17}_{-0.41}$ & 0.9 \\
GS21a & 1.910 & $2.08\pm0.65$ & 10.77 & 295 & $2.28^{+0.40}_{-0.90}$ & 322 \\
GS22a & 1.794 & $2.42\pm0.05$ & ... & ... & $2.39^{+0.51}_{-0.93}$ & 204 \\
{\bf GS23a} & ... & ... & ... & ... & $2.77^{+0.47}_{-1.07}$ & \\
{\bf GS23b} & 2.277 & $1.64\pm0.25$ & 11.35 & 33 & $2.36^{+0.44}_{-0.90}$ & 393 \\
GS24a & ... & $1.94\pm0.50$ & 9.37 & 130 & $3.04^{+0.74}_{-1.10}$ & 200 \\
{\bf GS25a} & 2.292 & $1.82\pm0.40$ & 10.11 & 467 & $1.52^{+0.32}_{-0.68}$ & 300 \\
GS26a & ... & $3.64\pm0.15$ & 9.61 & 47 & $>2.6$ & 150 \\
GS27a & 2.577 & $2.55\pm0.45$ & 10.02 & 383 & $2.47^{+0.57}_{-0.99}$ & 322 \\
GS28a & ... & $3.29\pm0.65$ & 9.95 & 1140 & $>2.6$ & 177 \\
{\bf GS29a} & 0.577 & $0.48\pm0.15$ & 10.45 & 4.0 & $>2.6$ & \\
GS29b & 2.340 & $2.34\pm0.15$ & 10.36 & 75 & $>2.6$ & 200 \\
GS30a & ... & $1.51\pm0.25$ & 11.02 & 6.0 & $2.29^{+0.56}_{-0.88}$ & 182 \\
{\bf GS31a} & 1.843 & $0.77\pm0.10$ & 11.96 & 0.3 & $2.68^{+0.66}_{-1.00}$ & 222 \\
GS33a & ... & $2.45\pm0.15$ & 10.08 & 6.0 & $2.46^{+0.60}_{-0.96}$ & 200 \\
{\bf GS35a} & ... & $2.96\pm0.35$ & 10.74 & 127 & $2.27^{+0.53}_{-0.91}$ & 508 \\
GS37c & ... & ... & ... & ... & $>2.6$ & \\
{\bf GS39a} & ... & ... & ... & ... & $2.01^{+0.51}_{-0.85}$ & \\
{\bf GS42a} & ... & $2.37\pm0.15$ & 9.56 & 4.1 & $>4.5$ & \\
{\bf GS43a} & ... & ... & ... & ... & $>4.0$ & \\
{\bf GS45b} & ... & $2.07\pm0.25$ & 10.20 & 12 & $>3.4$ & \\
{\bf GS45c} & ... & $2.91\pm0.35$ & 10.46 & 156 & $>3.4$ & \\
GS46a & ... & $1.67\pm0.10$ & 10.47 & 11 & $>3.5$ & \\
{\bf GS47a} & ... & ... & ... & ... & $2.59^{+0.55}_{-0.95}$ & \\
\hline
\end{tabular}
\label{tab:photoz}

(a) $z_{\mathrm{spec}}$ is a spectroscopic redshift. See \S~\ref{sec:notes} for the individual references;
(b) $z_{\mathrm{opt}}^a$ is a photometric redshift derived from the analysis of the optical/NIR SED -- see \S~\ref{sec:OIRSED}; and
(c) $z_{\mathrm{rm}}$ is a new photometric redshift derived from the radio and AzTEC 1100 \micron\ photometry -- see \S~\ref{sec:IRSED}.  The listed uncertainty corresponds to a redshift range that includes 68\% of acceptable fits.\\
\end{table*}

\begin{figure}
\includegraphics[width=7.4cm]{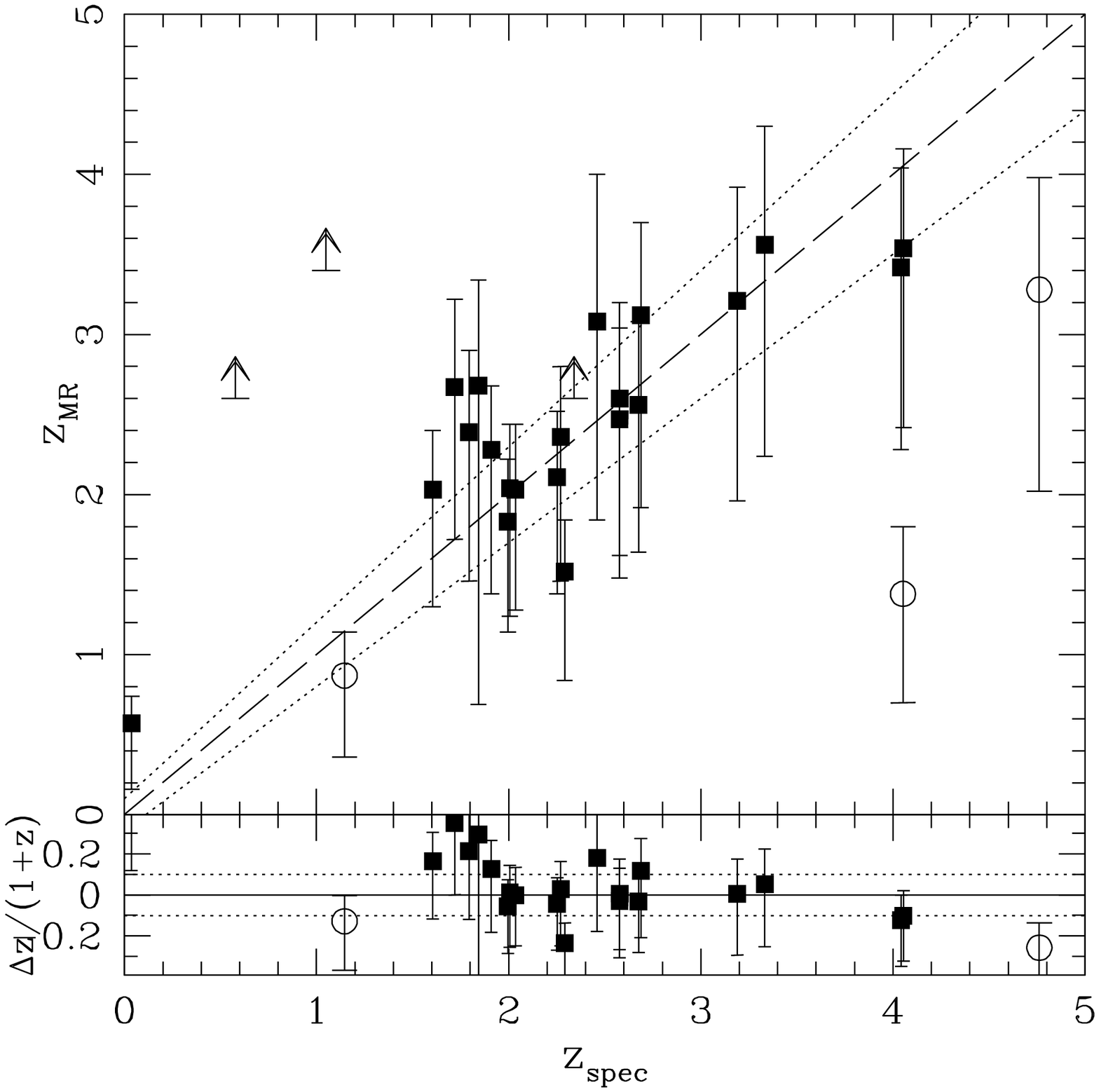}
\caption{Comparison of spectroscopic and millimeter-to-radio (MR) photometric redshifts for the AzTEC sources with a robust counterpart in the GOODS-South (Table~\ref{tab:ID}) and GOODS-North \citep{chapin09} fields. Three objects with clear evidence for AGN activity are identified as empty circles.  The two dotted curves represent redshift uncertainties of $\Delta z/(1+z) = 0.10$.  The distribution of $\Delta z/(1+z)$ is shown as a function of $z$ in the bottom panel.}
\label{fig:testz}
\end{figure}

\subsection{IR/mm/radio SED Analysis 
\label{sec:IRSED}}

We derive an independent estimate of photometric redshift, IR luminosity, and dust-obscured star formation rate ($SFR_{IR}$) by analyzing the observed IR/mm/radio part of the SED.  First, a photometric redshift is derived using an updated version of the photoz analysis method described by \citet{carilli99}.  Noting a remarkably tight correlation between radio and far-IR luminosity for all star forming galaxies \citep[see review by][]{condon92} and the rapid change in the observed flux density ratio between the 850 \micron\ band and the 20 cm radio continuum with redshift, \citet{carilli99} proposed this observed flux density ratio as a robust redshift indicator.  The success of this method rests on the fact that the Rayleigh-Jeans (R-J) part of the dust spectrum rises rapidly with frequency as $S\propto \nu^{3-4}$ while the radio part of the spectrum falls as $S\propto \nu^{-0.75}$, leading to more than two orders of magnitudes change in the observed flux density ratio between $z=0$ and $z\ge2$.  

Subsequent analysis by \citet{hughes02} and \citet{aretxaga03} have shown that incorporating additional photometric measurements in the far-IR to radio bands can improve the redshift estimate, but all of these methods are fundamentally limited by the intrinsic variation in the SED, arising from variations in the nature of the energy source and geometry of dust distribution.  To improve the accuracy of the derived redshift and its uncertainty, we adopted a three times larger set of SED templates, adding 34 new sources with two or more photometry measurements in the R-J part ($150\mu\ < \lambda < 1500\mu$) of the dust SED and at least one radio continuum measurement, mostly from the new study by \citet{clements10}.  We opted to use empirical templates of observed SEDs rather than a library of theoretical templates because there is growing evidence, such as the tightness of the radio-IR correlation, suggesting that nature favors a certain subset of SEDs.   

Another important addition is the use of Monte Carlo simulations to improve the handling of measurement errors, noise bias, and the template variations.  A notable outcome is that the derived redshift uncertainties, listed in Table~\ref{tab:photoz}, are asymmetric about the mean millimeter-to-radio photometric redshift ($z_{MR}$).  Citing the flattening of the IR part of the SED with increasing redshift, \citet{carilli00} have previously noted the asymmetry in the scatter of the ``mean galaxy model'', but {\em in the opposite sense} from the uncertainties in the derived $z_{MR}$.  This actually makes sense since the asymmetry in the mean template and the uncertainty in the derived $z_{MR}$ should be in the opposite sense.  This comparison thus shows that the common practice of quoting the redshift uncertainty using the (sub)mm-radio spectral index method based on the scatter in the Carilli \& Yun template \citep[e.g.,][]{aretxaga07,dannerbauer08,chapin09} is in error.  

The derived $z_{MR}$ is in good agreement with $z_{spec}$ in most cases as shown in Figure~\ref{fig:testz} with $\Delta z/(1+z) \sim 0.1$.  This is not surprising given that the well-know radio-FIR correlation appears to hold among high-redshift IR-selected galaxies \citep[see][]{ivison10b,lacki10}. This estimator may be more accurate for starburst-dominated SMGs since two outliers at $z=4.05$ and $z=4.76$ are known AGNs, and a similar ``radio-excess'' due to an AGN contribution in the radio wavelength has been previously seen among other high redshift QSOs \citep{yun00,yun02}.  In some cases (e.g., GS2.2a, GS29a) the derived $z_{MR}$ is completely inconsistent with their $z_{spec}$, primarily because of their radio non-detection.  Given that the radio-FIR correlation holds for all other objects, a likely explanation is that their optical counterparts are mis-identified, as is expected to happen in a small fraction ($\le5\%$) of cases (see \S~\ref{sec:methods}).

Once the redshift of an AzTEC counterpart is determined, its IR luminosity $L_{IR}$ can be estimated by adopting an SED template most consistent with the observed FIR/mm/radio photometry data.  For the ease of a direct comparison with optically derived $SFR_{UV}$, we fix the redshift of each source to $z_{spec}$ or $z_{opt}$.  Then we use this $L_{IR}$ to derive a dust-obscured star formation rate \citep[$SFR_{IR}$ -- e.g.,][]{kennicutt98}.  We adopt a set of theoretical SED templates for an ensemble of GMCs centrally illuminated by young star clusters by \citet{efstathiou00} that are shown to provide a good fit to a wide range of IR-selected sources at different redshifts \citep[e.g.,][]{efstathiou03,clements08,rowan-robinson10}.  Star formation rates derived from the IR SED fitting, $SFR_{IR}$, are computed directly from the best-fit model star formation history (also assuming the Salpeter IMF) and are summarized in the last column of Table~\ref{tab:photoz}.  Note that $z_{opt}$ and $z_{MR}$ do not always agree well, particularly for the cases where the counterpart is not secure.  Therefore $SFR_{IR}$ is derived primarily for the securely identified AzTEC sources with a $z_{spec}$ or a well-determined $z_{opt}$ only.

\begin{figure}
\includegraphics[width=7.6cm]{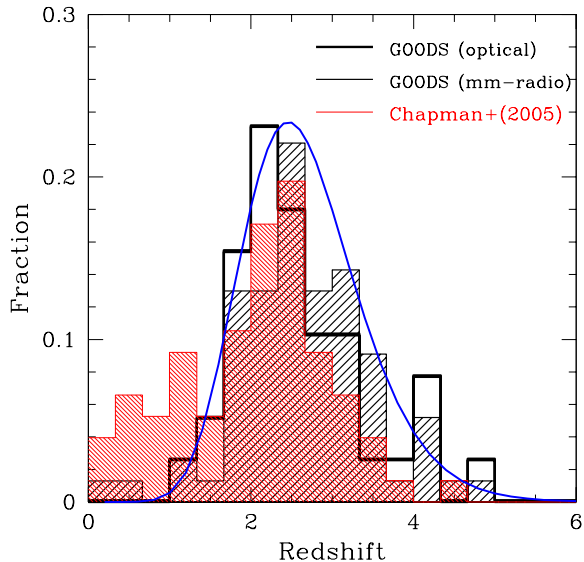}
\caption{Redshift distributions of AzTEC sources in both GOODS fields based on the optical photoz ($z_{opt}$, $N=38$; blank histogram) and the millimetric photoz ($z_{MR}$, $N=74$; hatched histogram) are compared with that of the 76 radio-identified SCUBA sources with spectroscopic redshifts by \citet[][shaded histogram]{chapman05}. The solid blue curve shown is a log-normal distribution as a function of $(1+z)$ with a mean redshift of $z_\mu =2.6$ and $\sigma=0.2$.}
\label{fig:comparez}
\end{figure}

\bigskip
\section{REDSHIFT DISTRIBUTION OF AZTEC-GOODS SOURCES}
\label{sec:redshifts}

\subsection{Derived redshift distribution}

The deep multiwavelength data and the extensive spectroscopic redshift surveys covering the two GOODS fields offer the best opportunity to identify millimetre- and submillimetre-bright galaxies and to construct the most complete redshift distribution yet.  By utilizing the analysis of AzTEC sources in both GOODS fields discussed in \S~\ref{sec:zdet}, we now have the opportunity to augment our understanding of the SMG redshift distribution with improved statistics.

\begin{figure}
\includegraphics[width=7.6cm]{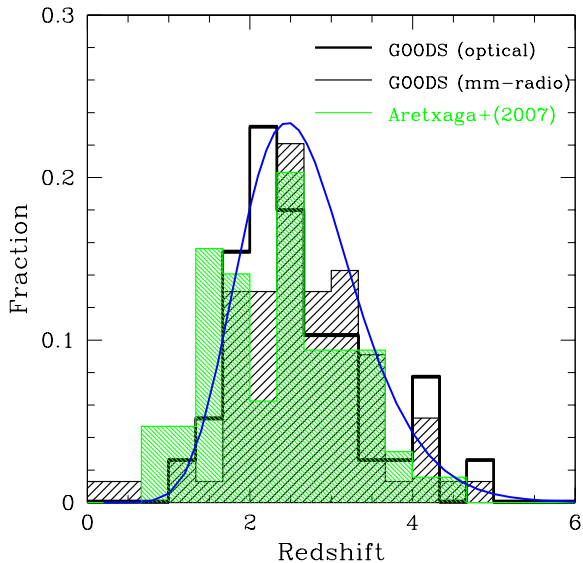}
\caption{Redshift distribution of AzTEC sources in both GOODS fields based on the optical photoz ($z_{opt}$; blank histogram) and the millimetric photoz ($z_{MR}$; hatched histogram) are compared with that of the 64 robustly detected 850 \micron\ sources in SHADES survey by \citet[][shaded histogram]{aretxaga07}.  Because of the shallow radio data in the Subaru/XMM Deep Field (SXDF), only the Lockman Hole sources are included for the Aretxaga et al. SHADES redshift distribution.  The solid blue curve shown is a log-normal distribution as a function of $(1+z)$ with a mean redshift of $z_\mu =2.6$ and $\sigma=0.2$.}
\label{fig:comparez2}
\end{figure}

In Figures~\ref{fig:comparez} and \ref{fig:comparez2}, we show the redshift distributions for the robust AzTEC counterparts in both GOODS fields. We use spectroscopic redshifts when available (22 robustly identified sources) and photometric redshifts otherwise, and we plot separately the distributions determined using $z_{opt}$ and $z_{MR}$ in both Figures.
The redshift distributions based on optical photoz ($z_{opt}$) and millimetric photoz ($z_{MR}$) are qualitatively in a good agreement with each other, while the two methods are each subject to potentially significant systematic uncertainties. These redshift distributions show that 80\% of sources are at $z\gtrsim2$, with 60\% just within the redshift range between $2.0 \lesssim z \lesssim 3.3$.  The relatively small number of robust counterparts with only a lower redshift limit (9 out of 74) assures that the median value of $z_{med}\approx 2.6$ is a robust estimate. In comparison, using a different redshift estimator and analyzing the properties of 29 AzTEC sources in the GOODS-North field only, \citet{chapin09} derived a median redshift of $z=2.7$, in good agreement.

The asymmetric redshift distribution of the AzTEC sources in the two GOODS fields shown in Figures~\ref{fig:comparez} and \ref{fig:comparez2} can be described reasonably well as a log-normal distribution of the form 
$$ f(z)=\frac{1}{(1+z)\sigma\sqrt{2\pi}}e^{-\frac{[ln(1+z)-ln(1+z_\mu)]^2}{2\sigma^2}}.$$  The solid curve shown in both Figures corresponds to a log-normal distribution with $z_\mu = 2.6$ and $\sigma=0.2$ in $ln(1+z)$.  No attempt is made to derive the best fit values of $z_\mu$ and $\sigma$ since some of the redshifts are only lower limits.  Nevertheless, these nominal parameters simultaneously describe the rapid drop-off on the low-$z$ side and the long tail on the high-$z$ side.

\subsection{Comparison with previous studies}

The redshift distribution of the SMG population (and thus their cosmic evolution) is still poorly understood.  A comparison of the redshift distribution derived from the AzTEC GOODS survey sources and those of of previous studies further illustrates this point.  A comparison of the AzTEC sources in the two GOODS fields with that of the radio-selected SCUBA 850 \micron\ sources by \citet[][Figure~\ref{fig:comparez}]{chapman05} gives an immediate impression that the two redshift distributions are substantially different.  In particular, the population of $z\le1.5$ sources present in the Chapman sample is missing in our sample while the AzTEC-GOODS sample shows a broader higher redshift tail.  It is important to understand the underlying causes of this difference since many studies have assumed that the redshift distribution derived by Chapman et al. is consistent with that of the SMG population as a whole \citep[e.g.][]{cooray10,dave10,narayanan10a,vieira10,almeida11,williams11}.  As noted by Chapman et al. and further discussed below (\S~\ref{sec:sample_selection}), these differences may be rooted in the use of radio selection for defining the Chapman sample.  

Another insightful comparison is made in Figure~\ref{fig:comparez2} by examining the redshift distribution of the GOODS AzTEC sources with the photometric redshifts of the 64 robustly detected 850 \micron\ sources in the SHADES survey by \citet{aretxaga07}. The Aretxaga et al. redshift distribution is also largely missing the low-z population, and the agreement with our redshift distribution is better.  If we take into account that 11 out 64 redshifts by Aretxaga et al. are only lower limits with $z_{lim}\le1.5$, the redshift distribution for the AzTEC-GOODS and the SHADES Lockman Hole sources is in fact very good.   

\begin{table*}
  \caption{
Comparison of mean redshifts and probabilities that two redshift distributions are drawn from the same parent distribution derived using the ASURV, as described in Section~\ref{sec:redshifts}. The first row shows the number of redshift data points ($N_d$) and lower limits ($N_l$) used. The second row shows the Kaplan-Meier estimator for the mean and standard deviation of the redshift distribution. The last three rows show, first, the Gehan's Generalized Wilcoxon test probability for each pair, and second, the Logrank test probability.}
  \begin{tabular}{lcccc}
    \hline
     & $z_{opt}$ & $z_{MR}$ & Chapman05 & Aretxaga07 \\
\hline
($N_d$, $N_l$) & (38, 0) & (74, 9) & (76, 0) & (64, 23) \\
$<z>$ & $2.516\pm0.129$ & $2.689\pm0.112$ & $2.000\pm0.104$ & $2.695\pm0.098$ \\
  \hline
$z_{opt}$  & -- & 0.19 / 0.21 & 0.030 / 0.026 & 0.06 / 0.30 \\
$z_{MR}$   &  & -- & 0.0001 / 0.0001 & 0.51 / 0.85 \\
Chapman05  &    &  & -- & 0.0000 / 0.0001 \\
\hline
\end{tabular}
\label{tab:asurv}
\end{table*}

\begin{table}
  \caption{
The K-S test probability that two redshift distributions are drawn from the same distribution, derived using the cumulative distributions of the same samples as in Table~\ref{tab:asurv} but excluding lower limits.}
  \begin{tabular}{l|cccc}
    \hline
          & $z_{opt}$  & $z_{MR}$ & Chapman05 & Aretxaga07 \\
  \hline
$z_{opt}$  &   --     & 0.845 & 0.250 & 0.098 \\
$z_{MR}$   &      &     --        & 0.014 & 0.362 \\
Chapman05  &   &            &      --      & 0.0014 \\
\hline
\end{tabular}
\label{tab:KS}
\end{table}

A more quantitative comparison of the derived redshift distributions is made using the Astronomy Survival Analysis \citep[ASURV;][]{feigelson85} package, which properly takes into account lower redshift limits.\footnote{This analysis assumes that the censored data follow a similar distribution to that of the measured population.}  As summarized in Table~\ref{tab:asurv}, the only pairs of redshift distributions showing non-negligible probability of being drawn from the same parent sample are between $z_{MR}$ and $z_{opt}$ and between $z_{MR}$ and the photometric redshifts of the 850 \micron\ sources in the SHADES survey by \citet{aretxaga07}.  On the other hand, both Gehan's Generalized Wilcoxon test and the Logrank test suggest that there is {\it at most} 3\% probability that the Chapman et al. redshift distribution is consistent with those of the GOODS AzTEC sources or the SHADES 850 \micron\ sources in the Lockman Hole region as analyzed by Aretxaga et al.  The Kaplan-Meier estimator gives the mean redshifts of $<z_{MR}>=2.689\pm0.112$ and $<z_{opt}>=2.516\pm0.129$.  The inclusion of lower limits in the redshift for radio-undetected sources appears to be the primary difference for these estimates (see below).  The mean redshift of the Aretxaga et al. sample is $2.695\pm0.098$, in a good agreement with that of $<z_{MR}>$.  In contrast, the mean redshift of the Chapman et al. sample is $2.000\pm0.104$, significantly lower.

The robustness of these analyses is further tested by conducting a K-S test for the same pairs of redshift distributions but excluding lower redshift limits.  Again, as summarized in Table~\ref{tab:KS}, the hypothesis that the Chapman et al. redshift distribution is identical to the $z_{MR}$ or Aretxaga et al. redshift distribution can be rejected with better than 98\% confidence while the same hypothesis for the other combinations of pairs cannot be rejected.  Nevertheless, both the survival analysis and the K-S tests are giving us consistent results in that the Chapman redshift distribution is substantially different from the SMG redshift distribution derived by us using the GOODS AzTEC sources and that of the Lockman Hole 850 \micron\ sources in SHADES survey by Aretxaga et al.

\subsection{Wavelength-dependent selection bias \label{sec:sample_selection}}

As discussed in some detail by \citet{chapman05}, their radio-selection for a spectroscopic redshift survey is intrinsically biased toward low redshift galaxies and those with cold dust temperature.  The observed radio flux density suffers from a strong {\it positive} k-correction, fading faster with increasing redshift than expected from the inverse square law.  This means the majority of faint ($S_{1.4GHz}\ge$ 30-40 $\mu$Jy) radio sources are at $z\le1$ and only extremely luminous ($P_{1.4GHz}\ge 10^{24}$W Hz$^{-1}$) radio sources are detectable at $z>2$ \citep[e.g.][]{smolcic08,strazzullo10}.  Chapman et al. discussed this foreground confusion and removed about 10\% of the sources from their sample, but the significant number of $z\le1.5$ sources remaining in their sample (but unseen in our sample and Aretxaga et al. sample) suggests that they under-estimated the foreground confusion.  Chapman et al. also considered the effects of rapidly declining sensitivity of the radio data with redshift, but they focused mainly on the dust temperature dependence on radio-IR correlation, assuming that the majority of sources undetected in the radio bands are at the same redshifts as those detected.  The significant high redshift ($z>3$) tail for the AzTEC and SCUBA sources seen in Figures~\ref{fig:comparez} and \ref{fig:comparez2} suggests that the highest redshift sources are missing in radio-selected samples like Chapman et al. because the depth of the available radio data is not sufficient to detect most SMGs at $z\gtrsim3$.  Mapping the true redshift distributions of SMGs will require future complete spectroscopic redshift surveys using instruments such as the Redshift Search Receiver \citep{erickson07} on the Large Millimeter Telescope or the Atacama Large Millimeter Array.

These discussions of k-correction and dust temperature raise another important question as whether SMGs selected at 850 \micron\ and 1100 \micron\ (and as a natural extension at 250-500 \micron\ by the \Herschel\ SPIRE instrument) are systematically different. Since the dust peak passes through these bands at different redshifts, some wavelength-dependent selection effects are expected.  For example, identification of ``drop-out'' objects in these bands has been suggested as a means to identify the highest redshift SMG population \citep[see][]{pope10}.  When we noted the low rate of coincidence between the LABOCA 870 \micron\ sources and our AzTEC survey initially (see S~\ref{sec:results}), this wavelength-dependent selection bias was one of the causes we explored, although we eventually concluded that the low completeness of the both surveys is the primary cause.  The general agreement between the redshift distributions of SCUBA 850 \micron\ selected sources by \citet{aretxaga07} and the AzTEC 1100 \micron\ sources seen in Figure~\ref{fig:comparez2} suggests that the redshift distribution and SEDs of the SMG population is such that the sources identified at these two wavelengths are indeed similar.

\begin{figure}
\includegraphics[width=8.2cm]{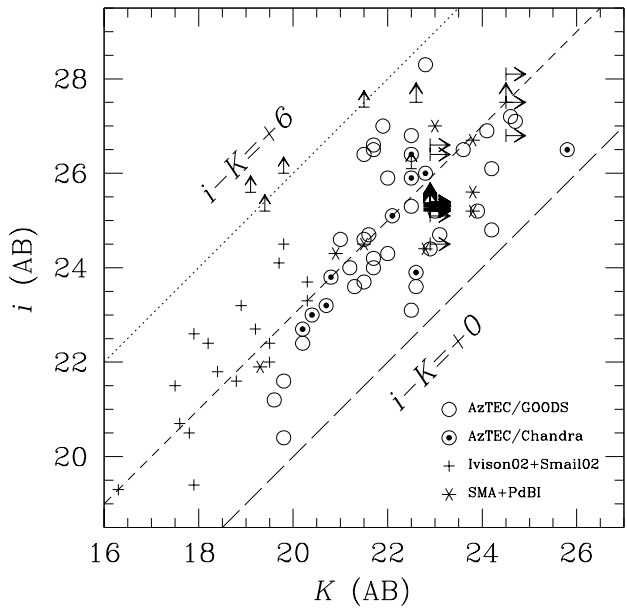}
\caption{Measured $K$-band versus $i$-band magnitudes of the robustly identified AzTEC counterpart sources (circles).  Lines of constant colors with $i-K$ = +0, +3, \& +6 are shown.  Circles with filled dots are AzTEC sources with \Chandra\ X-ray detection in the GOODS-North and GOODS-South fields while empty circles are the sources without X-ray detection.  Crosses represent the 17 ``securely'' identified counterparts to the SCUBA 8-mJy survey \citep{ivison02} and the SCUBA Lens Survey \citep{smail02} sources. Asterisks are SMGs securely identified by interferometric measurements at millimetre and submillimetre wavelengths \citep{iono06,younger07,younger08a,younger09,hatsukade10}.}
\label{fig:magmag}
\end{figure}

\section{OPTICAL AND IR LUMINOSITY AND STAR FORMATION}

\subsection{Not all SMGs are faint and red in the rest-frame UV and optical bands
\label{sec:optical}}

Although SMGs are a recently recognized class of galaxies, their relatively high density (0.1-0.5 arcmin$^{-2}$) and high luminosity ($L_{IR}\gtrsim 10^{12-13} L_\odot$) suggest that they represent a significant component of the general galaxy population and should play an important role in the overall galaxy evolution scenario.  In the ``down-sizing'' scenario \citep{cowie96}, more massive galaxies are thought to have been assembled earlier in cosmic history, presumably with a higher star formation rate (SFR).  Luminous infrared galaxies (LIRGs) and ultraluminous infrared galaxies (ULIRGs) with SFR $\gtrsim10$-100 $M_\odot$ yr$^{-1}$ are the dominant contributor to the cosmic star formation history at $z\sim1$ \citep{lefloch05,magnelli09,magnelli11}, and a significant contribution by SMGs with SFR of $\gtrsim10^{2-3} M_\odot$ yr$^{-1}$ would represent a natural progression at $z>1$. Massive galaxies with stellar mass $M_* \gtrsim 10^{11} M_\odot$ are thought to be already in place by $z\sim2$ \citep{vandokkum08}. Galaxies with even higher SFR might be found at higher redshifts.

What would these SMGs look like in the rest-frame UV and optical bands? And how do they fit into the larger population of high redshift galaxies identified in those more traditional bands?  Optical/UV size, morphology, and luminosity could provide an important test for their origin as merger-driven starbursts \citep[e.g.,][]{narayanan10a} or large disk systems \citep{efstathiou03,kaviani03,hayward11} fueled by a high rate of cosmological gas accretion \citep{keres05,dave10}.   Early studies in the optical and near-IR suggested a diverse population of bright, modest redshift ($z\lesssim1$) and faint, high-redshift ($z\sim2$) galaxies, as reported by \citet{lilly99}, \citet{barger99a,barger00}, \citet{ivison00a}, \citet{fox02}, and \citet{ivison02}.  However, high resolution interferometric imaging studies at millimeter wavelengths \citep{bertoldi00,frayer00,lutz01,dannerbauer02} have shown that the SMG counterparts are often undetected in the optical bands. A study of a large sample of radio-identified SMGs using deeper optical data by \citet{chapman01} showed that their counterparts are indeed quite faint ($I>25$), and Chapman et al. concluded that dust obscuration makes these galaxies essentially invisible in the ultraviolet bands.  This conclusion is not universally accepted, however -- see \citet{ivison02}.  Interestingly, \citet{chapman05} targeted their own sample for spectroscopy using the Keck Telescopes and successfully obtained emission and absorption line redshifts for about 50\% of their sample.

A major motivation for this work is to clear up the confusion about the rest-frame optical and UV properties of SMGs by examining a robustly identified large sample with significantly improved statistics by taking advantage of the deep multiwavelength data available in the GOODS fields.  In Fig.~\ref{fig:magmag} we examine the rest-frame UV and optical properties of SMGs by plotting the measured $i$-band and $K$-band photometry of robustly identified AzTEC sources in the GOODS-South field (this work) and the GOODS-North field \citep{perera08,chapin09} -- also see \citet{pope06}.  Sources identified by the SCUBA 8-mJy survey \citep{ivison02,smail04} and the SCUBA Lens Survey \citep{smail02} are also shown for comparison. A remarkable result is that SMGs span a very broad range of brightness in both $i$- and $K$-band (rest frame $\lambda$ = 240 nm \& 630 nm at $z=2.5$), spreading over 10 magnitudes, or a factor of $10^4$ in flux density.  The apparent brightness of the AzTEC GOODS sources by themselves span over 7 magnitudes with a median brightness of $i=25.3$ and $K=22.6$ when the upper limits are taken into account. Although there is some overlap with the sources identified by the earlier SCUBA 8-mJy survey and the SCUBA Lens Survey, the {\it AzTEC GOODS counterpart sources are systematically fainter by $\sim3$ magnitudes on average.}  Because earlier SMG identification studies relied on $K$-band data too shallow to detect the majority of the AzTEC counterparts in the GOODS-South field, this means earlier works may have missed or mis-identified the counterparts in many cases.  The few but highly secure SMG counterparts identified recently using deeper optical and near-IR data and high resolution interferometric imaging in the millimeter and submillimeter bands \citep[shown as asterisks --][]{iono06,younger07,younger08a,younger09,hatsukade10,wang07,wang11} have a brightness distribution more closely matching that of the AzTEC sources in both GOODS fields.

We also deduce from the observed scatter in Fig.~\ref{fig:magmag} that there is at least a factor of 10 variation in the intrinsic rest frame optical luminosity among these SMGs.    When viewed together with SCUBA-detected sources, these SMGs form a broad color track centered roughly around the $i-K=+3$ line (short-dashed line), which is quite red compared with field galaxies.  Some sources show a relatively flat color ($i-K=+1$) while there are others with extremely faint $i$-band upper limits and colors redder than $i-K=+6$.  The scatter about the mean relation appears to increase at $K\gtrsim22$, but the source density is also higher at these fainter magnitudes.  The full range of the scatter perpendicular to the mean relation is about 4 magnitudes in color.  This large spread in color substantiates the earlier suggestion that optical properties of SMGs are quite diverse \citep[e.g.,][]{ivison00a,ivison02}. However it cannot fully account for the up to $\sim10$ magnitude spread in their apparent brightness as a population.  In other words, there is an additional factor of $\gtrsim100$ variation (or $\gtrsim5$ mag) in the apparent brightness of SMGs on top of the apparent differences in color, which may be due to variations in SED and extinction.  Given their extreme luminosity, the SMG phase likely represents a brief, special moment during the rapid mass build-up phase \citep[e.g.,][]{narayanan10a,hayward11}.  If the majority of SMGs are seen in the redshift range between 1.5 and 4.0 (see \S~\ref{sec:redshifts}), then the spread in the luminosity distance can account for about a factor of 10 in the apparent brightness variation.  Therefore, the remaining factor of $\gtrsim10$ scatter in apparent brightness has to be accounted for by an intrinsic scatter in the rest-frame optical band luminosity.  

In the broader context of understanding star forming galaxies in the early universe, some of the AzTEC sources are bright in the rest-frame UV and optical bands and are already identified as star forming galaxies by past surveys.  For example, about 30\% are bright enough in the optical and NIR bands to be classified as ``BzK'' galaxies \citep{daddi04} using the existing photometry -- see discussions in Appendix.  Some of the $z\sim4$ SMGs have also been identified as ``Lyman break galaxies'' \citep[see][]{capak08}.  On the other hand, the majority of the AzTEC GOODS sources are too faint and red to have been identified in previous surveys of star forming galaxies and are likely to be entirely missed in the current accounting of the cosmic star formation history.  Future millimetre wavelength surveys with higher angular resolution are needed to probe deeper into the lower flux density (and lower luminosity) regime in order to bridge these populations and obtain a complete census of star forming galaxies.  

\begin{figure}
\includegraphics[width=8.5cm]{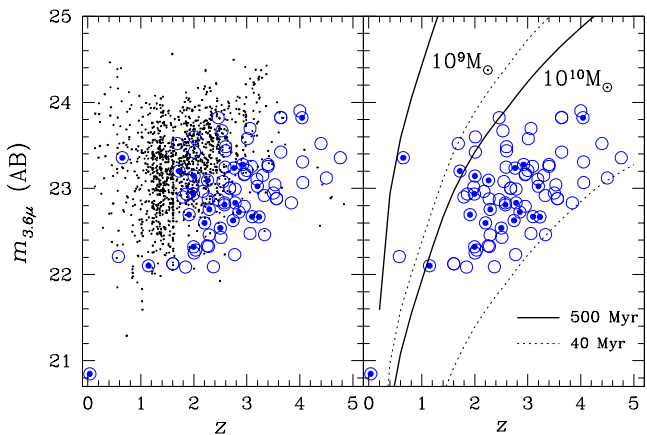}
\caption{\Spitzer\ IRAC 3.6 \micron\ band ``Hubble diagram''.  IRAC 3.6 \micron\ band brightness ($m_{3.6\mu}$) of the robust AzTEC counterparts is plotted as a function of redshift.  Symbols are the same as in Fig.~\ref{fig:magmag}.  Small dots on the left panel are $K-$band selected star forming galaxies with known spectroscopic redshifts in both GOODS fields.  The apparent brightness of a 40 Myr starburst population with a stellar mass of $10^9$ and $10^{10}M_\odot$ is shown in dotted lines, while the same population after passively evolving for 500 Myr is shown in solid lines.  An exponentially decaying starburst history with a 20 Myr e-folding time and solar metallicity is assumed for the models.
}
\label{fig:hubble}
\end{figure}

\bigskip
\subsection{SMGs as massive galaxies in a phase of rapid stellar mass build-up \label{sec:hubble}}

\subsubsection{Stellar luminosity of SMGs}

One constant in the high resolution interferometric millimetre and submillimetre observations of SMGs is the presence of a \Spitzer\ IRAC counterpart in the 3.6-8.0 \micron\ bands, and this is one of the key features we employ to identify AzTEC counterparts (see \S~\ref{sec:ID}).  A comparison of the apparent brightness in the 3.6 \micron\ band (rest-frame optical or near-IR) for the AzTEC counterparts and other K-band selected $z\sim2$ starforming galaxies with known spectroscopic redshifts in both GOODS fields is shown in Fig.~\ref{fig:hubble}a.  While there is some overlap between these two populations, the AzTEC counterparts are systematically brighter by $\gtrsim1$ magnitude on average and represent the most luminous galaxies at these redshifts.  This intrinsically high luminosity in the rest-frame optical and near-IR bands is clearly an important reason why these AzTEC sources are so readily detected by \Spitzer.  

Since AzTEC detection requires highly efficient dust-processing of the UV radiation from young stars, the high inferred luminosity in the rest-frame optical/NIR bands seems surprising.  After all, we just established in the previous section (\S~\ref{sec:optical}) clear evidence for severe attenuation of UV radiation among many of these objects.  A natural explanation for this apparent puzzle is found in the studies of the local ULIRG population.  An imaging study of local ULIRGs in the near- and far-UV bands by \citet{goldader02} has shown that activity traced in the UV bands is distributed over kiloparsec scales and is heavily obscured, particularly in the regions of the most intense starburst activities.  When observed at redshifts of $z\sim2$ to $z\sim4$, these ULIRGs are expected to be extremely red and faint in the observed optical and NIR bands ($R-K=$ 4-6, $K\approx$ 21-24), similar to the observed values for the AzTEC sources shown in Fig.~\ref{fig:magmag}.  At the same time, \citet{chen10} have shown that the stellar hosts of local ULIRGs are also extremely blue in rest-frame optical bands and are on average $\sim1$ magnitude {\em brighter} than the field star-forming population, owing to distributed star formation activity and the high intrinsic luminosity of young stellar clusters.  A Hubble Space Telescope NICMOS and ACS imaging study by \citet{swinbank10} has also found evidence for ongoing mergers and structured dust obscuration among $z\sim2$ SMGs, further supporting the parallel in the observed source luminosity and structure between the local ULIRGs and high-z SMG population.

One cannot automatically conclude from their large observed luminosity that these SMGs are also the most massive galaxies at their observed epochs if their luminosity is powered by a large population of widely distributed young stellar clusters.  The apparent 3.6 \micron\ brightness of a single stellar population starburst model with a total accumulated stellar mass ($M_*$) of $10^9 M_\odot$ and $10^{10} M_\odot$ is shown in Fig.~\ref{fig:hubble}b for two different scenarios: 40 Myr (dotted lines) and 500 Myr (solid lines) after the initial burst.  The observed brightnesses of AzTEC sources are well-bounded by the 40 Myr old starburst models with stellar masses of $10^9 M_\odot$ and $10^{10} M_\odot$, which are about {\it 10 times smaller} than the stellar masses derived for the $K-$band selected star forming galaxies shown in comparison \citep[$M_*=10^{10-11} M_\odot$,][]{daddi07a}.  However, after just 500 Myr of passive evolution, the same starburst systems fade by $\sim2$ magnitudes at $z\sim$ 2-4, bringing them back to a better agreement with the mass estimates for the $K-$band selected star-forming galaxies.  Alternatively, accounting for the observed 3.6 \micron\ band brightness of the AzTEC sources assuming a maturing stellar population would require stellar masses well in excess of $10^{11} M_\odot$.

\begin{figure}
\includegraphics[width=8.0cm]{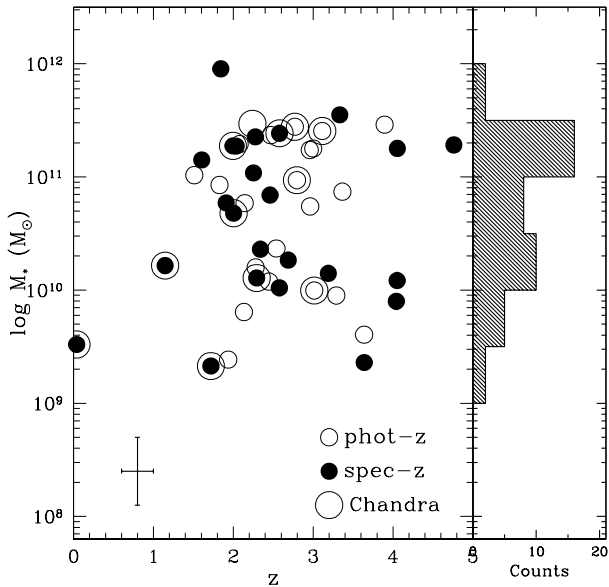}
\caption{The stellar masses of AzTEC counterparts derived from optical and NIR photometry as a function of redshift.  A histogram of stellar masses is shown on the right panel.  Sources with a spectroscopic redshift are shown as solid symbols while the ones with a photometric redshift are shown as empty circles, and they represent similar ranges of stellar mass.  Typical uncertainties for the $M_*$ and photo-z estimates are shown on the bottom left corner.  Those detected by the \Chandra\ in the X-ray are identified with a larger circle.  Only GOODS TFIT catalog sources with a proper stellar mass estimate are included.
}
\label{fig:mstar}
\end{figure}

\subsubsection{Stellar mass and star formation rate of SMGs}

We can get a better handle on the stellar mass by modeling the observed rest-frame UV and optical SED as discussed in \S~\ref{sec:OIRSED}.  The derived stellar masses from the modeling of the UV-optical SED, shown in Figure~\ref{fig:mstar}, range between $10^9$ to $10^{12} M_\odot$. The majority of the derived stellar masses are between 1 and 30 times $10^{10} M_\odot$, similar to those of the $K-$band selected massive star-forming galaxies at the same redshift, such as those discussed in \citet{daddi07a}. Similarly large stellar masses were found previously for SMGs \citep[e.g.,][]{dye08,tacconi08,daddi09b,hainline09,michalowski10}, and they are consistent with the idea that these submillimetre-bright galaxies are associated with the peak of the stellar mass function at $z=$ 2-3.  

Six out of 18 AzTEC counterparts with $M_*\ge 10^{11} M_\odot$ are also X-ray sources detected in deep \Chandra\ surveys.  The frequency of the \Chandra\ detected sources is about the same for the lower stellar mass galaxies (seven out of 25), and there is little evidence for any dependence on stellar mass.  The low number of $z\ge3$ sources detected by the 2-4 Ms \Chandra\ surveys may reflect the limiting sensitivity of the X-ray data, and the observed X-ray fraction is a lower limit.  Given the poor statistics, it is difficult to conclude whether the presence of a luminous AGN is influencing the modeling of the rest-frame optical SEDs.  

\begin{figure*}
\includegraphics[width=6.5cm,angle=270]{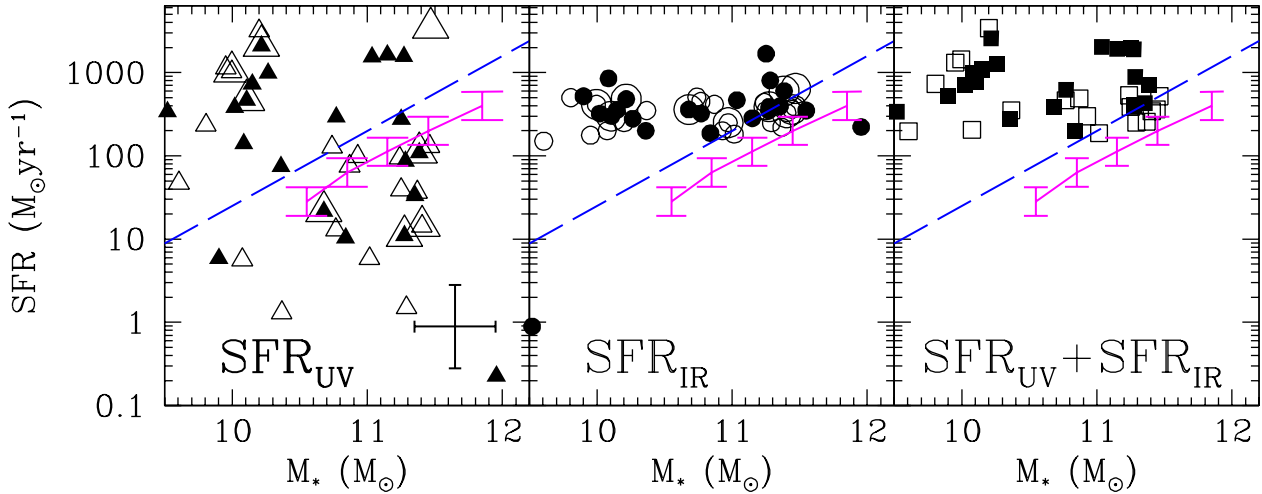}
\caption{Star formation rates for AzTEC GOODS sources estimated from the rest-frame UV ($SFR_{UV}$) and IR ($SFR_{IR}$) as a function of stellar mass.  Open and filled symbols represent photometric and spectroscopic redshifts, and those detected by the \Chandra\ in the X-ray are identified with a larger symbol.  A typical overall uncertainty for an object with photoz is shown on the bottom right corner of the left panel.  The mean $SFR-M_*$ relation for the $z=2.5$ model SMGs fueled by cold flow \citep{dave10} is shown by a solid line.  Long-dashed line is the observed mean $SFR-M_*$ relation for the $K$-band selected galaxies \citep{daddi07a}. The SMGs do not seem to follow either trends.
}
\label{fig:sfr}
\end{figure*}

The rest-frame optical SED modeling also yields a UV-derived star formation rate ($SFR_{UV}$). The derived $SFR_{UV}$ for the AzTEC-GOODS sources cover a broad range: $1-2000 M_\odot$ yr$^{-1}$ -- see the left panel of Figure~\ref{fig:sfr}. A surprising result is that the derived $SFR_{UV}$ is quite high, $\ge 100-1000 M_\odot$ yr$^{-1}$ for about 50\% of the cases.  The observed $SFR_{UV}$ distribution is also nearly independent of stellar mass.  The $SFR_{UV}$ distribution broadly overlaps the observed $SFR-M_*$ relation for the $K$-band selected galaxies studied by \citet[][dashed line]{daddi07a}, but there is little evidence that these AzTEC sources follow the same $SFR-M_*$ relation.  The AzTEC-GOODS sources also do not follow the $SFR-M_*$ relation predicted by the $z=2.5$ model SMGs fueled by cold flow accretion \citep[][solid line]{dave10}.  Some of the galaxies with the highest $SFR_{UV}$ are detected in the X-ray by \Chandra, raising the possibility that the UV light from the central AGN might contribute to these high values.  However, not all \Chandra\ detected sources are associated with a high $SFR_{UV}$, and neither the $SFR_{UV}/SFR_{IR}$ ratio nor the specific star formation rate discussed below offer any clear evidence to support this idea.  

For comparison, IR-derived star formation rates ($SFR_{IR}$) determined from modeling the IR SED are shown in the middle panel of Figure~\ref{fig:sfr}.  The $SFR_{IR}$ is uniformly high, $\ge 100-1000 M_\odot$ yr$^{-1}$, with a much smaller scatter and completely independent of stellar mass.  This is expected since these confusion-limited AzTEC surveys preferentially select sources with intrinsically large $L_{IR}$.  We note that the 1.1mm selection does not {\em guarantee} a high $L_{IR}$ or $SFR_{IR}$ if cold dust ($T_d=10-20$ K) emission dominates the millimetre spectrum.  On the other hand, our SED modeling does not find any cold dust dominated sources with $L_{IR}<10^{12} L_\odot$.  Since the IR luminosity accounts for the total amount of dust-processed UV radiation, a comparison of $SFR_{IR}$ with $SFR_{UV}$ should offer a crude measure of the geometry between the young stars and the obscuring dust.  The large $SFR_{UV}$ derived for a large fraction of AzTEC-GOODS sources is particularly interesting in this regard, and this result may indicate that star formation activities and dust distribution in these SMGs are not as concentrated as in the local ULIRGs, where $L_{IR} \approx L_{bol}$ \citep[see ][]{sanders96}.

\begin{figure}
\includegraphics[width=8.5cm]{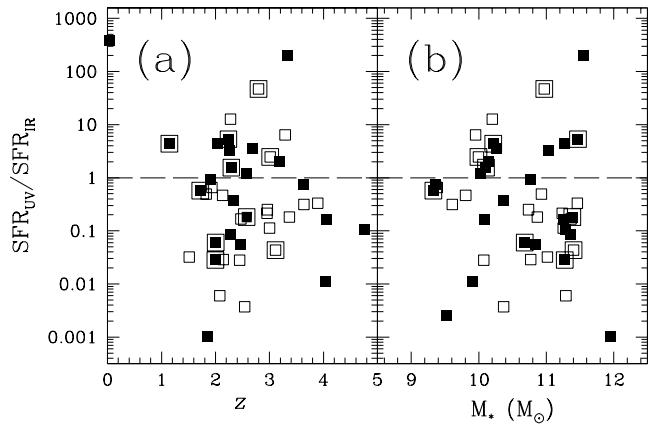}
\caption{The ratio of SFR derived from the rest-frame UV ($SFR_{UV}$) and the rest-frame IR ($SFR_{IR}$) as a function of (a) redshift
and (b) stellar mass ($M_*$). Galaxies with spec-z (photo-z) are shown in filled (empty) symbols, and those detected by the \Chandra\ in the X-ray are identified with a larger square. The dashed line marks the $SFR_{UV}=SFR_{IR}$ relation.
}
\label{fig:sfrratio}
\end{figure}

To explore the relationship between $SFR_{UV}$ and $SFR_{IR}$ further, their ratios are plotted as a function of redshift and $M_*$ in Figure~\ref{fig:sfrratio}.  This ratio varies widely from one source to another, spanning over 5 decades in total range, and it is independent of $z$ and $M_*$.  Finding a large number of sources with $SFR_{UV}$/$SFR_{IR}>1$ is particularly puzzling for these galaxies with a large stellar mass.  A mis-identification of the counterparts is also a plausible explanation, but the observed distribution would require the failure of counterpart identification in a large fraction of cases.  Either an under-estimate of $SFR_{IR}$ or an over-estimate of $SFR_{UV}$ (and possibly both) can provide an explanation, as the estimates of both $SFR$ and $M_*$ are subject to significant systematic uncertainties \citep[e.g., see][]{maraston10}.  If these galaxies represent young galaxies seen during their rapid mass build-up phase (see below), then the well-known mass-metallicity relation \citep{tremonti04} and the attenuation of UV light in the local universe may not be directly applicable.
The presence of an X-ray source detected by \Chandra\ is not correlated with the $SFR_{UV}$/$SFR_{IR}$ ratio, and the presence of an X-ray emitting AGN does not seem to contribute directly to the derived $SFR_{UV}$ in most cases.

\subsubsection{Specific star formation rate and mass build-up history}

Charting the star formation and stellar mass build-up history is one of the most powerful tests for galaxy evolution theories.  For example, the emergence of red sequence galaxies that are massive and passively evolving around $z\sim1$ and their increase in number with time are widely cited as important observational constraints that require additional complexities such as AGN feedback and ``dry'' mergers \citep{bell04,bell07,faber07}.  Statistical studies such as the Sloan Digital Sky Survey have shown that the bulk of stars now in massive galaxies formed at earlier epochs than stars in lower mass galaxies \citep[e.g.,][]{kauffmann03}, suggesting a strong link between galaxy mass and star formation history.  A particularly useful quantity to examine in this regard is the specific star formation rate (SFR per unit stellar mass; $SSFR\equiv SFR/M_*$). A systematic dependence of $SSFR$ on galaxy mass and a rapid increase of $SSFR$ with redshift have been established by several recent studies \citep{zheng07,damen09}.  Given the large stellar masses ($M_*=[1-30]\times 10^{10} M_\odot$) and SFRs ($>10^{2-3} M_\odot$ yr$^{-1}$) for these $z=$ 2-4 SMGs, examining their $SSFR$ in the context of the observed trends with stellar mass and redshift may provide a valuable new insight into the physical mechanisms driving the SMG phenomenon and massive galaxy formation.

\begin{figure}
\includegraphics[width=8.5cm]{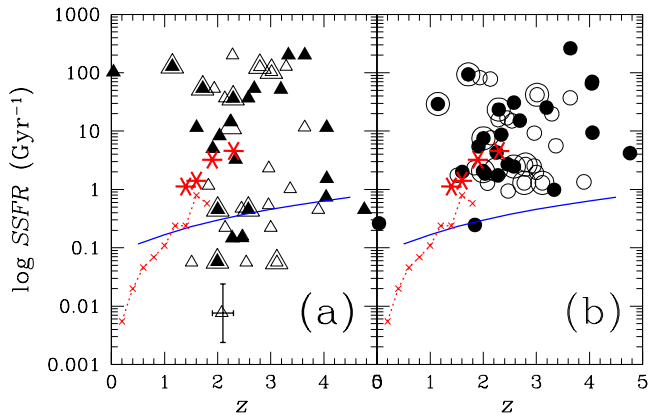}
\caption{Specific star formation rate of AzTEC counterpart candidates derived from the (a) rest-frame UV and optical SED fitting and (b) far-IR SED fitting using \citet{efstathiou00} dusty starburst SED templates.  Solid and empty symbols represent the sources with spectroscopic and photometric redshifts, respectively, and those detected by \Chandra\ in the X-ray are identified with a larger symbol.  A typical uncertainty is shown for one of the photo-z sources at the bottom of the left panel. The asterisks represent the radio-derived SSFRs for star forming galaxies with $M_*\sim 3\times 10^{10}M_\odot$ derived by \citet{pannella09}.  The solid line represents the inverse of the Hubble time, and sources above this line are in a starburst mode.  The dotted line connecting the crosses represents the average $SSFR$ for massive galaxies with $M_*\ge 10^{11}M_\odot$ found by \citet{damen09}.
}
\label{fig:ssfr}
\end{figure}

The computed $SSFRs$ for the AzTEC-GOODS sources are shown in Figure~\ref{fig:ssfr}. An immediately noticeable trend is that the derived $SSFR$s are uniformly quite high, $SSFR \approx$ 1-100 Gyr$^{-1}$.  Among the optically selected samples, galaxies with $SSFR \gtrsim10-100$ Gyr$^{-1}$ are generally associated lower stellar mass ($M_*\le 10^{10}M_\odot$) galaxies undergoing a starburst episode.  More massive galaxies in the local universe are associated with 1-2 orders of magnitudes lower $SSFR$ \citep{bauer05,feulner05,erb06,elbaz07}. The same trend also holds at higher redshifts as the $K-$band selected star forming galaxies at $z\sim2$ have on average $SSFR \approx 1$ Gyr$^{-1}$ \citep{daddi07a}, overlapping only at the bottom range of the $SSFR$ associated with the AzTEC sources.  \citet{pannella09} have also reported an average $SSFR \approx 5$ Gyr$^{-1}$ and $SFR\approx 100 M_\odot$ yr$^{-1}$ for their $z\sim2$ radio-identified star forming galaxies with an average $M_*=3\times 10^{10}M_\odot$ at $z\sim2$ \citep[also see][]{dunne09}. 

The $SSFRs$ for the AzTEC-GOODS sources are significantly higher than those of similar stellar mass galaxies in the local Universe, and they appear to follow the same broad trend of rapidly increasing $SSFR$ with redshift.  The dotted line shown in Figure~\ref{fig:ssfr} is the stellar mass-dependent $SSFR$ evolution mapped by \citet{damen09} for massive galaxies with $M_*\ge 10^{11}M_\odot$, and it shows a rapid rise as $SSFR \propto (1+z)^5$ between $z=0$ and $z=2$.  The radio-derived $SSFR$s for star forming galaxies with $M_*\sim 3\times 10^{10}M_\odot$ derived by \citet{pannella09}, shown in asterisks, extend this rapidly rising trend to $z\sim2.5$.  The AzTEC-GOODS sources extend this rise in $SSFR$ further to $z\sim4$, although there is significant scatter.  As noted by Damen et al. and others, there is a mass-dependence on the $SSFR$ evolution, and the spread in $M_*$ for the AzTEC sources likely contributes to some of the observed scatter.

The $SSFR$s derived for the AzTEC-GOODS sources provide the strongest evidence yet that SMGs are seen during the brief phase of rapid stellar mass build-up.  The solid lines in Figure~\ref{fig:ssfr} represent the inverse of the Hubble time, $1/t_H$: only galaxies with $SSFR$s above this line have sufficiently high $SFR$s to build up their current stellar masses within the Hubble time at their respective redshifts.  Considering the SSFRs derived from fitting the FIR SEDs (right panel of Figure~\ref{fig:ssfr}), all of the AzTEC GOODS SMGs are located at or above this critical line. We cannot exclude the possibility that these SMGs are rejuvenated galaxies, undergoing another episode of extreme luminosity, but it would require an even earlier episode of rapid stellar mass build-up.  Citing extremely high $SFR$ and similar density, previous studies have made plausible arguments for identifying SMGs as progenitors of present day massive elliptical galaxies \citep[e.g.,][]{blain04}.  Our new analysis of the $SSFR$ allows us to demonstrate quantitatively that these SMGs are seen {\em during} a phase of rapid stellar mass build-up.  

The absence of AzTEC sources with $SSFR$ below the $1/t_H$ line in the right panel of Figure~\ref{fig:ssfr} is primarily the result of AzTEC survey depth -- in fact, all existing confusion limited surveys carried out with, e.g., AzTEC and \Herschel, probe only the brightest end of the luminosity function.  Much of the cosmic IR background (CIRB) is expected to arise from fainter sources below the confusion limit, and their number counts can offer an important constraint to the evolution model for SMGs \citep{granato04,baugh05,rowan-robinson09}.  The location of the $1/t_H$ line in Figure~\ref{fig:ssfr} leaves a fairly limited parameter space for a lower luminosity dust-obscured starburst population that can contribute significantly to the CIRB -- e.g., $SFR \approx$ 20-100 $M_\odot$ yr$^{-1}$ for a (1-5) $\times 10^{10} M_\odot$ galaxy at $z=2$.  A $z\gtrsim2$ galaxy with $M_*\ge 10^{10-11}M_\odot$ can still appear with a $SSFR$ below this $1/t_H$ line, but the presence of a large population of such galaxies would have an important consequence in that the formation epoch of those massive galaxies has to be pushed to a much earlier time.  The decreasing $t_H$ with redshift also requires an even larger $SFR$ with increasing $z$, and in turn the submm/mm-detected fraction of galaxies with a stellar mass $M_*\ge 10^{10-11}M_\odot$ has to rise with increasing redshift.  The high detection rate of optically selected $z\ge4$ QSOs in the submm/mm continuum \citep[$\sim30\%$,][]{carilli01,wangr07,wangr08}, despite the selection bias against obscured systems, appears to be in line with this expectation.

The $SSFR$ analysis of the AzTEC-GOODS sources also suggests an intriguing idea that there may be two classes of SMGs, possibly driven by two different modes of star formation or observed at two different phases.  While the $M_*$ for these SMGs span over a factor of 30 (see Fig.~\ref{fig:mstar}), the $SFR_{IR}$ and $SFR_{UV}+SFR_{IR}$ show no dependence on $M_*$ in Figure~\ref{fig:sfr}.  This is in contrast to the finding by \citet{dave11a}, where a tight $M_*$-$SFR$ relation is a generic outcome of all of their cosmological hydrodynamic simulations incorporating galactic outflows.  One way to interpret our observational results summarized in Figure~\ref{fig:sfr} is that only SMGs with $M_*\gtrsim 10^{10.5}$ have properties similar to the objects modeled by Dav\'{e} et al., following the $M_*$-$SFR$ relation predicted for SMGs \citep[solid line in Figure~\ref{fig:sfr},][]{dave10}.  AzTEC-GOODS sources with  $M_* < 10^{10.5}$ may follow a different, currently unknown process that leads to 10 times larger $SSFR$. This is somewhat analogous to the mass-dependent $SSFR$ seen among galaxies in the local universe (``red'' and ``blue sequence'') with a similar range of $M_*$, although with 1-2 orders of magnitudes lower $SSFR$s.   A division by mass certainly seems somewhat arbitrary. On the other hand, these SMGs appear to show a sign of grouping by mass in the middle and right panel of Figure~\ref{fig:ssfr}. A hint of bimodality in the histogram of $M_*$ is also seen in Figure~\ref{fig:mstar}.  Future LMT and ALMA surveys of much larger samples with higher angular resolution and spectroscopic redshifts should provide a definitive test on this intriguing possibility.

\bigskip
\section{AGN AND STARBURST ACTIVITIES \label{sec:AGN}}

Determining the source of enormous luminosity associated with SMGs is an outstanding problem that has important implications on understanding the mass assembly history of galaxies.  To probe the nature of the heavily obscured power source, optically thin tracers in the X-ray, IR, or radio wavelengths are necessary.  Here, we examine the properties of AzTEC sources identified in the GOODS-South and GOODS-North fields using several well-established diagnostic tests utilizing these optically thin tracers.

\begin{figure}
\includegraphics[width=8.0cm]{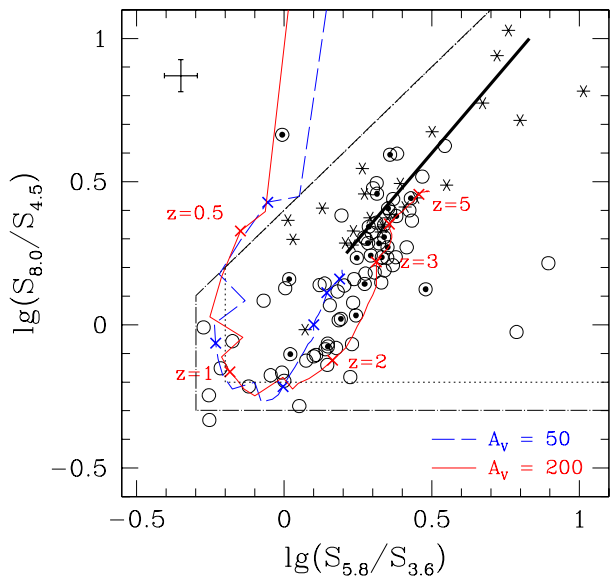}
\caption{A $S_{8.0\mu}/S_{4.5\mu}$ versus $S_{5.8\mu}/S_{3.6\mu}$  IRAC color diagnostic diagram for heavily obscured starbursts and AGNs based on the color combinations proposed by \citet{lacy04}. 
Areas occupied by dusty young starbursts as noted by \citet[][see their Fig.~1]{yun08} are outlined by a dot-dashed line while the area previously identified with power-law spectrum AGNs by Lacy et al. is outlined using a dotted line. The IRAC counterparts identified with the AzTEC sources in GOODS-South and GOODS-North with empty (undetected in X-ray) and dotted circles (detected in X-ray) cluster around the theoretical dusty starburst SED tracks with different amounts of dust extinction \citep[$A_V=50\ \& \ 200$,][]{efstathiou00}, as discussed by \citet{yun08}.  The thick solid line represents the theoretical track expected of purely power-law IR AGNs, while asterisks are power-law AGNs reported by Lacy et al. and \citet{martinez08}.  Typical uncertainties in the colors are shown by the cross in the upper left corner.}
\label{fig:irac_color1}
\end{figure}

\begin{figure}
\includegraphics[width=8.0cm]{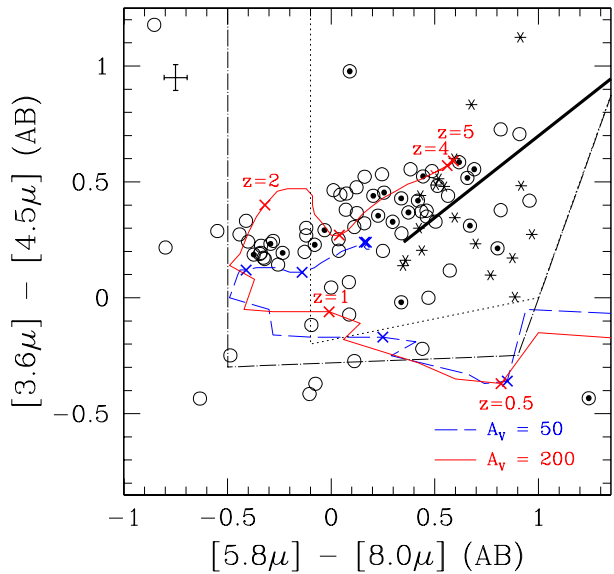}
\caption{A $[3.6\mu]-[4.5\mu]$ versus $[5.8\mu]-[8.0\mu]$  IRAC color diagnostic diagram for heavily obscured starbursts and AGNs based on the color combinations proposed by \citet{stern05}, adopted from Fig.~2 by \citet{yun08}. All symbols and models shown are identical to those in Fig.~\ref{fig:irac_color1}. }
\label{fig:irac_color2}
\end{figure}

\subsection{\Spitzer\ IRAC color-color diagram \label{sec:IRAC}}

The first set of diagnostic tests to examine are the IRAC color-color diagrams that are commonly used for identifying heavily obscured AGNs based on the color combinations proposed by \citet{lacy04} and \citet{stern05} as shown in Fig.~\ref{fig:irac_color1} and Fig.~\ref{fig:irac_color2}, respectively.  The majority of robustly identified AzTEC sources in both GOODS fields fall within the regions previously identified with power-law AGNs in both plots. As argued in detail below, these results are not a direct consequence of using a \Spitzer\ IRAC color selection for identifying AzTEC counterpart sources.  Interpreting these results should also require some care as a young starburst population at $z>1$ has a characteristic red SED in this part of the spectrum and should fall within the same color region \citep[see][]{yun08}.     

The sample size of AzTEC GOODS sources plotted in Figs.~\ref{fig:irac_color1}~\&~\ref{fig:irac_color2} is nearly 3 times larger than the sample previously analyzed by Yun et al., and these new plots show more clearly that these AzTEC sources cluster densely around the theoretical color tracks of 20-80 Myr old dusty starbursts at $z\gtrsim2$ by \citet{efstathiou00}.  The dispersion in the model tracks and the observed color are larger in Fig.~\ref{fig:irac_color2}, but the AzTEC sources again mostly occupy the region spanned by the starburst model tracks, rather than the area surrounding the IR power-law track.   It is particularly noteworthy that AzTEC sources identified with a \Chandra\ X-ray source (dotted circles) occupy largely the same area as those without X-ray detection (empty circles), and only a small fraction of sources (both with and without X-ray detection) overlap with the IR power-law AGN \citep[asterisks; see][]{yun08}.  Conversely, many of the AzTEC sources appearing among the power-law IR sources are undetected in the 2 \& 4 Ms \Chandra\ surveys \citep{alexander03b,luo08,johnson11}.   There is a weak trend of an increasing fraction of X-ray detected sources with redder color.   We can conclude from these diagnostic plots that nearly all of the AzTEC sources identified in the GOODS fields have IRAC SEDs consistent with that of a young starburst, while a small fraction ($\lesssim20\%$) show IR colors of a power-law AGN.

One thing to clarify is that the use of IRAC colors as a method to identify the AzTEC counterparts does not lead directly to these observed trends.  The adopted color selection, $[3.6\mu]-[4.5\mu]\ge0.0$, imposes no restriction on the 5.8 \micron\ and 8.0 \micron\ photometry.  This color selection is also only one of {\it three} independent criteria we examine jointly, and the radio data contribute overwhelmingly to the secure counterpart identification.  In fact, {\em none} of the robustly identified sources in Table~\ref{tab:ID} are based on the IRAC color selection alone (see \S~\ref{sec:methods}).  This color selection is not used to {\it reject} any counterpart candidates either, and indeed several robust counterparts shown in Fig.~\ref{fig:irac_color2} have a {\it blue} IRAC color ($[3.6\mu]-[4.5\mu]<0.0$).  In the context of the starburst SED model tracks shown, the color selection of $[3.6\mu]-[4.5\mu]\ge0.0$ effectively imposes a redshift bias against sources at $z\lesssim1$ such as AzTEC/GS20.  However, this bias is more than compensated by the radio and MIPS identification methods that systematically favor low-z candidate counterparts.

\subsection{Optical-IRAC-MIPS colors \label{sec:MIPS}}

\begin{figure}
\includegraphics[width=8.0cm]{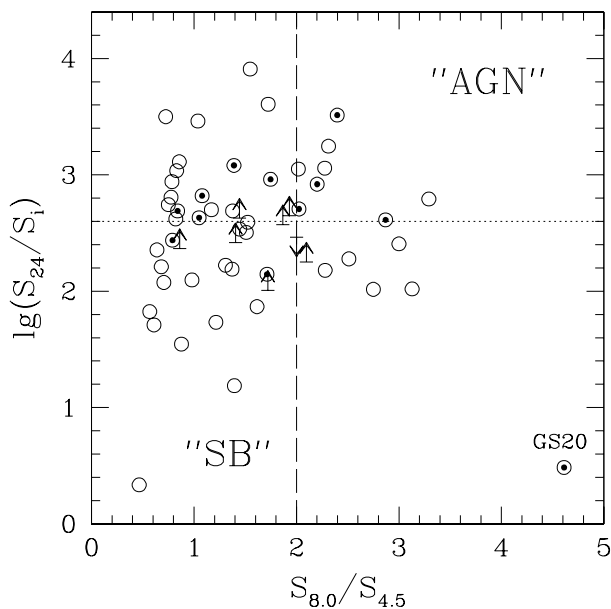}
\caption{A diagnostic color-color diagram using MIPS 24 \micron\ to optical $i$-band flux ratio versus $S_{8.0\mu}/S_{4.5\mu}$ IRAC band flux ratio.  Symbols are identical to those in Fig.~\ref{fig:irac_color1}.  The long-dashed line at $S_{8.0}/S_{4.5}= 2$ is the dividing line for AGNs and starbursts as proposed by \citet{pope08b}.  The dotted line near lg($S_{24}/S_i$)=2.6 is the equivalent division line for the $z\sim2$ ``dust-obscured galaxies'' with $F_\nu(24\mu m)/F_\nu(R)\gtrsim1000$ -- see \citet{dey08,fiore08}.}
\label{fig:DOGcolor}
\end{figure}

Another widely used AGN diagnostic diagram is the \textit{Spitzer} mid-IR color plot first introduced by \citet{ivison04}.  They noted that starburst and AGN color tracks as a function of redshift are well-separated in the plot of flux ratios $S_{24}/S_{8.0}$ vs. $S_{8.0}/S_{4.5}$ due to contributions by a power-law AGN and PAH emission in the 24 \micron\ band.    By analyzing \textit{Spitzer} Infrared Spectrograph (IRS) data on 24 \micron\ selected galaxies, \citet{pope08b} found that the main discriminatory information resides in the $S_{8.0}/S_{4.5}$ flux ratio, or the spectral slope in the rest-frame near-IR band.  This color selection is similar to the earlier IRAC color analysis proposed by \citet{lacy04} and \citet{stern05}, and it is again subject to the same confusion with young starburst systems as noted by \citet{yun08} and others.  The new AGN diagnostic condition of $S_{8.0}/S_{4.5}\ge 2$ proposed by \citet{pope08b} corresponds to $lg(S_{8.0}/S_{4.5})\ge +0.3$ in Fig.~\ref{fig:irac_color1}, and objects satisfying this criteria should also include dusty young starbursts at $z\le0.5$ and $z>3$ as well as power-law IR AGNs.  

Our version of the \citet{pope08b} diagnostic test is shown in Fig.~\ref{fig:DOGcolor}, which shows the plot of the MIPS 24 \micron\ to optical $i$-band flux ratio versus $S_{8.0\mu}/S_{4.5\mu}$ IRAC band flux ratio for AzTEC sources in both GOODS fields.  Among the 57 sources plotted, only 16 (28\%) have the $S_{8.0}/S_{4.5}$ flux ratio consistent with hosting an energetic AGN (right of the long-dashed line).  Again, X-ray detection (dotted circles) appears to have little bearing on whether an object falls on the starburst (``SB'') side or the ``AGN'' side.  The source with the highest ratio $S_{8.0\mu}/S_{4.5\mu}=4.6$ is AzTEC/GS20, which is a $z=0.0369$ galaxy whose $S_{8.0\mu}/S_{4.5\mu}$ ratio arises from the bright PAH line emission in the 8 \micron\ band, rather than due to a power-law AGN -- also see Fig.~\ref{fig:irac_color1}.

The choice of the flux ratio between the MIPS 24 \micron\ and the optical $i-$band for the vertical axis in Fig.~\ref{fig:DOGcolor} is motivated by the claim of a new class of faint MIPS 24 \micron\ sources that were missed by earlier optical studies.  These so-called ``dust-obscured galaxies'' \citep[``DOGs'';][]{dey08,fiore08} represent a population of infrared bright galaxies that are extremely faint in the optical bands, characterized by $F_\nu(24\mu m)/F_\nu(R)\gtrsim1000$.  These $z\sim2$ galaxies have similar projected density as SMGs and may account for as much as $\sim$1/4 of the IR luminosity density at this redshift \citep{dey08,pope08b}.  Evidence for both star formation and AGN activity has been reported for these objects.  Based on their stacking analysis of the X-ray hardness ratio, \citet{fiore08} proposed that as many as 80\% of these dust-obscured galaxies host a Compton-thick AGN.

Adopting a mean color of $r-i\approx+1.0$ for the radio-selected SMGs \citep[e.g.,][]{ivison02}, the ``$F_\nu(24\mu m)/F_\nu(R)\gtrsim1000$'' definition for DOGs translates to ``$lg (S_{24\mu}/S_i) \gtrsim 2.6$'' in Fig.~\ref{fig:DOGcolor}.  Among the 12 X-ray detected secure AzTEC counterpart sources with sufficient optical and Spitzer data to be included in this analysis, nearly equal numbers of SMGs fall on either side of this division line.  One significant difference is that 9 X-ray detected AzTEC sources fall on the optically faint side (above the dotted line), while only 3 X-ray detected AzTEC sources (including the low-z source AzTEC/GS20) are found on the optically bright side.  This trend is consistent with the suggestion by Fiore et al. that many of these dust-obscured galaxies host a Compton-thick AGN.  However, there are just as many AzTEC sources undetected in the X-ray above the division line.  Given that the observed X-ray emission can be largely accounted for by the starburst activity in many cases \citep[see below and][]{alexander05}, the significance of a higher frequency of X-ray detection among these optically faint SMGs is not entirely clear.

\subsection{X-ray \label{sec:xray}}

The 2 \& 4 Ms \Chandra\ X-ray surveys of the two GOODS fields \citep{alexander03b,luo08,xue11,johnson11} are some of the deepest X-ray data available and thus offers the best opportunity to determine X-ray properties of all types of extragalactic sources.  A cross-correlation of the AzTEC and \textit{Chandra} X-ray catalogue has shown that 16 (out of 48) and 8 (out of 40) AzTEC sources in the GOODS-South and North fields have an X-ray source within 6\arcsec\ of the AzTEC centroid positions \citep{johnson11}. Given the low density of X-ray sources, these coincidence are highly statistically significant.  On the other hand, only a subset of these X-ray sources are robust counterparts, and another physical link such as the clustering of massive galaxies \citep[e.g.,][]{almaini03} must play a role.  

The derived 2-10 keV X-ray luminosity of these \Chandra\ sources favor the AGN-origin for the observed X-ray emission.   Examining the X-ray properties of faint radio sources in the Hubble Deep Field (North), \citet{bauer02} found that the linear correlation between X-ray luminosity and 1.4 GHz radio luminosity density of late type galaxies extends to luminous X-ray detected emission line galaxies at intermediate redshift, suggesting both the X-ray and radio processes are associated with star formation activities.  \citet{persic04} have shown that the integrated emission from high-mass X-ray binaries (HMXBs) can offer a natural explanation for the observed correlation, and given their short lifetime the measured X-ray luminosity can offer an {\it instantaneous snapshot} of the ongoing star formation rate.  Since HMXBs also display a characteristic hard X-ray spectrum, the hardness ratio of the observed X-ray emission does not provide a unique probe of AGN activity \citep[e.g.,][ -- see \S~\ref{sec:MIPS}]{fiore08}.  On the other hand, X-ray luminosity of these \Chandra\ sources associated with AzTEC detection ranges between $L_X(2.0-10~keV)=10^{42}$ and $10^{43}$ erg s$^{-1}$. When converted to a SFR using the relation given by \citet{persic04}, their inferred SFR ranges between $10^3$ and $10^4 M_\odot$ yr$^{-1}$, exceeding the SFR derived from their UV and IR properties (see Fig.~\ref{fig:sfr}).  A fainter X-ray source with 2-10 keV luminosity of $10^{42}$ ergs s$^{-1}$ can be either a low luminosity AGN or an SMG with a $SFR=10^3 M_\odot$ yr$^{-1}$, but any source with a higher X-ray luminosity would require a significant AGN contribution \citep[see ][]{johnson11}.

One intriguing trend found is that the fraction of robust AzTEC counterparts that are also \Chandra-detected X-ray sources is higher for the brighter AzTEC sources.  The 30\% (7 and 8 out of 25) of the brightest AzTEC sources in the GOODS-South and GOODS-North fields are detected individually as a \Chandra\ X-ray source, and  an increasing AGN activity may be associated with the most luminous AzTEC sources.  In the same vein, the number of candidate \Chandra\ counterpart in the GOODS-South field did {\em not} change from the 2 Ms catalog to 4 Ms catalog, and the greater depth of the X-ray data had curiously little impact.  
The higher frequency of X-ray counterpart in the GOODS-South field (16/48 vs. 8/40) may reflect the cosmic variance in these two relatively small size fields. A more detailed discussion of the X-ray properties of AzTEC sources is presented elsewhere \citep{johnson11}.  

The high detection rate for the AzTEC sources in the X-ray bodes well for coeval mass growth scenarios for the stellar component and the central supermassive black hole (SMBH) designed to explain the apparent correlation between the central blackhole mass and stellar velocity dispersion \citep[``M-$\sigma$ relation'' --][]{magorrian98,ferrarese00,gebhardt00}.  For example, through detailed numerical modeling, \citet{narayanan10a,narayanan10b} have shown that a rapid build-up of stellar mass and the growth of the central supermassive black hole can be achieved through a merger-driven starburst, and can reproduce the observed properties of SMGs and dust-obscured QSOs.  Winds driven by the starburst and the AGN activity can effectively disrupt the central concentration of gas and dust, driving the evolution of such objects from an SMG phase to a QSO phase \citep{narayanan08}.

A natural consequence of such a scenario is that a massive stellar galaxy with a maturing young stellar population would emerge unobscured as the central AGN begins to dominate the overall energetics.  As the feedback process starts to clear out the obscuring dust and gas, the central AGN would also become more detectable in the X-ray, UV, and optical bands, marking the beginning of the classical QSO phase.  However, the X-ray detected AzTEC sources in the optical and near-IR bands span the entire observed range of brightness, indistinguishable from the X-ray undetected sources in Fig.~\ref{fig:magmag}.  The $i-K$ colors of the X-ray detected sources are also indistinguishable from the others, suggesting that either (1) the X-ray detection does {\it not} signal the emergence of the central AGN as the dominant energy source or (2) additional complexity is required in the SMG-QSO evolution model.

By analyzing the properties of a sample of $z\approx2$ SMGs exhibiting broad H$\alpha$ and H$\beta$ emission lines, \citet{alexander08} have estimated their black hole mass to be $\gtrsim3$ times smaller than those found in comparable mass normal galaxies in the local universe, and $\gtrsim10$ times smaller than those predicted for $z\approx2$ luminous quasars and radio galaxies.  Based on this evidence, they argued that the growth of the black hole lags that of the host galaxy in SMGs.  We find only marginal evidence for AGN contribution to the near-IR (Figs.~\ref{fig:irac_color1} \& \ref{fig:irac_color2}) and mid-IR (Fig.~\ref{fig:DOGcolor}) SEDs for the robust AzTEC sources, even among those detected in the X-ray.  The spectral decomposition and the analysis of the emission, absorption, and continuum features in the \textit{Spitzer} IRS spectra of SMGs by \citet{pope08}, \citet{menendez09}, and \citet{murphy09} have found that a starburst dominates the luminosity in the large majority of cases, even when the sample is selected to have AGN-like colors \citep{coppin10a}.   A consistent trend emerging from these multiwavelength data analyses is that dust-obscured starburst activity can account for most of the luminosity in SMGs, with little or only a minor contribution from AGNs.

\section{SUMMARY AND DISCUSSION}

Taking advantage of some of the deepest imaging and photometry data and extensive spectroscopic information in the GOODS-South field, we searched for counterparts to the 48 AzTEC sources found in the deep 1.1\,mm wavelength survey by \citet{scott10}, using a $P$-statistic analysis involving VLA 1.4\,GHz, Spitzer/MIPS 24\,$\mu$m, and IRAC catalogs, combined with cross identification with LABOCA 870\,$\mu$m sources.  Robust ($P\le 0.05$) and tentative ($0.05 < P \le 0.20$) counterpart candidates are found for 27 and 14 AzTEC GOODS-South sources, respectively.  Five of the sources (10\%) have two robust counterparts, supporting the idea that these SMGs are strongly clustered and/or confused.  A spectroscopic redshift is available for 12 robust counterparts and 12 tentative counterparts while photometric redshifts based on rest-frame UV-to-optical and radio-millimetric SED analysis are available for the remainder.  Stellar mass ($M_*$) and $SFR_{UV}$ are derived by modeling the observed optical and \Spitzer\ IRAC photometry while $SFR_{IR}$ is derived by analyzing the IR, (sub)millimetre, and radio photometry using theoretical templates.  To improve the statistics of the subsequent analysis, we applied the same counterpart identification and SED analysis to the AzTEC 1.1mm sources identified in the GOODS-North field \citep{perera08,downes11}.

Estimates of the redshift distribution of AzTEC-GOODS sources are constructed by combining spectroscopic redshifts with UV+optical and radio-millimetric photometric redshifts, and these two redshift distributions agree well with each other as shown in Figure~\ref{fig:comparez}.  Our analysis shows that 80\% of AzTEC sources are at $z\ge2$, with a median redshift of $z_{med}\sim 2.6$, and there is a significant high-redshift tail with 20\% of AzTEC sources at $z\ge 3.3$.   These distributions are quite different from the commonly cited SMG redshift distribution of \citet{chapman05}, primarily at the low redshift end.  The SHADES survey redshift distribution by \citet{aretxaga07} is in better agreement with our redshift distribution derived from the AzTEC GOODS surveys, and like ours, is missing the low-redshift tail seen in \citet{chapman05}.  Complete CO spectroscopic redshift surveys using the LMT and ALMA will be able to accurately determine the SMG redshift distribution by overcoming the large number of systematic biases inherent in all of these analyses.

An examination of the rest-frame UV and optical photometry for the securely identified AzTEC sources shows a nearly 10 magnitude (a factor of $10^4$ in flux density) spread in the $i-$ and $K-$band photometry and extremely red colors spanning $i-K$ color between 0 and +6.  There are a small minority of SMGs that are bright in the rest frame UV bands, overlapping with star forming galaxy population previously identified in the rest-frame UV searches.  On the other hand, AzTEC GOODS sources are on average quite red and faint, with a median brightness of $i=25.3$ and $K=22.6$, and a large fraction of AzTEC sources are entirely missed by previous surveys of star forming galaxies.    Examining the observed scatter in the $i-K$ color, we deduce that there is at least a factor of 10 variation in the intrinsic rest frame optical luminosity among these SMGs.

A Hubble diagram of the observed IRAC 3.6 \micron\ flux density shows that these AzTEC-GOODS sources are some of the most luminous galaxies in the rest-frame optical bands at $z\ge2$, offering a good explanation as to why nearly every SMG identified with interferometric observations shows a relatively bright IRAC counterpart.  Modeling of the observed rest-frame UV and optical SEDs shows that the stellar masses are rather large, $M_*=$ (1-30) $\times 10^{10} M_\odot$, with a surprisingly large $SFR_{UV} \gtrsim 100 - 1000 \, M_\odot$ yr$^{-1}$ for about 50\% of these galaxies.  In comparison, $SFR_{IR}$ derived from modeling the IR to radio SED covers a relatively tight range of 200-2000 $M_\odot$ yr$^{-1}$, independent of the redshift or stellar mass.  Whether a galaxy has been detected in the X-ray by \Chandra\ appears to have no influence on the derived $M_*$, $SFR_{UV}$, $SFR_{IR}$, and $SFR_{UV}/SFR_{IR}$ ratio, and the presence of an X-ray bright AGN appears to have relatively little influence on these quantities.

These AzTEC-GOODS sources have a specific star formation rate 10-100 times higher ($SSFR \approx$ 1-100 Gyr$^{-1}$) than similar stellar mass galaxies at $z=0$, and they extend the previously observed rapid rise of $SSFR$ with redshift \citep[$SSFR\propto (1+z)^5$,][]{damen09} to $z=2-5$.  More importantly, {\em all} of the AzTEC-GOODS sources have a $SSFR$ above the inverse Hubble time line, indicating that they have a current $SFR$ high enough to have built up their entire stellar mass within the Hubble time at their observed redshift.  This might be the best quantitative evidence yet that we are witnessing these galaxies during their rapid mass build-up phase.  The flat $SSFR$ as a function of redshift we deduce contradicts the model prediction of a tight $M_*$-$SFR$ relation based on cosmological hydrodynamic simulations incorporating galactic outflows \citep{dave10,dave11a}.  However, AzTEC sources with $M_* \gtrsim 10^{10.5} M_\odot$ appear to follow this model prediction, and one plausible explanation is that a different mechanism is operating for the lower mass SMGs, leading to a 10 times higher $SSFR$.  Alternatively, they are seen at a different phase of rapid mass build-up.  Much of the cosmic IR background (CIRB) is expected to be associated with fainter sources below the confusion limit of our AzTEC surveys, and their number counts can offer an important constrain to the evolution model for SMGs \citep{granato04,baugh05,rowan-robinson09}.  It still remains to be shown whether these are young, lower mass galaxies seen in their rapid formation epoch or the simmering activities in more massive galaxies that have already undergone an SMG-like rapid build-up phase in even earlier epochs.  

Lastly, we examine the evidence for luminous AGNs in these systems using three different diagnostic tests: (1) the \Spitzer\ IRAC color-color diagram; (2) optical-IRAC-MIPS colors; and (3) X-ray luminosity.  We find only marginal evidence for AGN contribution to the near-IR (Figs.~\ref{fig:irac_color1} \& \ref{fig:irac_color2}) and mid-IR (Fig.~\ref{fig:DOGcolor}) SEDs for the robust AzTEC sources, even among those detected in the X-ray.  A consistent trend emerging from this multiwavelength data analysis and similar studies by other groups is that dust-obscured starburst activity can account for most of the luminosity in submm/mm-selected galaxies, with little or only a minor contribution from AGNs.

\section*{acknowledgments}
The authors would like to thank M. Dickinson and
R.-R. Chary for use of the \Spitzer\ GOODS-S IRAC/MIPS catalogs. This work has also benefited from valuable discussions with C. Carilli, E. Chapin, R. Ivison, A. Pope, D. Sanders, N. Scoville and many others.  We
would also like to thank everyone who helped staff and support the
AzTEC/ASTE 2007 operations, including K. Tanaka, M. Tashiro,
K. Nakanishi, T. Tsukagoshi, M. Uehara, S. Doyle, P. Horner, J. Cortes,
J. Karakla, and G. Wallace.  Support for this work was provided in part by the National Science Foundation grants AST-0838222 and AST-0907952.


\bibliography{references}

\appendix

\section{Notes on individual sources} \label{sec:notes}

\medskip
\noindent{\bf AzTEC/GS1.}
There is one clear, robust counterpart (GS1a in Table~\ref{tab:fluxes}), which is a radio source found 4.8\arcsec\ north of AzTEC/GS1 ($P_{1.4}=0.045$).  This faint radio source has an IRAC/MIPS counterpart ($P_{24\mu}=0.161$), which is also a \Chandra/X-ray source. This source has a red IRAC color with $[3.6\mu$m$]$ $-$ $[4.5\mu$m$]$ = +0.37 ($P_{color}=0.133$), similar to the AzTEC sources identified using submillimetre interferometry \citep{younger07,yun08}. The 870 \micron\ LABOCA source LESS J033211.3$-$275210 ($S_{870\mu}=9.2\pm1.2$ mJy) position is only 2.1\arcsec\ away from GS1a, 
and \citet{biggs11} also identify GS1a as the robust counterpart.  No spectroscopic redshift is available for this extremely faint optical source ($i>23.5$, $K>22.9$), and it is likely a high redshift ($z\ge3$) source with $z_{opt}=2.96\pm0.45$ and $z_{MR}=3.56^{+0.66}_{-1.20}$.
\medskip

\noindent{\bf AzTEC/GS2.1.}
There are two radio sources (GS2.1a \& GS2.1b) within the 4.5\arcsec\ search radius, each with a high likelihood of being the AzTEC counterpart. The western-most source GS2.1a with a red IRAC color ($[3.6\mu$m$]$ $-$ $[4.5\mu$m$]$ = +0.38) is a robust identification ($P_{1.4}=0.001$).  The 870 \micron\ LABOCA source LESS J033219.0$-$275219 ($S_{870\mu}=9.1\pm1.2$ mJy) position is only 4.8\arcsec\ away from GS2.1a. The adjacent, second faint radio source GS2.1b is an extremely rare example of a faint radio source without any counterpart in the IRAC and MIPS images. Both radio sources are formally considered robust counterparts by our analysis and by \citet{biggs11}.  There are two additional faint radio sources just outside the search radius, making this an exceptionally crowded field in the radio band. These two more distant radio sources have the same spectroscopic redshift of $z = 1.097$ \citep{lefevre04,norris06}, and their blue IRAC color, $[3.6\mu$m$]$ $-$ $[4.5\mu$m$]$ = $-$0.27 suggests that they are indeed foreground sources.  The catalog position for the BLAST 250 \micron\ source 59 \citep{dunlop10} is located near the peak of the AzTEC/GS2 contours but between the two deconvolved components AzTEC/GS2.1 and AzTEC/GS2.2. A third potential counterpart, GS2.1c, identified by MIPS 24 \micron\ detection is only 3.9\arcsec from the AzTEC centroid; however it has a blue IRAC color ($[3.6\mu$m$]$ $-$ $[4.5\mu$m$]$ = $-$0.32) with a spectroscopic redshift of $z=0.644$, and is likely a foreground object.
\medskip

\noindent{\bf AzTEC/GS2.2.}
There are no radio sources within 15\arcsec\ of AzTEC/GS2.2.  The MIPS 24 \micron\ source GS2.2a is a potential counterpart with $P_{24\mu}=0.212$ with a blue IRAC color ($[3.6\mu$m$]$ $-$ $[4.5\mu$m$]$ = $-0.23$) and a spectroscopic redshift of $z=1.046$ \citep{popesso09}.  Therefore, it is likely a member of the foreground $z=1.10$ cluster GCL~J0332.2$-$2752 \cite[$\sigma_v=433$ km/s;][]{diaz-sanchez07} whose center is located only $\sim20$\arcsec\ to the northeast, at $\alpha=03^h 32^m 17.5^s$ and $\delta = -27^\circ 52' 32''$.  This MIPS source is blended with a second IRAC source located 3\arcsec\ to the southwest, GS2.2b, which has a red IRAC color ($[3.6\mu$m$]-[4.5\mu$m$]$ = +0.26; $P_{color}=0.390$).  GS2.2b is also a BzK galaxy and thus is an actively star forming system. The IRAC source GS2.2c is another BzK galaxy with a red IRAC color ($[3.6\mu$m$]$ $-$ $[4.5\mu$m$]$ = +0.10) and is an interesting alternative counterpart, though $P_{color} = 0.439$. The closest 870 \micron\ LABOCA source LESS J033217.6$-$275230 ($S_{870\mu}=6.3\pm1.3$ mJy) position is 15\arcsec\ northeast of the AzTEC centroid, nearly centered on the $z=1.1$ cluster. The position of AzTEC/GS2.2, however, is very uncertain as it is blended with AzTEC/GS2.1, so the counterpart identification is highly problematic.
\medskip

\noindent{\bf AzTEC/GS3.}
The faint IRAC source GS3a ($S_{1.4GHz}=40.7\pm6.5$ $\mu$Jy; $P_{1.4}=0.045$) is also a MIPS 24 \micron\ source with a red IRAC color ($[3.6\mu$m$]-[4.5\mu$m$]$ = +0.37; $P_{color}=0.174$). The 870 \micron\ LABOCA source LESS J033248.1$-$275414 ($S_{870\mu}=8.8\pm1.2$ mJy) is only 2.4\arcsec\ away from GS3a, leading \citet{biggs11} to conclude this source as a secure counterpart as well.  \citet{dunlop10} also identified GS3a as the counterpart to the 250 \micron\ BLAST source 593 and estimated a redshift $z>2.5$ for this optically invisible source. Our radio-mm photometric redshift of $z_{MR}=3.09^{+0.55}_{-1.11}$ supports this high-z hypothesis. There is a faint red IRAC source, GS3b, which is is also a tentative detection that cannot be ruled out.
\medskip

\noindent{\bf AzTEC/GS4.}
There is only one red IRAC source (GS4a; $P_{color}=0.070$) within the 6.5\arcsec\ search radius.  It is a faint radio emitter ($S_{1.4GHz}=25.4\pm6.5$ $\mu$Jy, $P_{1.4}=0.021$).  There are two other radio sources within 17\arcsec\ from the AzTEC position, but GS4a is the only source falling within the error circle of the 870 \micron\ LABOCA source LESS J033249.2$-$274246 ($S_{870\mu}=8.8\pm1.2$ mJy). Therefore GS4a is a robust counterpart for the AzTEC source although it is only a tentative counterpart for the LABOCA source \citep{biggs11}.  This is another high-z candidate source with $z_{opt}=3.37\pm0.25$ and $z_{MR}=3.53^{+0.57}_{-1.27}$.
\medskip

\noindent{\bf AzTEC/GS5.}
There is a single tentative counterpart within the 7.1\arcsec\ search radius from the AzTEC/GS5 position. However, the AzTEC contours are elongated in the east-west direction, joining the two VLA radio sources GS5a \& GS5b.  Both radio sources have red IRAC counterparts, and both sources may contribute to the AzTEC emission.  The 870 \micron\ LABOCA source LESS~J033150.8$-$274438 ($S_{870\mu}=3.9\pm1.4$ mJy) is located only 4.1\arcsec\ away from GS5a (also a \Chandra\ X-ray source), and \citet{biggs11} and \citet{chapin11} also identify GS5a as the secure counterpart to the LABOCA source. \citet{casey11} reported a spectroscopic redshift of $z=1.599$ for GS5a based on some absorptions features, and this redshift is further supported by the 9 hr long integration VLT spectrum by \citet{silverman10}.  
\medskip

\noindent{\bf AzTEC/GS6.}
The IRAC/MIPS source GS6b is located only 0.4\arcsec\ from the AzTEC/GS6 centroid.  However, its blue IRAC color ($[3.6\mu$m$]$ $-$ $[4.5\mu$m$]$ = $-$0.23) and spectroscopic redshift of $z=1.102$ (Stern et al., in prep) suggest that GS6b is likely a foreground object.  At a distance of 12.4\arcsec, the IRAC/MIPS source GS6a is located outside the 7.5\arcsec\ formal search radius for a counterpart, but it has very red IRAC color ($[3.6\mu$m$]$ $-$ $[4.5\mu$m$]$ = +0.45) and is a faint radio source ($S_{1.4GHz}=31.0\pm6.3$ $\mu$Jy; $P_{1.4}$ = 0.268). \citet{biggs11} identify GS6a as the robust counterpart for the 870 \micron\ LABOCA source LESS J033225.7$-$275228 ($S_{870\mu}=5.8\pm1.4$ mJy) located 6.2\arcsec\ away, and this galaxy is likely the primary counterpart to the AzTEC source as well.
\medskip

\noindent{\bf AzTEC/GS7.}
The red IRAC/MIPS source GS7a is the only radio source within the 8.7\arcsec\ search radius ($S_{1.4GHz}=51.2\pm6.4$ $\mu$Jy; $P_{1.4}$ = 0.126).  Therefore, it is considered a tentative counterpart to AzTEC/GS7.  The spectroscopic redshift of GS7a, which is also identified as a \Chandra\ X-ray source, is $z=2.676$, in excellent agreement with its radio-mm photometric redshift of $z_{MR}=2.56^{+0.52}_{-0.92}$.  The 870 \micron\ LABOCA source LESS J033213.6$-$275602 ($S_{870\mu}=9.1\pm1.2$ mJy) is located only 4.2\arcsec\ away, and \citet{biggs11} also identify GS7a as a robust counterpart. A second possible counterpart, GS7b, is a red IRAC/MIPS source ($P_{24\mu}=0.151$, $P_{color}=0.168$) located on the other side of the AzTEC centroid and may contribute to the observed 1100 \micron\ emission.
\medskip

\noindent{\bf AzTEC/GS8.}
There is a single robust radio counterpart GS8a, 4.4\arcsec\ from the AzTEC centroid position with $P_{1.4}=0.038$.  The IRAC/MIPS counterpart GS8a also has a red IRAC color ($[3.6\mu$m$]$ $-$ $[4.5\mu$m$]$ = +0.33) and relatively bright MIPS 24 \micron\ emission ($S_{24\mu}=620$ $\mu$Jy; $P_{24\mu}=0.203$).  Both \citet{chapin11} and \citet{biggs11} have identified this $z=2.252$ galaxy as the counterpart to the 870 \micron\ LABOCA source LESS J033205.1$-$274652 ($S_{870\mu}=7.5\pm1.2$ mJy), located only 7.7\arcsec\ away from AzTEC position.  The second IRAC/MIPS source GS8b, located 2.9\arcsec\ from the AzTEC position, is also a plausible MIPS 24 \micron\ counterpart ($P_{24\mu}=0.089$) with red IRAC color and is also a \Chandra\ X-ray source.  
\medskip

\noindent{\bf AzTEC/GS9.}
The single tentative radio counterpart ($S_{1.4}=86.8\pm6.6$ $\mu$Jy; $P_{1.4}=0.070$) is associated with a red IRAC/MIPS source GS9a, located 6.2\arcsec\ from the AzTEC centroid position.  It is also a \Chandra-detected X-ray source and should be considered a strong candidate for the AzTEC counterpart.  Located only 3.5\arcsec\ away from the AzTEC centroid, the IRAC/MIPS source GS9b is an intriguing alternate counterpart candidate given its red IRAC color ($[3.6\mu$m$]$ $-$ $[4.5\mu$m$]$ = +0.29) and MIPS 24 \micron\ emission.  If GS9b is the primary source of 1100 \micron\ continuum emission, then it is likely to be a high redshift system as its radio non-detection requires $z_{MR}>3.3$.  Slightly further away is GS9c, also a tentative red IRAC source.  No nearby source is found in the 870 \micron\ LABOCA catalog, but the LABOCA map shows a $S/N=3.0$ peak ($3.4\pm1.2$ mJy) at the position of GS9a.
\medskip

\noindent{\bf AzTEC/GS10.}
There is a single robust radio counterpart to AzTEC/GS10, located 5.3\arcsec\ from the AzTEC centroid ($S_{1.4}=89.3\pm6.4$ $\mu$Jy; $P_{1.4}=0.053$).  Its red IRAC/MIPS counterpart GS10a ($[3.6\mu$m$]$ $-$ $[4.5\mu$m$]$ = +0.14) has a reported spectroscopic redshift of $z=0.0338$ in the GOODS/ESO VIMOS DR1 catalog \citep{popesso09}, but the same group revised its redshift to $z=2.035$ using new data \citep{balestra10}. This revised spectroscopic redshift is in excellent agreement with our photometric redshift, $z_{MR}=2.03^{+0.41}_{-0.75}$. Another red IRAC source GS10b ($[3.6\mu$m$]$ $-$ $[4.5\mu$m$]$ = +0.04; $P_{color}=0.077$), located only 3.2\arcsec\ away from the AzTEC position, is not detected in the radio or by MIPS.  The 870 \micron\ LABOCA source LESS J033207.6$-$275123 ($S_{870\mu}=7.6\pm1.3$ mJy) is located only 4.3\arcsec\ away from GS10a, and it is also identified as a robust LABOCA counterpart by \citet{biggs11}. 
\medskip

\noindent{\bf AzTEC/GS11.}
There is a single tentative VLA radio counterpart for this source located 6.5\arcsec\ from the AzTEC centroid ($S_{1.4}=46.0\pm6.4$ $\mu$Jy; $P_{1.4}=0.081$), and it is also a \Chandra\ X-ray source.  Its IRAC/MIPS counterpart GS11a has a slightly blue color ($[3.6\mu$m$]$ $-$ $[4.5\mu$m$]$ = $-$0.02), but the VLT ISAAK $K-$band image \citep{retzlaff10} shows that this IRAC source is a blend of an optically bright ($i=21.7$) $z=0.246$ foreground source and an optically faint source second source, which is the radio source.  The 250 \micron\ BLAST source 109, located $\sim$30\arcsec\ southeast of the AzTEC centroid, is undetected at longer wavelength bands and is identified with a $z=0.124$ foreground disk galaxy \citep{dunlop10}.  Our photometric redshift for GS11a, $z_{MR}=2.50^{+0.52}_{-0.88}$, is completely inconsistent with this identification, and the proposed BLAST counterpart is unlikely to be related to the AzTEC source.  No nearby source is found in the 870 \micron\ LABOCA catalog, but the LABOCA map shows a $S/N\sim3$ peak ($3.5\pm1.2$ mJy) at the position of GS11a.
\medskip

\noindent{\bf AzTEC/GS12.}
The most likely counterpart candidate is a faint radio, red IRAC source GS12a located 4.0\arcsec\ away from the AzTEC position with $[3.6\mu$m$]$ $-$ $[4.5\mu$m$]$ = +0.10.  This $z=4.762$ galaxy was also identified as the counterpart to the LABOCA 870 \micron\ survey source LESS J033229.3$-$275619 ($S_{870\mu}=5.1\pm1.4$ mJy) by \citet{coppin09} based on its proximity to the LABOCA position and the presence of a $\sim3\sigma$ radio source.  Redshifted CO emission at $z_{CO}=4.755$ has been reported by \citet{coppin10b}, lending further support for the counterpart identification.
\medskip

\noindent{\bf AzTEC/GS13.}
There is a high concentration of IRAC/MIPS sources with spectroscopic redshifts between 1.0 and 1.6 in the region surrounding AzTEC/GS13.  The most likely counterpart for AzTEC/GS13 is a red IRAC/MIPS source GS13a ([3.6$\mu$m] $-$ [4.5$\mu$m] = +0.24), located only 2.1\arcsec\ away from the AzTEC centroid.  This source is also a faint radio source with no known spectroscopic redshift. A second IRAC/MIPS sources GS13b is a \Chandra\ X-ray source at $z=1.039$ \citep{mignoli05}, but its statistical likelihood of being the AzTEC counterpart is lower (see Table~\ref{tab:ID}). A third candidate counterpart, GS13c, is a faint radio source, but it has a blue IRAC color ([3.6$\mu$m] $-$ [4.5$\mu$m] = -0.25) and is therefore likely a foreground source. The 250 \micron\ BLAST source 193 is located $\sim$25\arcsec\ south of the AzTEC position.  Although the low density of the AzTEC and BLAST sources make the chance coincidence of these two sources even at such a substantial separation small, few plausible candidates are found within the BLAST position error circle.  No nearby source is found in the 870 \micron\ LABOCA catalog, but the LABOCA map shows a $S/N\sim3$ peak ($3.6\pm1.2$ mJy) near the AzTEC centroid position.
\medskip

\noindent{\bf AzTEC/GS14.}
There are no radio sources within the 9.0\arcsec\ search radius. The only tentative counterpart is a faint, red IRAC source GS14a ([3.6$\mu$m] $-$ [4.5$\mu$m] = +0.04; $P_{color}=0.083$) located only 3.1\arcsec\ from AzTEC/GS14. Although no nearby source is found in the 870 \micron\ LABOCA catalog, the LABOCA map shows a $S/N\sim3$ peak ($3.3\pm1.2$ mJy) nearly coincident with this position. The previously unpublished redshift of $z=3.640$ (Keck LRIS spectrum, H. Spinrad, priv. comm.) for GS14a is consistent with its photoz $z_{opt}=3.50\pm0.30$ and the non-detection in the radio and the MIPS 24 \micron\ bands, similar to the $z=4.762$ AzTEC/LABOCA source GS12a.     
\medskip

\noindent{\bf AzTEC/GS15.}
There are several faint, red IRAC/MIPS sources within the 9.0\arcsec\ search radius, although they are individually not particularly compelling.  The faint radio source GS15a, although 12.7\arcsec from the AzTEC centroid, is also a \Chandra\ X-ray source and is located 1.4\arcsec from the centroid of the LABOCA 870 \micron\ source LESS J033151.5$-$274552 ($S_{870\mu}=4.2\pm1.4$ mJy), and GS15a is the secure LABOCA counterpart \citep{biggs11}. 
\medskip

\noindent{\bf AzTEC/GS16.}
There are two faint radio sources within the counterpart search radius of 10.5\arcsec.  The red IRAC/MIPS source GS16a ([3.6$\mu$m] $-$ [4.5$\mu$m] = +0.53) is 6.1\arcsec\ away from the AzTEC position, and it is a tentative radio counterpart with $P_{1.4}=0.072$.  GS16a is also an X-ray source detected by \Chandra\ and has a spectroscopic redshift of $z=1.719$ (Silverman et al., in prep.).  The second radio source GS16b is located slightly further away, 7.9\arcsec.  Although it is a brighter MIPS 24 \micron\ source, it has a blue IRAC color ([3.6$\mu$m] - [4.5$\mu$m] = $-$0.29) and a spectroscopic redshift of $z=1.017$ \citep{mignoli05}, suggesting it is a foreground source.  No nearby source is found in the 870 \micron\ LABOCA catalog, but the LABOCA map shows a $S/N\sim2$ peak near the position of G16a.  
\medskip

\noindent{\bf AzTEC/GS17.}
Two plausible IRAC/MIPS sources are found within the 10.5\arcsec\ search radius. The IRAC/MIPS source GS17a is a faint radio source with $P_{1.4} = 0.007$, making it formally a robust identification.  However, it has a blue IRAC color with [3.6$\mu$m] $-$ [4.5$\mu$m] = $-$0.27 and is thus likely a foreground object($z_{opt}=1.01\pm0.10$). The IRAC/MIPS source GS17b is another robust identification based on the MIPS detection ($P_{24\mu}=0.026$) and red IRAC color ([3.6$\mu$m] $-$ [4.5$\mu$m] = +0.19, $P_{color}=0.021$), and this \Chandra\ detected X-ray source has a photometric redshift of $z=2.66$ \citep{silverman10}. Though slightly further from the AzTEC centroid, GS17c is an interesting alternative possibility: this red IRAC/MIPS source ([3.6$\mu$m] $-$ [4.5$\mu$m] = +0.36) has optical colors satisfying the BzK criteria for star forming galaxies at high redshifts. There is also a close pair of radio sources, GS17d and GS17e, located 12\arcsec north of the AzTEC position; though they are outside the nominal search radius, both have red IRAC colours and remain plausible counterparts to the AzTEC source.  No nearby source is found in the 870 \micron\ LABOCA catalog, but the LABOCA map shows a $S/N\sim3$ emission peak between GS17b and the two radio sources GS17d \& GS17e.
\medskip

\noindent{\bf AzTEC/GS18.}
There are two faint radio sources within the 9.3\arcsec\ search radius of AzTEC/GS18. The closest IRAC/MIPS source GS18a is formally a robust counterpart owing to its proximity to the AzTEC position and its very red IRAC color ([3.6$\mu$m] $-$ [4.5$\mu$m] = +0.47, $P_{color}=0.113$). Additionally, GS18b is a $z=2.688$ galaxy which is an X-ray source detected by \Chandra\ \citep{akiyama05} and has a red IRAC color ([3.6$\mu$m] $-$ [4.5$\mu$m] = +0.31), making it a plausible candidate for the counterpart to AzTEC/GS18. A third radio-faint IRAC/MIPS source GS18c is a tentative candidate based on its MIPS 24 \micron\ emission, but it has very blue colour ([3.6$\mu$m] $-$ [4.5$\mu$m] = $-$0.39) and is likely a foreground object. The 870 \micron\ LABOCA source LESS J033243.6$-$274644 ($S_{870\mu}=6.4\pm1.3$ mJy) is located $\sim$8.4\arcsec\ south of the AzTEC position, putting GS18a between the AzTEC and LABOCA centroids.  \citet{biggs11} identified GS18a as a tentative (and only) counterpart to the LABOCA source.  This extremely faint optical source ($i=28.1$, $K>24.5$) is another high-z candidate with $z_{MR}= 3.00^{+0.56}_{-1.14}$.
\medskip

\noindent{\bf AzTEC/GS19.}
A pair of radio sources, GS19a \& GS19b (with $P_{1.4} = 0.102$ and $P_{1.4} = 0.143$, respectively), are strongly favored as the counterparts to AzTEC/GS19 when all statistical measures are taken into account. They form a blended source in the MIPS 24 \micron\ band, and both have red IRAC colors. It is possible that these two sources are physically related and both contribute to the AzTEC emission.  No spectroscopic redshift is available for either sources while their photometric redshifts are quite similar ($z_{opt}= 1.83\pm0.35~\&~ 2.08\pm0.25$).  At $z=2.0$, their projected separation of 3.2\arcsec\ corresponds to 27 kpc. There is a third IRAC/MIPS counterpart, GS19c, that is a tentatively counterpart with a lower likelihood ($P_{24\mu}=0.181$). No nearby source is found in the 870 \micron\ LABOCA catalog, but the LABOCA map shows a $S/N\sim2.4$ emission peak near GS19a \& GS19b.
\medskip

\noindent{\bf AzTEC/GS20.}
The AzTEC contours are well centered and follow the light profile of the $z=0.0369$ galaxy GS20a, which is also a bright and well resolved radio, MIPS 24 \& 70 \micron, and X-ray source.  This source is quite blue ([3.6$\mu$m] $-$ [4.5$\mu$m] = $-$0.43) and is obviously a bright foreground galaxy.  It is the brightest BLAST 250 \micron\ source within the GOODS-South survey field proper, and \citet{dunlop10} argue that this foreground galaxy is the correct counterpart for the BLAST source. It is possible that the 1.1~mm emission originates from the cold dust associated with this spiral galaxy (as seen in the HST images), but it is difficult to reproduce the entire measured spectral energy distribution from $\lambda=1$ \micron\ to 20 cm using a reasonable set of assumptions on dust temperature, IR luminosity, and star formation rate for this low redshift galaxy.  Instead, the AzTEC emission may originate from a background object, possibly lensed by this foreground galaxy, similar to AzTEC J100008.05+022612.2 imaged at high angular resolution by \citet{younger07}. 
\medskip

\noindent{\bf AzTEC/GS21.}
The single tentative radio counterpart (GS21a) has [3.6$\mu$m] $-$ [4.5$\mu$m] = +0.24 and a spectroscopic redshift of $z=1.910$ \citep{vanzella08}.  The position centroid of the BLAST 250 \micron\ source 861 is displaced by $\sim$7\arcsec\ to the northeast, but GS21a is still within the error-circle of the BLAST source. No nearby source is found in the 870 \micron\ LABOCA catalog, but the LABOCA map shows a $S/N\sim1.8$ emission peak near GS21a. A second tentative counterpart GS21b is a red IRAC/MIPS galaxy located just 2.5\arcsec away from the AzTEC centroid.
\medskip

\noindent{\bf AzTEC/GS22.}
The faint radio source GS22a, located 7.8\arcsec\ away from the AzTEC centroid, is the most likely counterpart ($S_{1.4}=34.6\pm6.5$ $\mu$Jy).  This galaxy is also a red IRAC/MIPS source ([3.6$\mu$m] $-$ [4.5$\mu$m] = +0.30) and has a spectroscopic redshift of $z=1.794$ \citep{wuyts09}. The 250 \micron\ BLAST source 552 position centroid is $\sim$10\arcsec\ east of the AzTEC position, and \citet{dunlop10} also identified the radio source GS22a (located $\sim$7\arcsec\ away from the BLAST position) as the likely counterpart. A second red IRAC/MIPS candidate counterpart, GS22b, is closer to the AzTEC centroid and is an interesting alternative.  No nearby source is found in the 870 \micron\ LABOCA catalog, but the LABOCA map shows a $S/N\sim3.1$ emission peak near this red IRAC/MIPS source GS22b.
\medskip

\noindent{\bf AzTEC/GS23.}
The faint radio source GS23a ($S_{1.4}=23.4\pm6.5$ $\mu$Jy; $P_{1.4}=0.030$) is a robust counterpart, and its red IRAC color ([3.6$\mu$m] $-$ [4.5$\mu$m] = +0.45) adds to the high likelihood of being the correct counterpart. A second tentative radio counterpart, GS23b is only slightly further from the AzTEC centroid, and it is a red IRAC/MIPS source with a spectroscopic redshift of $z=2.277$ \citep{chapin11}. Both GS23a \& GS23b are within the beam area of the 870 \micron\ LABOCA source LESS J033221.3$-$275623 ($S_{870\mu}=4.7\pm1.4$ mJy) and the 250 \micron\ BLAST source 158.  \citet{dunlop10} and \citet{biggs11} have identified GS23b as the counterpart consisting of ``an extremely complex faint system'' at optical wavelengths.
\medskip

\noindent{\bf AzTEC/GS24.}
There is a single tentative faint radio counterpart within the 12.2\arcsec\ counterpart search radius, GS24a. This red IRAC/MIPS source is the most likely counterpart to AzTEC/GS24. The 250 \micron\ BLAST source 104 straddles GS24a and a $z=2.578$ type~2 QSO J033235.78$-$274916.82 \citep[][also detected at MIPS 70 \micron]{rigopoulou09}, and both sources likely contribute to the 250 \micron\ continuum.  \citet{dunlop10} instead identified the $z=0.547$ radio source located at the edge of the AzTEC and BLAST beam based on the radio $P-$statistic.  No nearby source is found in the 870 \micron\ LABOCA catalog, but the LABOCA map shows a $S/N\sim3$ emission peak near the $z=2.578$ type~2 IR QSO J033235.78$-$274916.82 and a secondary $S/N\sim2$ emission peak near GS24a.  The AzTEC contours are elongated in the north-south direction, and this may be another example of a blended source. 
\medskip

\noindent{\bf AzTEC/GS25.}
The red IRAC/MIPS source associated with radio emission GS25a is located only 6.8\arcsec\ away from the AzTEC centroid position. This galaxy, also detected in the X-ray by \Chandra, has a spectroscopic redshift of $z=2.292$ \citep{popesso09}. The 870 \micron\ LABOCA source LESS J033246.7$-$275120 ($S_{870\mu}=5.9\pm1.3$ mJy) is well-centered on GS25a, and \citet{biggs11} also identify the same galaxy as the robust LABOCA counterpart. 
\medskip

\noindent{\bf AzTEC/GS26.}
AzTEC/GS26 has no radio counterpart within the 12.2\arcsec\ search radius of the AzTEC position, but there are five red IRAC sources.  Of these, GS26a ([3.6$\mu$m] $-$ [4.5$\mu$m] = +0.22; $P_{color}=0.237$) located 5.5\arcsec\ away is the most probable counterpart to this AzTEC source. 
GS26b and GS26c are other red IRAC sources located slightly further away ($\sim7$\arcsec). No nearby source is found in the 870 \micron\ LABOCA catalog, but the LABOCA map shows a $S/N\sim2.9$ emission peak near the $z=2.331$ MIPS galaxy J033216.3$-$274343.4 with a red IRAC color ([3.6$\mu$m] $-$ [4.5$\mu$m] = +0.35), located $\sim7.5$\arcsec\ away from the AzTEC position.
\medskip

\noindent{\bf AzTEC/GS27.}
The red IRAC/MIPS source GS27a ([3.6$\mu$m] $-$ [4.5$\mu$m] = +0.38) is located 13.0\arcsec\ away from the AzTEC centroid, just within the counterpart search radius of 13.0\arcsec, and is associated with weak radio emission ($S_{1.4}=23.6\pm6.5$ $\mu$Jy) from a $z=2.577$ \citep{popesso09} galaxy. However, formally this is not a secure identification ($P_{1.4} > 0.20$) owing to its large separation from the AzTEC centroid. Still, this is the most plausible counterpart within the proximity of AzTEC/GS27.  No nearby source is found in the 870 \micron\ LABOCA catalog, but the LABOCA map shows a $S/N\sim3.1$ emission peak near this red IRAC/MIPS source GS27a.
\medskip

\noindent{\bf AzTEC/GS28.}
AzTEC/GS28 has no radio source within the 13.0\arcsec\ counterpart search radius.  The red IRAC sources GS28a ([3.6$\mu$m] $-$ [4.5$\mu$m] = +0.55; $P_{color}=0.073$) and GS28b ([3.6$\mu$m] $-$ [4.5$\mu$m] = +0.07; $P_{color}=0.136$) are the most likely counterparts.
No nearby source is found in the 870 \micron\ LABOCA catalog, but the LABOCA map shows an extended source with a $S/N\sim3.0$ emission peak near the AzTEC source position, and this may be another blended source.
\medskip

\noindent{\bf AzTEC/GS29.}
There are no compelling counterparts to AzTEC/GS29. The closest IRAC/MIPS source GS29a (3.8\arcsec\ away) is very blue ([3.6$\mu$m] $-$ [4.5$\mu$m] = $-$0.43) and has a spectroscopic redshift of $z=0.577$, so this is likely a foreground source.  The next closest source GS29b is a $z=2.340$ galaxy located 4.6\arcsec\ away, but it is also blue ([3.6$\mu$m] $-$ [4.5$\mu$m] = $-$0.10). There are two radio sources located just outside the 13.0\arcsec search radius, but 
they both have relatively high probabilities of being false associations.
No nearby source is found in the 870 \micron\ LABOCA catalog, but the LABOCA map shows an extended source with two $S/N\sim$2-3 emission peaks near the red IRAC/MIPS sources J033158.65$-$274516.3 and J033200.13$-$274453.2.
\medskip

\noindent{\bf AzTEC/GS30.}
The brightest MIPS 24 \micron\ source within the counterpart search radius of 13.5\arcsec\ ($S_{24\mu}=459\pm 6 \mu$Jy; $P_{24\mu}=0.44$) is also a faint radio source ($S_{1.4}=37.2\pm6.2\mu$Jy; $P_{1.4}=0.082$) with a red IRAC color ([3.6$\mu$m] $-$ [4.5$\mu$m] = +0.19). Therefore, GS30a is a tentative but the most plausible counterpart for AzTEC/GS30.  The spectroscopic redshift of this galaxy is yet unknown. A second radio source (GS30b) and a radio-faint IRAC source (GS30c) are also tentative counterparts to AzTEC/GS30, both with red IRAC colors. No nearby source is found in the 870 \micron\ LABOCA catalog, but the LABOCA map shows an extended source with a $S/N=2.9$ centered on the AzTEC centroid position.
\medskip

\noindent{\bf AzTEC/GS31.}
Both of the two bright MIPS 24 \micron\ sources (GS31b) within the 13.6\arcsec\ search radius region are associated with faint radio emission. The western source GS31a however is closer (2.7\arcsec) to the AzTEC centroid, making it a robust counterpart candidate ($P_{1.4}=0.015$).  The eastern source GS31b is located 7.9\arcsec\ away and is a tentative counterpart ($P_{1.4}=0.118$). It is possible that both sources contribute to the 1.1~mm emission detected by AzTEC. They both have {\it blue} IRAC colors ([3.6$\mu$m] $-$ [4.5$\mu$m] = $-$0.25 and $-$0.36 for GS31a and GS31b, respectively) however, and the true counterpart may be a much fainter source located between or behind these sources. The spectroscopic redshift of GS31a is $z=1.843$ \citep{wuyts09}. 
No nearby source is found in the 870 \micron\ LABOCA catalog, but the LABOCA map shows an elongated north-south structure with a $S/N=2.8$ peak, similar to the AzTEC source morphology, and it is likely another blended source.
\medskip

\noindent{\bf AzTEC/GS32.}
There are three radio counterparts within the 13.5\arcsec\ search radius of the AzTEC centroid, and all three are also MIPS 24 \micron\ sources.  The nearest radio source GS32a ($S_{1.4}=30.3\pm6.8~\mu$Jy, $P_{1.4}=0.162$; and $S_{24\mu}=371.1\pm11.7~\mu$Jy, $P_{24\mu}=0.421$), located 9.6\arcsec\ away from the AzTEC centroid, is the primary candidate by its proximity.  The source GS32b is similarly bright in radio and MIPS 24\micron\ bands, but it is slightly further away, 10.5\arcsec, from the AzTEC centroid position. Both GS32a and GS32b are quite blue, however; we include GS32c, a faint radio galaxy with red IRAC color, in the catalog as an alternative, although it is located 13.4\arcsec\ from the AzTEC centroid. No nearby source is found in the 870 \micron\ LABOCA catalog, and no significant emission peak is found at this location in the LABOCA map.
\medskip

\noindent{\bf AzTEC/GS33.}
The faint IRAC source GS33a located 7.4\arcsec\ from the AzTEC centroid position is associated with a weak radio source ($S_{1.4}=28.6\pm6.2~\mu$Jy). It is also a weak MIPS 24 \micron\ source and has a red IRAC color ([3.6$\mu$m] $-$ [4.5$\mu$m] = +0.34; $P_{color}=0.390$).  Therefore this galaxy is a tentative candidate counterpart for AzTEC/GS33. No nearby source is found in the 870 \micron\ LABOCA catalog, but the LABOCA map shows an isolated emission peak with $S/N\sim2.3$ centered near GS33a.
\medskip

\noindent{\bf AzTEC/GS34.}
There is a high density of IRAC sources in this field, but few are located in the immediate vicinity of the AzTEC centroid position.  There are no robust or tentative counterparts to AzTEC/GS34: $P > 0.20$ for all sources within the 13.5\arcsec search radius.  No nearby source is found in the 870 \micron\ LABOCA catalog, but the LABOCA map shows an emission peak with a $S/N\sim3.5$ centered near the $z=1.356$ (Silverman et al., in prep) faint radio and MIPS source GS34a.
\medskip

\noindent{\bf AzTEC/GS35.}
The red IRAC source GS35a, located 2.0\arcsec\ away from the AzTEC source centroid position, has a robust radio counterpart ($S_{1.4}=41.3\pm 6.7 \mu$Jy; $P_{1.4}=0.008$).  Its MIPS emission ($P_{24\mu}=0.027$) and red IRAC color ([3.6$\mu$m] $-$ [4.5$\mu$m] = +0.37; $P_{color}=0.022$) make this galaxy a robust counterpart to the AzTEC source. Its spectroscopic redshift is unknown. The second red IRAC source GS35b located 4.9\arcsec\ away from the AzTEC position ($P_{color}=0.191$) is undetected at radio or MIPS 24 \micron\ bands.  No nearby source is found in the 870 \micron\ LABOCA catalog, and the LABOCA map shows an emission peak with a $S/N\sim2.1$ centered near the $z=0.734$ radio-loud QSO J033227.00$-$274105.0.
\medskip

\noindent{\bf AzTEC/GS36.}
No radio source is detected within the 13.5\arcsec\ counterpart candidate search radius, and the galaxy GS36a is the only IRAC source with a red IRAC color, [3.6$\mu$m] $-$ [4.5$\mu$m] = +0.68 ($P_{color}=0.304$). So this IRAC source is the most promising counterpart candidate, although it has a high probability of false association. Few other potential candidates are present in this field. No nearby source is found in the 870 \micron\ LABOCA catalog, and the LABOCA map does not show any significant emission peak within the search radius.
\medskip

\noindent{\bf AzTEC/GS37.}
The red IRAC/MIPS source GS37a with [3.6$\mu$m] $-$ [4.5$\mu$m] = +0.08 is associated with faint radio emission ($S_{1.4}=20.5\pm6.4 \mu$Jy), though it is not formally a tentative counterpart candidate for AzTEC/GS37.  There is a single tentative red IRAC source 4.3\arcsec\ from the AzTEC centroid, which is possibly at high redshift ($z>3$) given its non-detection at 1.4\,GHz.  No nearby source is found in the 870 \micron\ LABOCA catalog, but the LABOCA map shows a ridge of 2.1-2.7$\sigma$ peaks running along the similar ridge found in the AzTEC map (including GS37a), suggesting this is likely another blended source.  
\medskip

\noindent{\bf AzTEC/GS38.}
There is a single tentative radio counterpart (GS38a; $P_{1.4}=0.116$), which is also a \Chandra\ X-ray source.  It has a very blue IRAC color ([3.6$\mu$m] $-$ [4.5$\mu$m] = $-$0.55) and a spectroscopic redshift of $z = 0.735$ \citep{vanzella08}.  Therefore GS38a is likely a foreground object.
No nearby source is found in the 870 \micron\ LABOCA catalog, but the LABOCA map shows a 2.3$\sigma$ peak near the red IRAC/MIPS source GS38b, located 7.4\arcsec\ away from the AzTEC centroid ($P_{24\mu}=0.453$, $P_{color}=0.386$). 
\medskip

\noindent{\bf AzTEC/GS39.}
There is a single tentative radio counterpart (GS39a; $P_{1.4}=0.083$) which is also a red IRAC source ([3.6$\mu$m] $-$ [4.5$\mu$m] = +0.55). The position centroid of the 870 \micron\ LABOCA source LESS J033154.4$-$274525 ($S_{870\mu}=3.8\pm1.4$ mJy) is offset from GS39a by only 4.9\arcsec.  \citet{biggs11} also identify GS39a as a tentative LABOCA counterpart, and GS39a is a strong counterpart candidate for AzTEC/GS39. 
Another tentative counterpart, GS39b, is a MIPS source located only 2.8\arcsec from the AzTEC centroid, but given its blue color, it is likely a foreground object.
\medskip

\noindent{\bf AzTEC/GS40.}
There are no compelling counterparts to AzTEC/GS40. We list a single IRAC source within the 15.0\arcsec search radius. GS40a is 10.2\arcsec away, and is slightly blue with a high probability of being a false association.
No nearby source is found in the 870 \micron\ LABOCA catalog, and the LABOCA map shows only a $2\sigma$ emission peak located $\sim10$\arcsec\ northwest of the AzTEC position.
\medskip

\noindent{\bf AzTEC/GS41.}
The radio sources GS41a and GS41b are both promising counterparts to AzTEC/GS41 with $P_{1.4}=0.120$ \& 0.132.  They are also red IRAC sources with [3.6$\mu$m] $-$ [4.5$\mu$m] = +0.23 \& +0.30 although they are just far enough away from the AzTEC centroid to make them tentative counterparts only.  Two other red IRAC sources GS41c and GS41d are slightly closer to the AzTEC centroid and are tentative counterparts as well.  The centroid of the 870 \micron\ LABOCA source LESS J033302.5$-$275643 ($S_{870\mu}=12.0\pm1.2$ mJy) is located closest to GS41b and GS41d, and \citet{biggs11} identify GS41b as the robust counterpart to the LABOCA source.    
\medskip

\noindent{\bf AzTEC/GS42.}
There are no radio counterparts within the 6.9\arcsec\ search radius of AzTEC/GS42. The only tentative counterpart is GS42a, a red IRAC source ([3.6$\mu$m] $-$ [4.5$\mu$m] = +0.04) that is also well-matched to the position of the LABOCA source LESS J033314.3$-$275611 ($S_{870\mu}=14.5\pm1.2$ mJy). The AzTEC contours are extended in the north-south direction, and \citet{weiss09} modeled this source with a second, fainter component (LESS J033313.0$-$275556, $S_{870\mu}=4.3\pm1.4$ mJy). 
\citet{biggs11} identified adjacent red IRAC/MIPS source J033314.41$-$275612.0 as the robust LABOCA counterpart.
  
\medskip

\noindent{\bf AzTEC/GS43.}
The only tentative counterpart is a red IRAC source GS43a, which has a red IRAC color ([3.6$\mu$m] $-$ [4.5$\mu$m] = +0.23) and is also centered on the 870 \micron\ LABOCA source LESS J033302.9$-$274432 ($S_{870\mu}=6.7\pm1.3$ mJy). Since this object is undetected in the radio and MIPS 24 \micron, it is also likely a high redshift object ($z_{MR}>4.1$). \citet{biggs11} did not find any robust or tentative counterpart for the LABOCA source. The AzTEC contours are significantly elongated in the east-west direction, suggesting this is a blend of more than one source. \citet{weiss09} model this source with a fainter second component, LESS J033303.9$-$274412 ($S_{870\mu}=5.3\pm1.4$ mJy).  
\medskip

\noindent{\bf AzTEC/GS44.}
There is only one VLA radio source found within the 10.4\arcsec\ search radius centered on the AzTEC peak position.  This source GS44a has a flat IRAC color, [3.6$\mu$m] $-$ [4.5$\mu$m] = $-$0.07 and is a modest MIPS 24 \micron\ source with $S_{24\mu}=143.1\pm9.3$ $\mu$Jy.  The 870 \micron\ LABOCA source LESS J033240.4$-$273802 ($S_{870\mu}=5.0\pm1.4$ mJy) is found within the AzTEC search radius, but its centroid is offset from GS44a by 12\arcsec, just outside the nominal beam area of LABOCA.  Another radio source J033239.14$-$273810.5 at $z=0.830$ \citep{lefevre04} is a better candidate for the LABOCA counterpart, but this radio source is located 24\arcsec\ away from the AzTEC centroid, which is well outside the nominal search radius for the AzTEC source. We list the LABOCA flux in the Table, but note the large separation between this source and GS44a.
\medskip

\noindent{\bf AzTEC/GS45.}
Two red IRAC/MIPS sources (GS45b \& GS45c) are found within 2\arcsec\ of the AzTEC centroid, and they are the most likely candidate counterparts to the AzTEC source, primarily by their proximity. The 870 \micron\ LABOCA source LESS J033218.9$-$273738 ($S_{870\mu}=8.1\pm1.2$ mJy) is found within the AzTEC search radius, half way between GS45b and GS45a, the latter of which is a bright MIPS 24 \micron\ source and a VLA radio continuum source with a rather blue IRAC color ([3.6$\mu$m] $-$ [4.5$\mu$m] = $-$0.32). A higher resolution K-band image shows that the brighter radio peak is associated with a fainter component, suggesting this is a blended source.  GS45b is 5.8\arcsec\ away from the LABOCA centroid, and it is a robust counterpart for both AzTEC and LABOCA sources \citep{biggs11}.
\medskip

\noindent{\bf AzTEC/GS46.}
No radio source is found within the 13.0\arcsec\ search radius.  The IRAC/MIPS source GS46a located 6.2\arcsec\ away from the AzTEC centroid is the tentative counterpart based on the combination of its red IRAC color ([3.6$\mu$m] $-$ [4.5$\mu$m] = $+$0.17; $P_{color}=0.255$) and MIPS 24 \micron\ emission ($P_{24\mu}=0.226$). The LABOCA source LESS J033157.2$-$275633 ($S_{870\mu}=4.8\pm1.4$ mJy) is located 23\arcsec from GS46a and thus also far from the AzTEC centroid. We list the LABOCA flux in the table but note this large separation.
\medskip

\noindent{\bf AzTEC/GS47.}
The faint VLA 1.4 GHz radio source GS47a ($S_{1.4}=43.2\pm7.0$ $\mu$Jy, $P_{1.4}=0.105$) is found 7.6\arcsec\ away from the AzTEC centroid, and it is also a red IRAC source ([3.6$\mu$m] $-$ [4.5$\mu$m] = $+$0.46) as well as a MIPS 24 \micron\ source. The 870 \micron\ LABOCA source LESS J033208.1$-$275818 ($S_{870\mu}=7.3\pm1.2$ mJy) coincides in position with AzTEC/GS47 within 3.4\arcsec\ of each other, making it a robust LABOCA counterpart as well \citep{biggs11}.
\medskip

\end{document}